\documentclass{aa}  

\usepackage{graphicx}
\usepackage{txfonts}
\usepackage[colorlinks]{hyperref}

\def\aap{A\&A}
\def\apj{ApJ}
\def\apjs{ApJS}

\def \hi {\ion{H}{i}}
\def\h2{H$_2$}

\def\kms{km\,s$^{-1}$}

\def\arcmin{\hbox{$^\prime$}}

\def\fdg{\hbox{$.\!\!^\circ$}}
\def\farcm{\hbox{$.\mkern-4mu^\prime$}}

\begin{document}

\title{Local \hi\ filaments driven by a small-scale dynamo}

\subtitle{Update on the velocity decomposition algorithm}

   \author{P.\ M.\ W.\ Kalberla \inst{1},  J.\ Kerp \inst{1} \and U.\ Haud \inst{2} }

\institute{Argelander-Institut f\"ur Astronomie,
           Auf dem H\"ugel 71, 53121 Bonn, Germany \\
           \email{pkalberla@astro.uni-bonn.de}
           \and
           Tartu Observatory, University of Tartu,
           61602 T\~oravere, Tartumaa, Estonia }

   \authorrunning{P.\,M.\,W. Kalberla, J. Kerp \& U.\ Haud } 

   \titlerunning{Coherence in \hi\ filaments}

   \date{Received 1 January 2021 / Accepted 7 August 2021}

  \abstract 
{\hi\ filaments are closely related to dusty magnetized structures that
  are observable in the far infrared (FIR). Recently it was proposed
  that the coherence of oriented \hi\ structures in velocity traces the
  line of sight magnetic field tangling. }
{We study the velocity-dependent coherence between FIR emission at 857
  GHz and \hi\ on angular scales of 18\arcmin.  }
{We use HI4PI \hi\ data and {\it Planck} FIR data and apply the Hessian
  operator to extract filaments. For coherence, we require that local
  orientation angles $\theta$ in the FIR at 857 GHz along the filaments
  be correlated with the \hi.}
{ We find some correlation for \hi\ column densities at $
  |v_{\mathrm{LSR}}| < 50 $ \kms,\ but a tight agreement between FIR and
  \hi\ orientation angles $\theta$ exists only in narrow velocity
  intervals of 1 \kms. Accordingly, we assign velocities to FIR
  filaments. Along the line of sight these \hi\ structures show a high
  degree of the local alignment with $\theta$, as well as in velocity
  space. Interpreting these aligned structures in analogy to the
  polarization of dust emission defines an \hi\ polarization.  We
  observe polarization fractions of up to 80\%, with averages of
  30\%. Orientation angles $\theta$ along the filaments, projected 
  perpendicular to the line of sight, are fluctuating systematically and
  allow a characteristic distribution of filament
  curvatures to be determined. }
{ Local \hi\ and FIR filaments identified by the Hessian analysis are
  coherent structures with well-defined radial
  velocities. \hi\ structures are also organized along the line of sight
  with a high degree of coherence. The observed bending of these
  structures in the plane of the sky is consistent with models for
  magnetic field curvatures induced by a Galactic small-scale turbulent
  dynamo.  }


  \maketitle
%

\section{Introduction}
\label{Intro}

The interstellar medium (ISM) is shaped by many processes. Energy is
injected by sources such as supernovae and stellar winds. This energy
affects the dynamics of the ISM: It stimulates turbulence that cascades
down from large to small scales. These processes have an
imprint on the dynamics and the distribution of diffuse neutral hydrogen
(\hi) as well as on the physical state of the \hi, which is characterized by the
composition in different phases.  Gas traced by \hi\ emission is not
completely neutral; as a consequence, the neutral phase is coupled to
ions and the dynamics of the \hi\ is linked to the magnetic field.\ This
process is usually described as flux freezing \citep{Heiles2005}. Gas
and dust are well mixed (\citealt{Clark2019a} and \citealt{Kalberla2020}).
One of the most important tracers of the magnetic field is polarization
from aligned dust grains, which emit the absorbed starlight in the
far infrared \citep[FIR;][]{Heiles2005}. A close relation between \hi\ and
FIR emission and the magnetic field is expected \citep{Clark2019b}.

Recent investigations have shown that the \hi\ at high Galactic
latitudes is organized in thin filamentary structures, sometimes
described as fibers \citep{Clark2014}. These filaments are associated
with FIR filaments \citep{Kalberla2016}. The morphology of these
filaments suggests that the structures have been shaped by the magnetic
field, and it has been shown that they are indeed extremely well aligned
with the plane-of-sky magnetic field as probed by both starlight
polarization \citep{Clark2014} and polarized dust emission
(\citealt{Clark2014} and \citealt{Kalberla2016}). Thus, \hi\ and FIR
structures are coupled to the magnetic field, and the morphology of these
structures may be used to study magnetism in this 
environment. The most recent publications in this field have shown that
\hi\ filaments are coherent structures, indicating ordered polarized
emission analogous to FIR polarization (\citealt{Clark2018} and
\citealt{Clark2019b}). In turn, these coherent FIR and \hi\ structures can
be used to constrain models for the three-dimensional orientation of the interstellar
magnetic field and the twisting of the magnetic flux tubes.

Magnetohydrodynamics (MHD) simulations by \citet{Planck2015b} are based
on a large-scale anisotropic component of the magnetic field, with an
additional turbulent component due to velocity perturbations imposed on
converging flows that drive turbulence. A different assumption is that
there is a uniform magnetic field with isotropic turbulent fields in
different three-dimensional layers \citep{Planck2020a}. We observe a
systematic bending of the filaments that can be used to probe models
on turbulence.  Magnetic fields exert a tension force that opposes
bending; they behave like elastic strings threading the
fluid. Accordingly, the field strength and the field-line curvature are
affected. Field tangling in the case of a small-scale turbulent dynamo, also
called a fluctuation dynamo \citep{St-Onge2018}, leads to distinct
predictions for the curvature distribution (\citealt{Schekochihin2002} and
\citealt{Schekochihin2004}) that can be verified.

Previous investigations of the coupling between the magnetic field and
FIR and \hi\ emission by \citet{Clark2019b} made use of the Rolling
Hough Transform (RHT), a technique from machine vision for detecting and
parameterizing linear structures \citep{Clark2014}. This method, as
applied by \citet{Clark2019b}, measures the linearity of structures in a
region of diameter 75\arcmin \ around each pixel, and so the effective
resolution is rather limited. These authors considered angular
resolutions between 80\arcmin\ and 160\arcmin.  Here we study the
relations between \hi\ and FIR structures at the highest possible
resolution. We use HI4PI survey data \citep{HI4PI2016} that have beam
widths between 10\farcm8 and 14\farcm5, the best currently available
resolutions for all-sky data. However, our Hessian analysis is pixel
based and therefore by construction limited to a resolution of
18\arcmin. This is still an improvement by about a factor of four to
nine, and we present, for the first time, an all-sky high resolution
study of \hi\ filaments. In Sect. \ref{Observations} we explain our data
reduction, the Hessian analysis, and the comparison of filamentary
structures in \hi\ and FIR at 857, 545, and 353 GHz.  We find coherence
between \hi\ and FIR and explore the physical conditions of these
\hi\ filaments in Sect. \ref{Coherence}.  Curvatures in the case of
magnetized filaments imply a magnetic tension with back reactions on the
filament shape. We determine the distribution of filament curvatures in
Sect. \ref{Curvature} and relate our results to predictions for the
small-scale turbulent dynamo. Changes in curvatures between the central
parts of the filaments and their environment are discussed in
Sect. \ref{Curve_dev}. We summarize our results in Sect. \ref{Summary}.
{\bf In the Appendix \ref{VDA} we consider the velocity
  decomposition algorithm (VDA) recently developed by \citet{Yuen2021}. }

\section{Observations and data reduction }
\label{Observations}

We use HI4PI \hi\ observations \citep{HI4PI2016}, combining data from
the Galactic all sky survey (GASS; \citet{Kalberla2015}), measured with
the Parkes radio telescope and the Effelsberg-Bonn \hi\ Survey (EBHIS;
\citealt{Winkel2016}) with data from the 100 m telescope. The \hi\ data
from both surveys have been gridded to independent nside = 1024 HEALPix
databases, retaining the original resolution for each of the telescopes
(14\farcm5 for the GASS and 10\farcm8 for the EBHIS). Subsequently
these \hi\ profiles were decomposed into Gaussian components
(\citealt{Haud2000}, \citealt{Kalberla2015}, and \citealt{Kalberla2018}).

To compare the \hi\ data with the FIR emission we used the most recent
maps from the Public Data Release 4 (PR4) \citep{Planck2020} at
frequencies of 857, 545, and 353
GHz\footnote[1]{HFI\_SkyMap\_857\_2048\_R4.00\_full.fits and
  HFI\_SkyMap\_545\_2048\_R4.00\_full.fits, and
  HFI\_SkyMap\_353-field-IQU\_2048\_R4.00\_full.fits from
  \url{https://irsa.ipac.caltech.edu/data/Planck/release_3/ancillary-data/HFI_Products.html}}.
These maps have been generated using the NPIPE data processing pipeline,
a natural evolution of previous Planck analysis efforts. PR4 data need
to be corrected for the Solar dipole. We used parameters from
\citet{Planck2020} with longitude 263\fdg986 $\pm$ 0\fdg035 and latitude
48\fdg247 $\pm$ 0\fdg023 and amplitude of 3366.6 $\pm$ 2.7 $\mu$K. The
maps are published on an nside = 2048 HEALPix grid. We downgraded the
maps to an nside = 1024 grid using the ud\_grade software from the
HEALPix software distribution. As detailed later, we smoothed some of
the maps using the HEALPix smoothing program.

Our aim is to characterize filamentary structures that are common to the
\hi\ and FIR emission. Several different methods have previously been
used to detect filamentary structures.  \citet{Clark2014} applied
  RHT to data from the Galactic Arecibo L-Band Feed Array \hi\ survey
  (GALFA-\hi\ \citealt{Peek2018}) and to GASS data. \citet{Clark2019b}
  applied the same approach to HI4PI data. \citet{Kalberla2016} applied
  unsharp masking (USM) to suppress large-scale structures in HI4PI data
  on scales below 30\arcmin. In addition they used the Hessian matrix
to work out the all-sky distribution of filamentary structures in
analogy to similar investigations by \citet{Schisano2014} and
\citet{Planck2016} in the FIR and submillimeter range.  Recently
\citet{Soler2020} used the Hessian analysis to analyze the filamentary
structure in the disk of the Milky Way. These authors have shown that
results from alternative algorithms, FilFinder \citep{Koch2015} and RHT
\citep{Clark2014}, are consistent with those from the Hessian matrix. In
the following we describe the Hessian analysis and the data processing
as applied by us.

\subsection{Hessian analysis }
\label{Hesse}

We classified filamentary structures as enhancements in intensity
that are organized as elongated regions with a preferred spatial
orientation. Such features on scales of arc minutes up to more than 20
degrees are often approximated as one-dimensional structures and termed
fibers \citep{Clark2014}, but they may also be bent, resembling worms
(\citealt{Heiles1984} and \citealt{Kalberla2016}). Filaments may also be caused
by projection effects from two-dimensional structures, such as sheets or
walls, seen edge on.

As a tool to classify structures as filament-like, we used the Hessian
operator $H,$ which is based on partial derivatives of the intensity
distribution
\begin{equation}
     \label{eq:hessI} 
        H(x,y)\, \equiv \, \left ( \begin{array}{cc} H_{xx} & H_{xy }\\
            H_{yx} & H_{yy} \end{array} \right ).
\end{equation}  
Here {\it x} and {\it y} refer to true angles in longitude
$x=l\,\cos{b}$ and latitude $y=b$ . The second-order partial derivatives
are $H_{xx}=\partial^2 I / \partial x^2$, $H_{xy}=\partial^2
I / \partial x \partial y$, $H_{yx}=\partial^2 I /
\partial y \partial x$, and $H_{yy}=\partial^2 I / \partial y^2$.

The eigenvalues of H, 
\begin{equation}
\label{eq:lambda}
\lambda_{\pm}=\frac{(H_{xx}+H_{yy}) \pm \sqrt{(H_{xx}-H_{yy})^2+4H_{xy}H_{yx}}}{2},
\end{equation}
describe the local curvature of the features; $\lambda_- < 0 $ is in
direction of least curvature and indicates filamentary structures or
ridges. Alternatively $\lambda_{+} < 0 $ highlights blobs. Equation
\ref{eq:lambda} with $\lambda_- < \lambda_{+} $ implies that the blobs
are not independent but located as enhancements along the filaments.

The Hessian operator $H$ has first been used for the analysis of
\hi\ data by \citet{Kalberla2016} but for consistency with
\citet{Clark2014}, \citet{Clark2019b}, \citet{Jow2018} and
\citet{Soler2020} we follow here the definition $x=l\,\cos{b}$. The
derivatives are calculated by us from symmetric differential
quotients. Due to the interleaved structure of the HEALPix database it
is necessary to interpolate the HEALPix data in Galactic longitudes and
we use a cubic spline interpolation for this purpose. The Hessian matrix
is symmetrical, $H_{xy} = H_{yx}$ but to minimize possible numerical
artifacts we calculate both ways independently. All calculations are
done in double precision. The numerical procedure takes only a few
seconds for an nside = 1024 data slice.

The local orientation of filamentary structures relative to the Galactic
plane is given by the angle
\begin{equation}\label{eq:theta}
\theta =
\frac{1}{2}\arctan\left[\frac{H_{xy}+H_{yx}}{H_{xx}-H_{yy}}\right],
\end{equation}
in analogy to the relation
\begin{equation}\label{eq:theta2}
\theta_S =\frac{1}{2}\arctan \frac{U}{Q},
\end{equation}
that can be derived from polarimetric observations that provide the 
Stokes parameters $U$ and $Q$.

\begin{figure*}[th] 
   \centering
   \includegraphics[width=18cm]{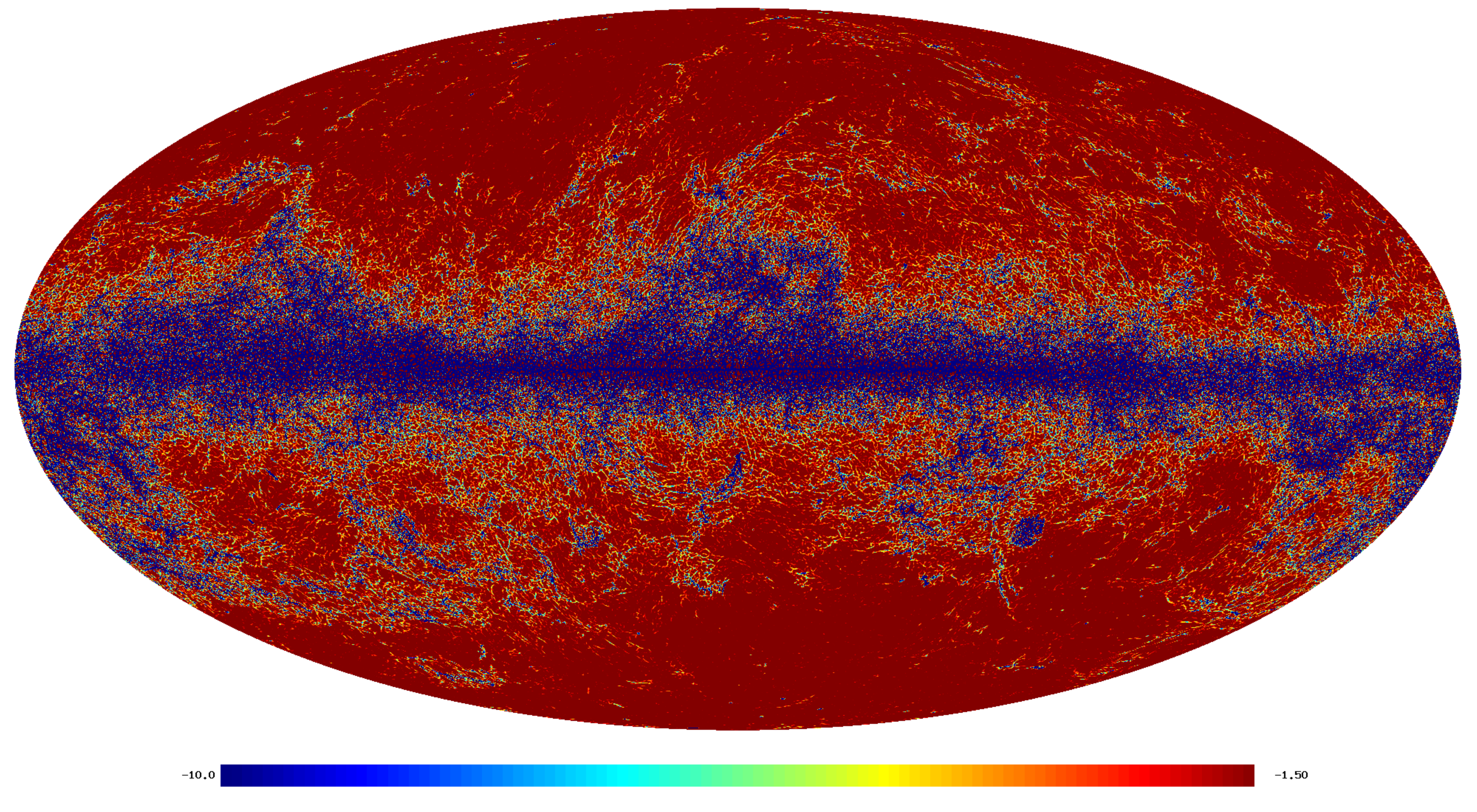}
   \includegraphics[width=18cm]{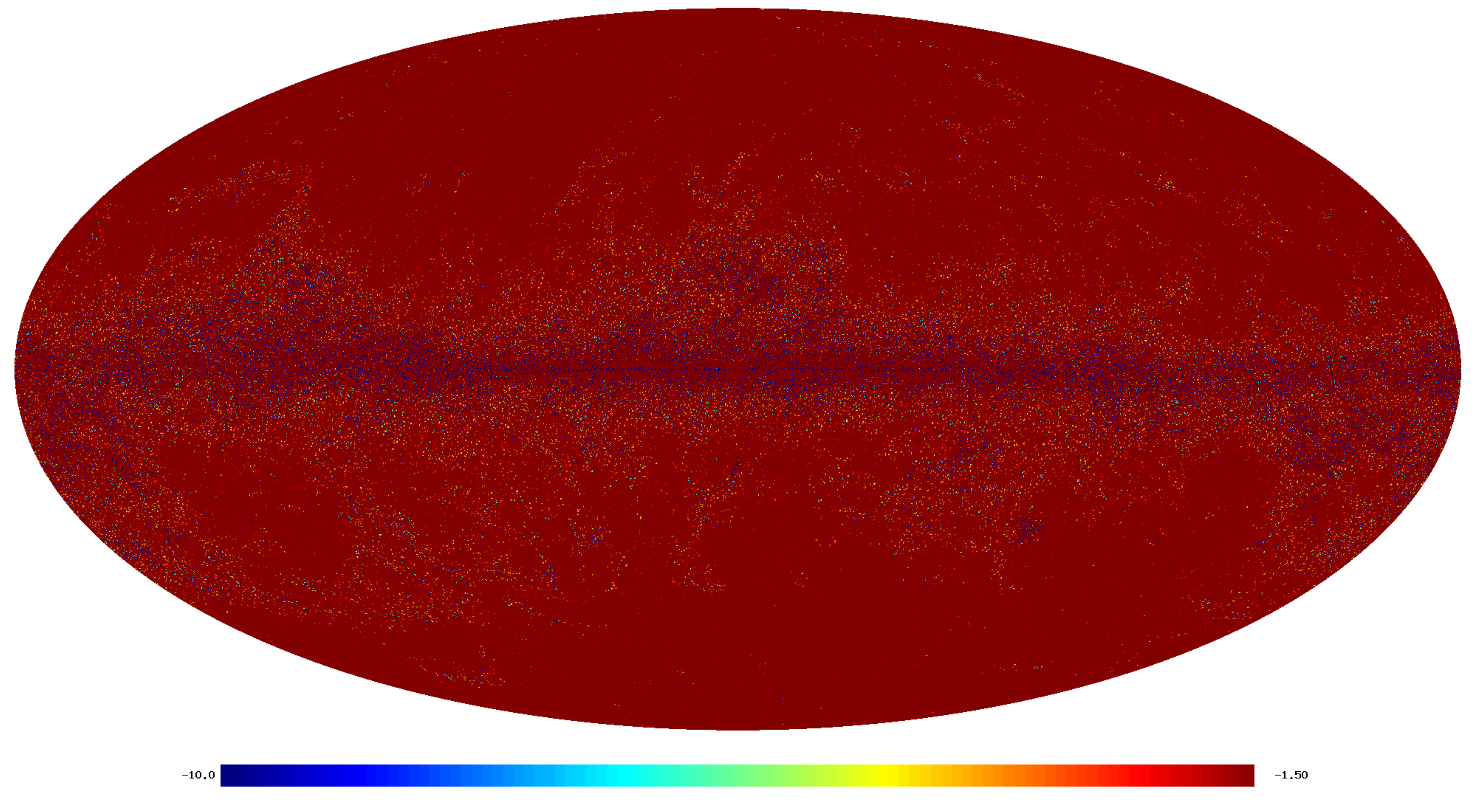}
   \caption{ All-sky Mollweide displays of derived FIR
       structures at 857 GHz.  Top: Distribution of eigenvalues
       $\lambda_- $ that characterize filamentary structures. Bottom:
       Distribution of $\lambda_+ $, characterizing local enhancements
       (blobs) along filaments.  }
   \label{Fig_lam}
\end{figure*}

This analogy between these two relations is intriguing but before we use
angular orientations $\theta$ from Eq. (\ref{eq:theta}) we need to
understand the meaning of this measure in comparison to
$\theta_S$. Results according to Eq. (\ref{eq:theta2}) need independent
observations of $U$ and $Q$, two of the four Stokes parameters. In the
case of Eq. (\ref{eq:theta}) we are limited to a single observation of the
Stokes parameter $I$. The orientation angle $\theta$ is derived merely
from the morphology of the intensity distribution. The advantage in this
case is however the high signal-to-noise ratio (S/N) of $I$, while the
polarization information entering Eq. (\ref{eq:theta2}) is typically much
closer to the noise level.

The derivatives in Eq. (\ref{eq:hessI}) are calculated over a fixed grid
of 5x5 pixels, centered on $x,y$.  This implies that $\theta$ from
Eq. (\ref{eq:theta}) is a measure for the local intensity fluctuations
within this grid only. The Stokes parameters $U$ and $Q$ from
polarimetric measurements are integrated along the line of sight and
depend on the telescope beam. The 5x5 pixel filter of the Hessian
operator is only meaningful if it is matched to the spatial scale under
consideration, in our case defined by the resolution of the data
product. The sensitivity to structures on scales that are large compared
to the 5x5 pixel filter is low. For a more general application a
multi-scale morphology approach may be necessary \citep{Aragon2007}.

We analyzed HEALPix databases with nside = 1024; therefore, the effective
pixel resolution is 3\farcm44 \citep{Gorski2005} and, accordingly, the
$H(x,y)$ filter acts on a scale of 18\arcmin. The FIR data, in
comparison to the \hi\ channel maps, are sensitivity limited (see,
e.g., Figs. 3 and 4 of \citealt{Kalberla2016}). To avoid unnecessary noise
amplification when deriving first and second derivatives, we smoothed the
{\it Planck} 857 GHz data to an effective resolution of 18\arcmin. Thus
we matched the full width at half maximum (FWHM) of the smoothing filter
to the 5 pixels of the grid used by the Hessian analysis. Such an
adapted smoothing leads to significant reduction of the uncertainties (see Sect. \ref{proof}). In Fig. \ref{Fig_lam} we display the
  resulting distribution of eigenvalues that characterize filaments and
  blobs for FIR structures at 857 GHz. 
  
For the \hi\ data we used Gaussian components from \citet{Kalberla2019}
with a FWHM beamwidth of 14\farcm5 for the GASS and 10\farcm8 for the
EBHIS. In comparison to the FIR databases there is a slight mismatch in
spatial resolution but we intend to analyze correlations between FIR
emission and \hi\ at the highest possible resolution. A smoothing would
disable the use of the Gaussian database since the fitting is a
nonlinear process. Our analysis is hierarchical in the sense that we
always use the 857 GHz FIR data as a template to search for structures
at other frequencies. A slightly smaller FWHM width for the \hi\ data
does not violate the Nyquist theorem since both databases are on the
same HEALPix grid.  Tests with several different smoothing kernels
indicated anyway that the derived eigenvalues do not depend critically
on the smoothing kernel used.

\begin{table*}
\caption{Filamentary alignment measures.}             
\label{table:1}      
\centering          
\begin{tabular}{c c c c c c c c c c }     
\hline\hline       
Data 1 & Data 2 & latitude & $f$ & \multicolumn{2}{c}{$\delta \theta$}  & $\xi$ & PRS & $\sigma_{\mathrm{PRS}}$ & Fig. \\ 
& & range & & $\sigma_{\mathrm{Gauss}}$ & $\sigma_{\mathrm{Voigt}}$ & & & \\
\hline                    
  {\it Planck} 857 GHz & {\it Planck} 545 GHz & all  & .71  & 5\fdg3 & 4\fdg3 & 0.80 & 3361.87 & .67 \\  
18\arcmin\ FWHM &545 GHz 18\arcmin\ FWHM &  all & .39  & 1\fdg7 & 1\fdg2 & 0.99 & 3088.46 & .10 &  \\  
18\arcmin\ FWHM &353 GHz I, 18\arcmin\ FWHM & all & .39  & 4\fdg1 & 3\fdg4 & 0.89 & 2792.08 & .41 & \\  
18\arcmin\ FWHM &353 GHz Q, U, 18\arcmin\ FWHM & all & .39  & 54\fdg3 & 43\fdg3 & 0.13 & 400.04 & 1.0 & \\  
  \hline                  
18\arcmin\ FWHM & \hi, $ |v_{\mathrm{LSR}}| < 50 $ \kms &  all & .39 & 28\fdg1 & 23\fdg9 & 0.34 & 1051.99 & .96 &  \\  
18\arcmin\ FWHM & \hi\ in 1 \kms\ channels &  all & .37 & 2\fdg6 & 1\fdg9
& 0.94 & 2873.67 & .17 & \ref{Fig_Aligne_857} center left\\  
18\arcmin\ FWHM & \hi\ in 2 \kms\ channels &  all & .36 & 5\fdg7 &
4\fdg3 & 0.92 & 2769.24 & .21 &  \\  
18\arcmin\ FWHM & \hi\ in 4 \kms\ channels &  all & .36 & 8\fdg1 & 6\fdg6
& 0.90 & 2741.70 & .22 &  \\  
18\arcmin\ FWHM & \hi\ in 8 \kms\ channels &  all & .33 & 13\fdg3 & 11\fdg2
& 0.85 & 2469.78 & .28 &  \\  
18\arcmin\ FWHM & \hi\ in 16 \kms\ channels &  all & .26 & 17\fdg6 & 15\fdg8
& 0.82 & 2102.51 & .32 &  \ref{Fig_Aligne_857} bottom left \\  
\hline
\hline                    
  {\it Planck} 857 GHz & {\it Planck} 545 GHz & $|b| > 20 \degr$ & .69  & 8\fdg3 & 7\fdg5 & 0.75 & 2526.12 & .71 \\  
18\arcmin\ FWHM &545 GHz 18\arcmin\ FWHM & $|b| > 20 \degr$ & .26  & 2\fdg4 & 1\fdg9 & 0.98 & 2026.24 & .13 &  \\  
18\arcmin\ FWHM &353 GHz I, 18\arcmin\ FWHM &  $|b| > 20 \degr$ & .26  & 8\fdg5 & 6\fdg9 & 0.81 & 1679.71 & .54 &  \\  
18\arcmin\ FWHM &353 GHz Q, U, 18\arcmin\ FWHM &  $|b| > 20 \degr$ & .26  & 46\fdg3 & 38\fdg2 & 0.17 & 356.14 & 0.99 & \ref{Fig_Aligne_857} top left \\  
  \hline                  
18\arcmin\ FWHM & \hi, $ |v_{\mathrm{LSR}}| < 50 $ \kms &  $|b| > 20 \degr$ & .26 & 23\fdg0 & 22\fdg2 & 0.48 & 994.25 & .87 & \ref{Fig_Aligne_857} top right \\  
18\arcmin\ FWHM & \hi\ in 1 \kms\ channels &  $|b| > 20 \degr$ & .24 &
4\fdg1 & 3\fdg1 & 0.92 & 1817.64 & .23 & \ref{Fig_Aligne_857} center
right \\
18\arcmin\ FWHM & \hi\ in 2 \kms\ channels &  $|b| > 20 \degr$ & .23 & 8\fdg8 & 6\fdg9 & 0.89 & 1727.46 & .26 & \\
18\arcmin\ FWHM & \hi\ in 4 \kms\ channels &  $|b| > 20 \degr$ & .23 & 10\fdg6 & 8\fdg9 & 0.88 & 1728.04 & .26 & \\
18\arcmin\ FWHM & \hi\ in 8 \kms\ channels &  $|b| > 20 \degr$ & .20 & 15\fdg0 & 13\fdg7 & 0.83 & 1532.95 & .31 & \\
18\arcmin\ FWHM & \hi\ in 16 \kms\ channels &  $|b| > 20 \degr$ & .15 &
16\fdg9 & 16\fdg4 & 0.82 & 1275.89 & .33 & \ref{Fig_Aligne_857} bottom
right  \\
\hline   
\end{tabular}
\end{table*}
%

\subsection{Comparison of filamentary structures in different databases } 
\label{Comparison}

To compare different data sets we apply simultaneously a Hessian analysis
to both data sets. According to our definition of filaments as intensity
enhancements that are organized and have a preferred spatial orientation
we demand that both databases have at the same position significant
eigenvalues $\lambda_- < 0 $. If this is the case we calculate and
compare the orientation angles $\theta$.

\citet{Clark2019b} and \citet{Jow2018} have previously considered
similar comparisons and developed different strategies to measure the
alignment of filamentary structures. For a deeper discussion we refer
to the original publications; we use these measures but define
them below only briefly.

According to \citet{Planck2016a}  and \citet{Clark2019b}, the angular difference between the
orientation angles $\theta_{1}$ and $\theta_{2}$ at each position of two
data sets can be calculated by
\begin{equation}
\delta \theta = \frac{1}{2}
\mathrm{arctan}\left[\frac{\mathrm{sin}(2\theta_{1})\mathrm{cos}(2\theta_{2})
    - \mathrm{cos}(2\theta_{1})\mathrm{sin}(2\theta_{2})
  }{\mathrm{cos}(2\theta_{1})\mathrm{cos}(2\theta_{2}) +
    \mathrm{sin}(2\theta_{1})\mathrm{sin}(2\theta_{2}) }\right]
\label{eq:angdif}
,\end{equation}
and a mean degree of alignment, 
\begin{equation}
\label{eq:xi}
\mathrm{\xi} = \left< \mathrm{cos} \phi \right>,
\end{equation}
can be defined for  
\begin{equation}
\phi = 2 \delta \theta.
\end{equation}

A different metric, called projected Rayleigh statistic (PRS), was
advocated by \citet{Jow2018},
\begin{equation}
\mathrm{PRS} = \sqrt{\frac{2}{N}} \sum_i \mathrm{cos}\, \phi_i,
\label{eq:PRS}
\end{equation}
and the uncertainty of this measure can be estimated from
\begin{equation}
\sigma^2_{\mathrm{PRS}} = \frac{2 \sum_i \mathrm{cos}^2\, \phi_i -
  (PRS)^2 } {N}.
\label{eq:sig_PRS}
\end{equation}

\subsection{Proof of concept }
\label{proof}

To test our data processing and the performance of the different
alignment measures we use two neighboring {\it Planck} frequency
channels. Both probe essentially the same dust foreground spectral
energy distribution at slightly different frequencies. Data collected
close in frequency should have the highest degree of similarity and thus
the best spatial correlation of two independently measured data sets.
For a quantitative comparison we compare first FIR data at 857 and 545
GHz.  These data have been processed with the NPIPE pipeline and are
claimed to have currently the best internal consistency between the
various frequency channels \citep{Planck2020}. However, despite noise
and differences in instrumental performance, we expect deviations when
comparing 857 and 545 GHz data. For frequencies $\nu > 217$ GHz the
contribution of the Galactic foreground is dominated by the ISM, at
lower frequencies the Galactic synchrotron and free-free radiation
contribute significantly. The cosmic microwave background (CMB)
contribution is increasing for lower frequencies and the relative
  contribution from the cosmic infrared background (CIB) can be
  considered as a significantly fainter and spatially unrelated offset
  to the \hi\ emission \citep[][their Fig.\,3]{Odegard2019}. In
general, for {\it Planck} 545 GHz and 857 GHz frequency channels the
Galactic foreground dominates across the sky. The total emission for the
different grain models is highest at 857 GHz
\citep[][Fig. 7]{Draine2009}. Despite the fact that across the sky some
fluctuations in the spectral energy distributions are expected, a good
agreement is observed at 545 and 857 GHz
\citep[][Fig. A1]{Irfan2018}. These two data sets are best suitable to
test how far the application of the Hessian operator $H$ and the
subsequent determination of the angular orientation of filamentary
structures (Eq. \ref{eq:theta}) lead to coherent results at both
frequencies.

\begin{figure}[th] 
   \centering
   \includegraphics[width=9cm]{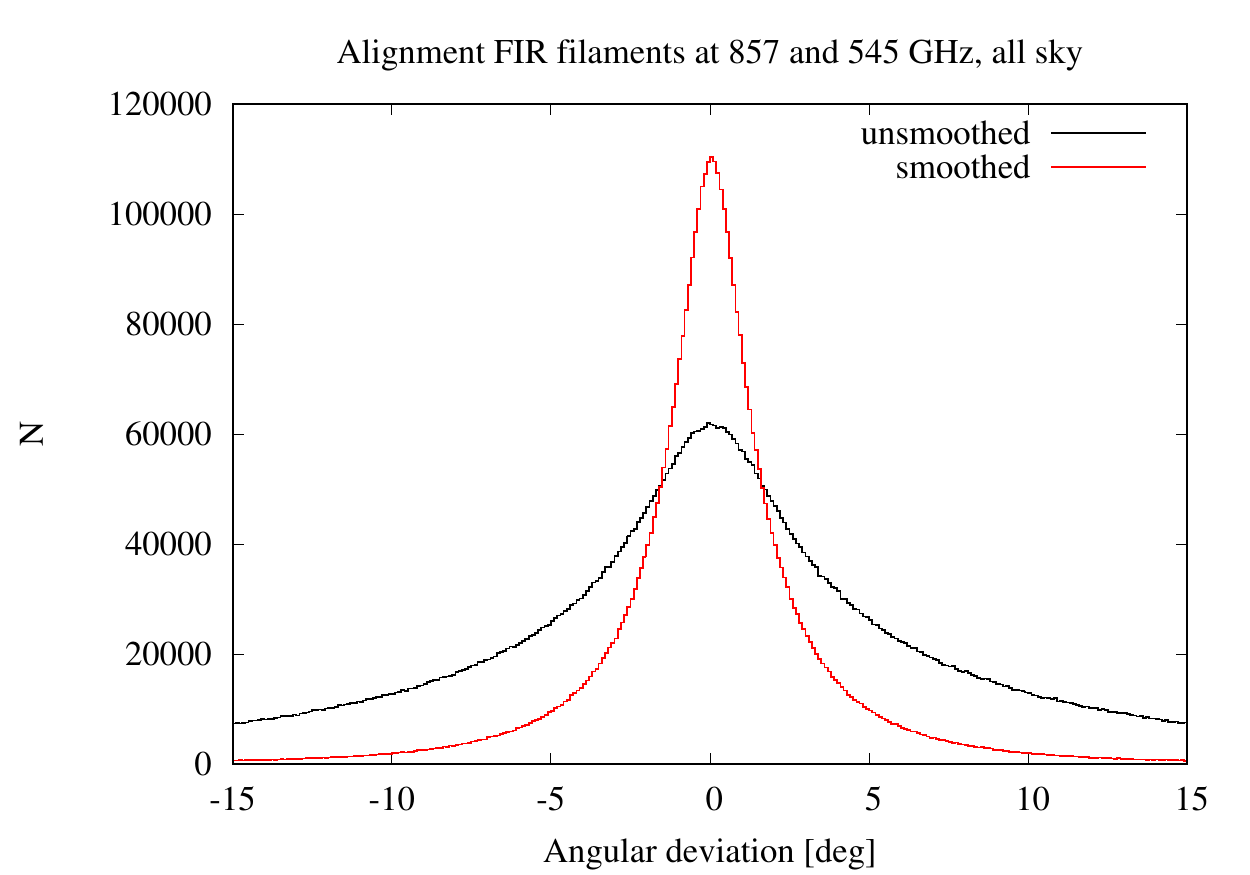}
   \caption{Histograms of angular alignment deviations for filamentary
     structures according to Eq. (\ref{eq:angdif}). We compare 857 and 545
     GHz data, unsmoothed as observed and smoothed to a resolution of
     18\arcmin\ FWHM. }
   \label{Fig_Aligne_test}
\end{figure}

\begin{figure*}[thp] 
   \centering
   \includegraphics[width=9cm]{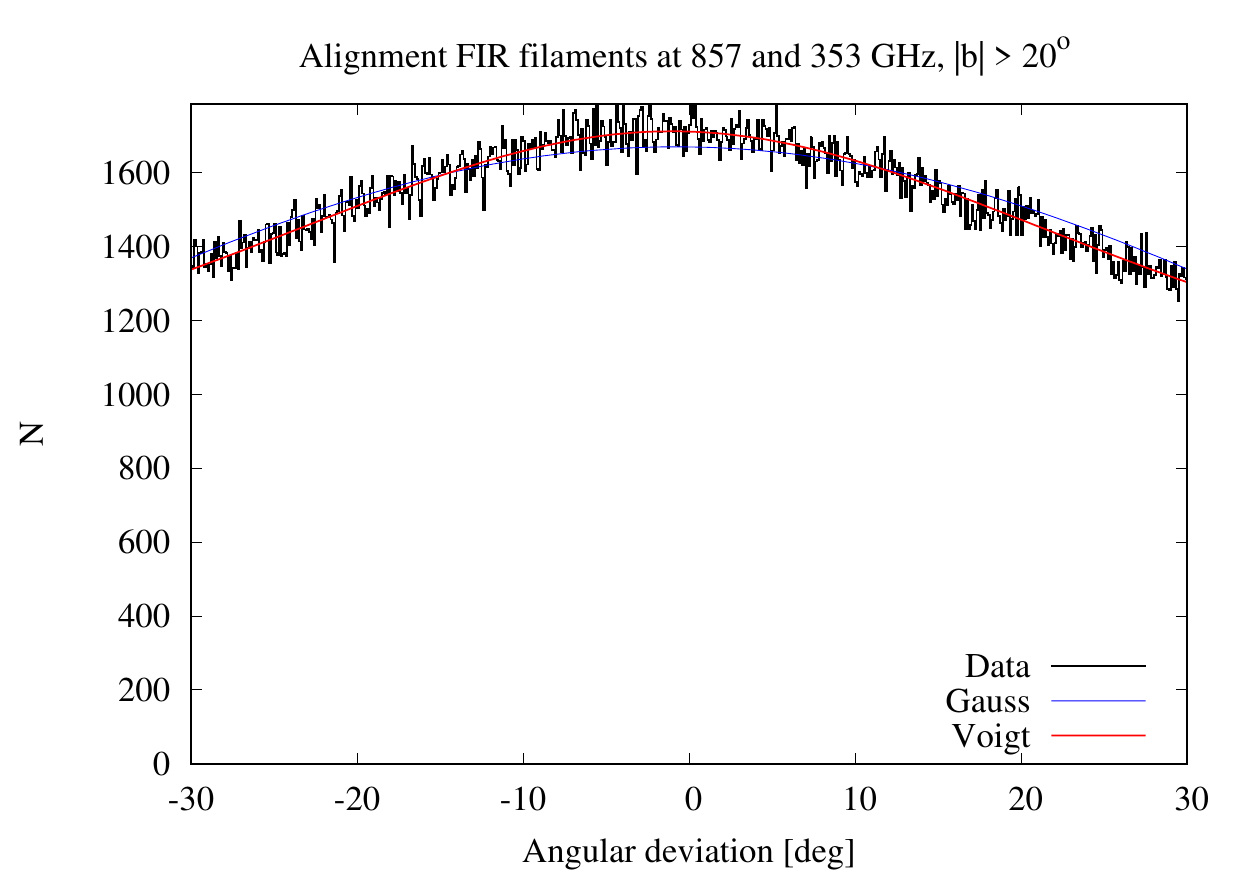}
   \includegraphics[width=9cm]{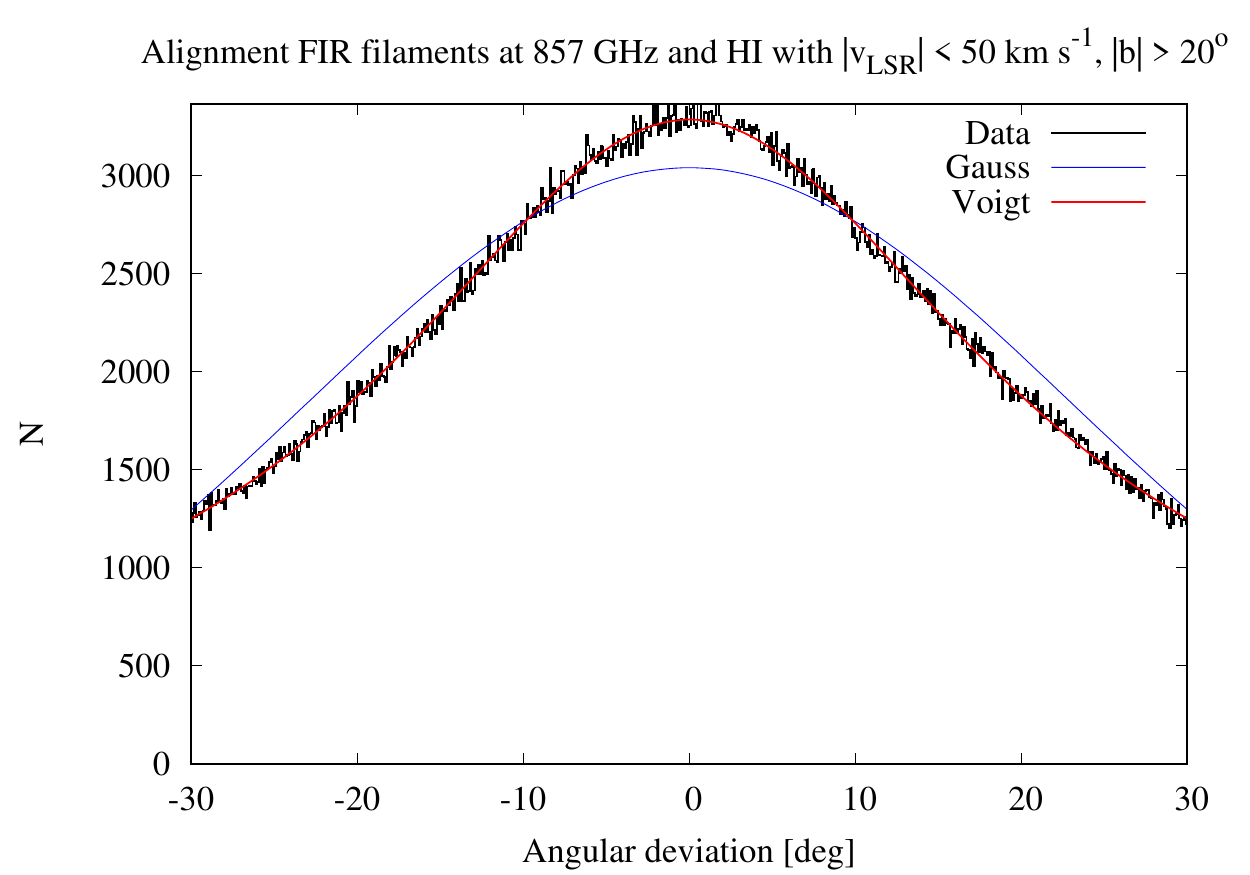}
   \includegraphics[width=9cm]{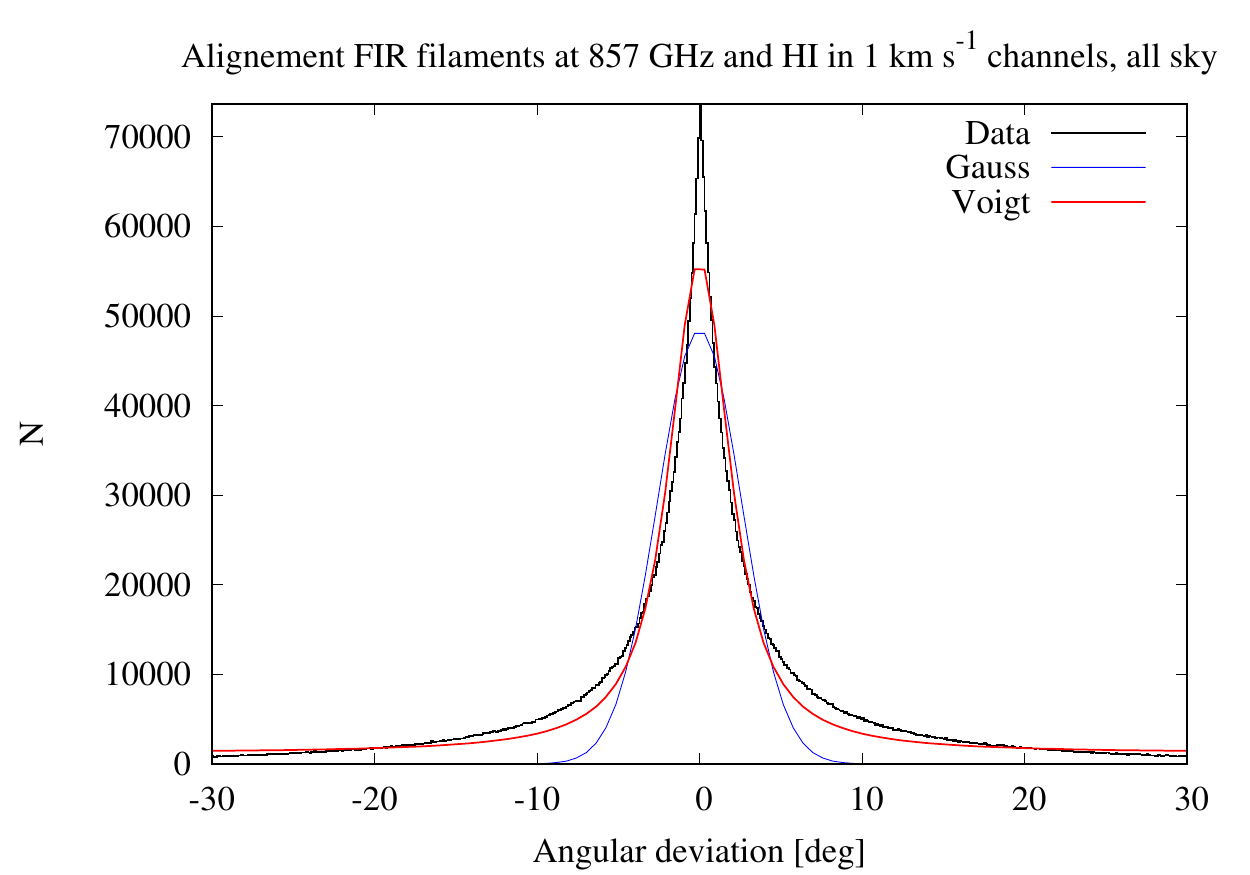}
   \includegraphics[width=9cm]{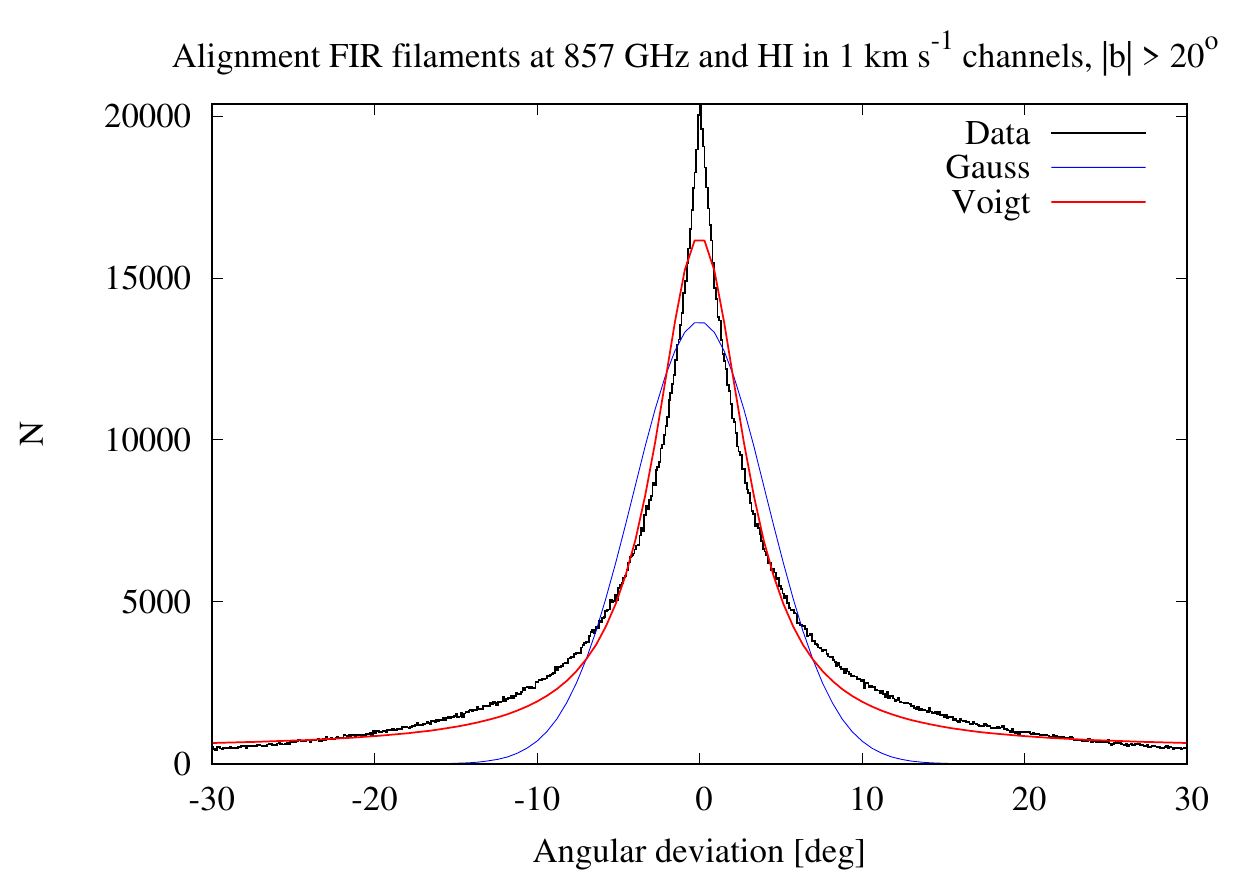}
   \includegraphics[width=9cm]{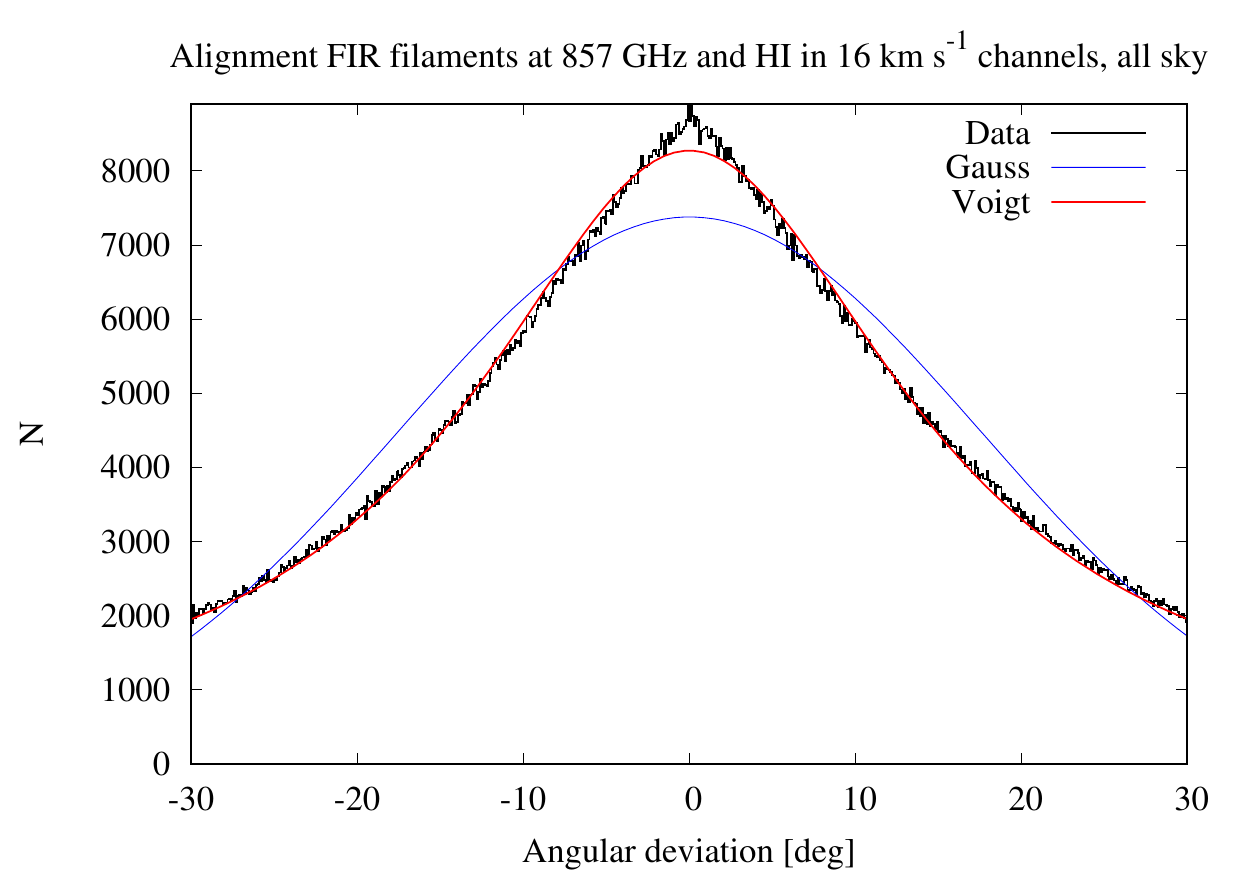}
   \includegraphics[width=9cm]{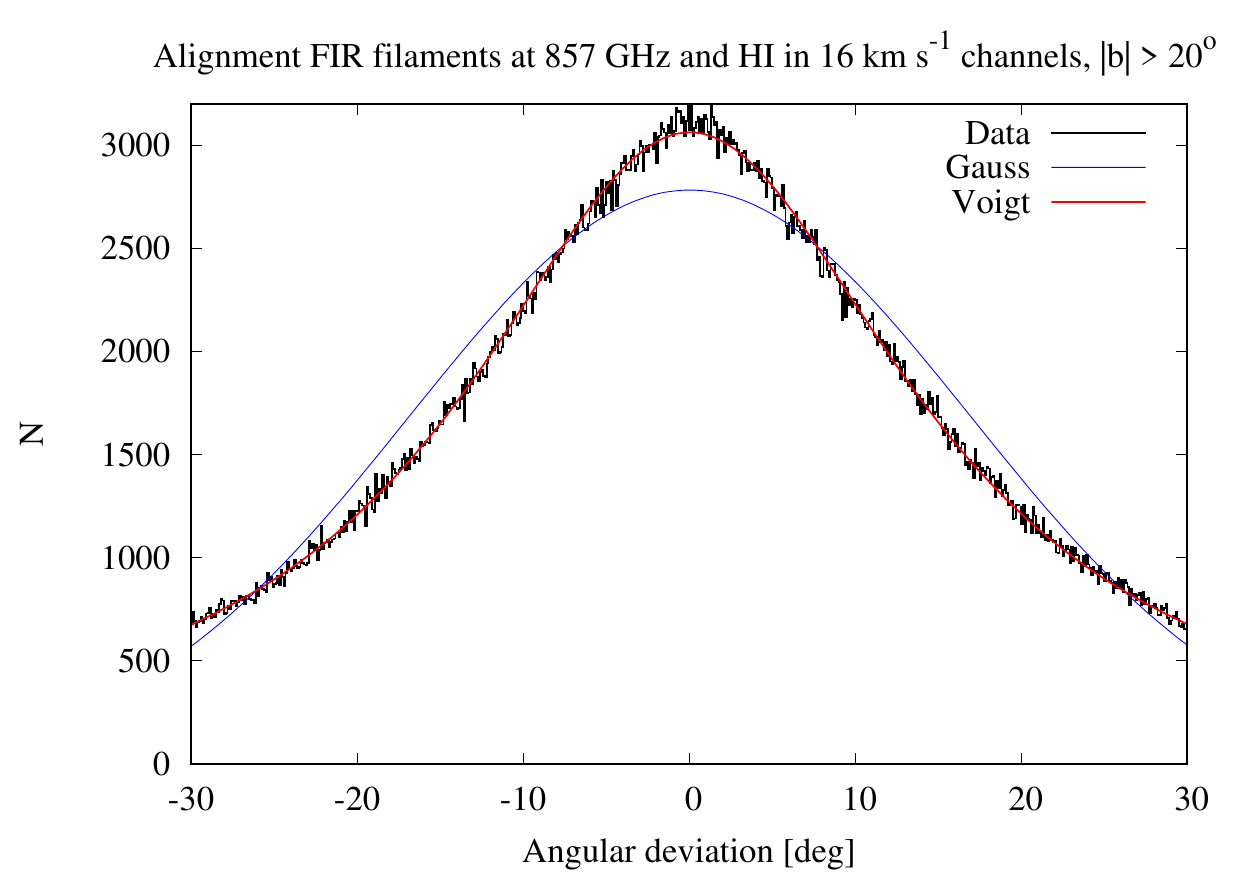}
   \caption{Histograms of angular alignment deviations according to
     Eq. (\ref{eq:angdif}) for filamentary structures. Top left: {\it
       Planck} 857 GHz compared with {\it Planck} 353 GHz and angular
     orientations according Eq. (\ref{eq:theta2}), $|b| > 20 \degr$. Top
     right: {\it Planck} 857 GHz compared with \hi\ column densities for
     $ |v_{\mathrm{LSR}}| < 50 $ \kms, $|b| > 20 \degr$. Center left:
     {\it Planck} 857 GHz compared with best fit single channel
     \hi\ filaments, all sky. Center right: {\it Planck} 857 GHz
     compared with best fit single channel \hi\ filaments, $|b| > 20
     \degr$. Bottom left: {\it Planck} 857 GHz compared with best fit
     \hi\ filaments at a channel width of 16 \kms, all sky. Bottom
     right: {\it Planck} 857 GHz compared with best fit \hi\ filaments
     at a channel width of 16 \kms, $|b| > 20 \degr$.  }
   \label{Fig_Aligne_857}
\end{figure*}

For a basic test of our data analysis we use first unsmoothed 857 and
545 GHz data, downgraded to a HEALPix grid with nside = 1024. We select
structures with eigenvalues $\lambda_- < -1.5\ \mathrm{K/deg}^{-2}$ at 857
GHz. To determine coherence we apply the measures according
Eqs. \ref{eq:angdif} to \ref{eq:sig_PRS}, the results are listed in
Table \ref{table:1}. We repeat analysis and comparison for 857, 545, and
353 GHz data after Gaussian smoothing of all data sets to an effective
resolution of 18\arcmin\ FWHM.

The distribution of angular differences $\delta \theta$
(Eq. \ref{eq:angdif}) for all sky 857 and 545 GHz data is shown in
Fig. \ref{Fig_Aligne_test}. From this plot it is obvious that for the
original unsmoothed data the angular differences $\delta \theta$ are
limited by noise. However, applying to the data a Gaussian smoothing
with a kernel that is matched to the 5x5 pixel Hessian operator $H$ leads
to significant improvements with a lower spread for the angular
  alignment deviations. 

The probability density distribution for angular differences
  $\delta \theta$ is a circular distribution in the range $-\pi$
  to $+\pi$. A normal distribution is in this case approximated by the 
  von Mises distribution \citep[e.g.,][]{Jow2018}
\begin{equation}
        \mathcal{F}(\delta \theta \vert \mu,\kappa) \equiv \frac{e^{\kappa\cos{\delta \theta-\mu}}}{2\pi I_0(\kappa)},
        \label{eq:vonMises}
\end{equation}
with $\kappa$ approaching for narrow distributions the reciprocal of the
dispersion of a normal distribution, $\kappa \sim 1/\sigma^2$. $I_0$ is
the modified Bessel function of order 0. Fitting $\mathcal{F}$ we found
in cases with narrow $\delta \theta$ distributions virtually identical
distributions (within the thickness of the lines plotted) in comparison
to a Gaussian fit. In the case of broad distributions the $\mathcal{F}$ fit
was found to diverge while the Gaussian fit was still stable. For this
reason we plot Gaussian approximations and cite Gaussian parameters
only. The shapes of the $\delta \theta$ distribution in
Figs. \ref{Fig_Aligne_test} and \ref{Fig_Aligne_857} are difficult to
fit. We also considered a simple estimate by measuring the width of the
$\delta \theta$ distributions at half maximum without fitting, but such
a measure seriously underestimates the width in the case of a strong central
peak (central panels of Fig. \ref{Fig_Aligne_857}). Central peaks and
extended wings are best taken into account by a Voigt function and we
found that the Voigt function approximates noisy data
best\footnote[2]{For fitting the von Mises, Gauss, and Voigt/Faddeeva
  profiles, we used implementations provided by gnuplot; version 5.4 or
  higher is needed.}.

To compare the distribution of angular differences between different
data sets in a quantitative way we list in Table \ref{table:1} the
surface filling factor $f$ for common filamentary structures, the 
dispersions $\sigma_{\mathrm{Gauss}}$ and $\sigma_{\mathrm{Voigt}}$ and
further alignment indicators $\xi$, PRS, and $\sigma_{\mathrm{PRS}}$
according to Eqs. \ref{eq:angdif} to \ref{eq:sig_PRS}.


\subsection{Alignment between FIR and \hi\ column density structures} 

We aim to determine the degree of alignment between FIR and
\hi\ structures. In the following we use only smoothed 857 GHz data for
the FIR since these have among the {\it Planck} observations the best
S/N \citep[][Fig. C.1]{Planck2014}. In contrast to the \hi\ distribution
the 857 GHz data are noise limited \citep{Kalberla2016} but, as shown in
Fig. \ref{Fig_Aligne_test}, smoothing helps to improve our analysis
significantly. For the \hi\ we chose first a velocity range of $
-50 < v_{\mathrm{LSR}} < 50 $ \kms. This range comprises about 75\% of
the total HI emission.  It also accounts for the phenomenon of dusty
intermediate-velocity clouds \citep{Roehser2016} and thus is complete in
the sense that it samples all \hi\ emission that is associated with FIR
emission. For a first comparison we calculate the integral across this
radial velocity range (the moment zero map) and derive column densities
$N_{\mathrm{HI}}$. As mentioned before we do not smooth \hi\ data.

Table \ref{table:1} shows the resulting alignment measures for two
cases, using all sky data and restricting the analysis to high Galactic
latitudes, $|b| > 20 \degr$. The high latitude comparison shows the best
agreement, the distribution function for angular deviations is best fit
with a Voigt profile with a dispersion of about 20\degr. We display this
distribution in Fig. \ref{Fig_Aligne_857} (top right) for comparison
with alignment deviations between FIR at 857 and 353 GHz
(Fig. \ref{Fig_Aligne_857} top left). There is some alignment between
FIR filaments and \hi\ but apparently the $N_{\mathrm{HI}}$ filaments
fail a detailed correlation with the FIR. 

\subsection{Alignment between FIR and \hi\ in single channels}
\label{v_fil}

Cold \hi\ filaments are observed to have average FWHM line widths of 3
\kms\ (e.g., \citet{Clark2014} or \citet{Kalberla2016}). This empirical
result, typical for a cold neutral medium (CNM), may imply that
coherence between FIR and \hi\ can exist only for a restricted range in
radial velocity.  The filamentary CNM structures are embedded in a
  more diffuse and warmer environment, the warm neutral medium (WNM).
  Dust and gas are well mixed and the CNM is in any case found to be
  associated with WNM \citep[e.g.,][]{Kalberla2018}. The question arises
  whether we can distinguish between different \hi\ phases that can be
  related to the FIR filaments. Hence it is mandatory to consider the
  line-width of possible \hi\ counterparts of FIR filaments. For a
  turbulent medium it is in addition a matter of debate whether
  filaments are real density structures or just velocity caustics,
  fluctuations imprinted by the turbulent velocity field
  \citep{Clark2019a}.  In the case of velocity caustics it is assumed that
  density and velocity fields are completely uncorrelated. Sufficient
  large velocity channel widths of $\sim 17$ \kms, typical for a WNM,
  need accordingly to be used to characterize the density field.

To verify the different hypotheses we modify our analysis and probe the
alignment between FIR and \hi\ structures depending on both, radial
velocities and velocity channel widths.  In five independent runs we
generate separate HEALPix databases with 101, 51, 25, 13, and 7 velocity
slices in the range $ -50 \la v_{\mathrm{LSR}} \la 50 $ \kms, each slice
with a width of 1, 2, 4, 8, and 16 \kms, respectively. We apply the
Hessian operator individually to each of these slices and calculate
$\lambda_{\pm}(l,b,v_{\mathrm{LSR}})$ and associated angles
$\theta(l,b,v_{\mathrm{LSR}})$. Next we compare for all HEALPix
positions the angular alignment between FIR at 857 GHz and \hi\ at each
of the velocity channels with different channel widths. The Hessian
  analysis is performed in all cases over a 5x5 pixel region centered on
  ($l,b$).  For each of the five different setups independently the
velocity with the minimum scatter in the $\delta \theta$ distribution
(or alternatively the best alignment) defines the best fit filament
velocity $v_{\mathrm{fil}}$ at this position. Selecting then
$\lambda_{\pm}(l,b,v_{\mathrm{fil}})$ we reduce the three-dimensional
$\lambda_{\pm}(l,b,v_{\mathrm{LSR}})$ distributions to two
dimensions. For each of the five velocity channel widths we obtain this
way an \hi\ distribution that can be tested for internal coherence in
velocity (discussed in Sect. \ref{Vel_disp}) but also for consistency
with FIR at 857 GHz.

The derived alignment measures are summarized in Table
\ref{table:1}. During our investigations we preferred to inspect the
distributions of the angular alignment deviations, characterized by
$\sigma_{\mathrm{Voigt}}$, but found none of the listed alignment
measures to be superior over the others. However, all of these parameters
show a common trend. To our understanding the collapsed filamentary
\hi\ distribution that was extracted for the velocity slice with a width
of 1 \kms\ results in the best fit. However, we also note opinions
  that the 8 and 16 \kms\ integrated maps might be most appropriate for
  this kind of analysis and should be preferentially analyzed.  In the
following we consider both solutions, for 1 and 8 or 16
\kms\ channel widths in some detail.

  Figure \ref{Fig_Aligne_857} shows examples of the derived distribution
  functions for angular alignment deviations, all sky and restricted to
  high latitudes $|b| > 20 \degr$. Approximating the alignment
  distributions with Voigt functions we obtain $\sigma_{\mathrm{Vogit}}
  < 3\degr $ at high latitudes and $\sigma_{\mathrm{Voigt}} < 2\degr $
  all sky in the case of 1 \kms\ broad channels (Fig. \ref{Fig_Aligne_857}
  center). This comes even close to the correlation of the two
  neighboring {\it Planck} frequency bands, inspected in
  Sect. \ref{proof}. We conclude that there is a very well-defined
  coherence between FIR and \hi\ filaments at the narrowest velocity
  interval. Figure \ref{Fig_Aligne_857} shows at the bottom the
  distribution of angular alignment deviations for a channel width of
  $\delta v_{\mathrm{LSR}} = 16$ \kms\ with significantly larger
  dispersions.

\begin{figure}[th] 
   \centering
   \includegraphics[width=9cm]{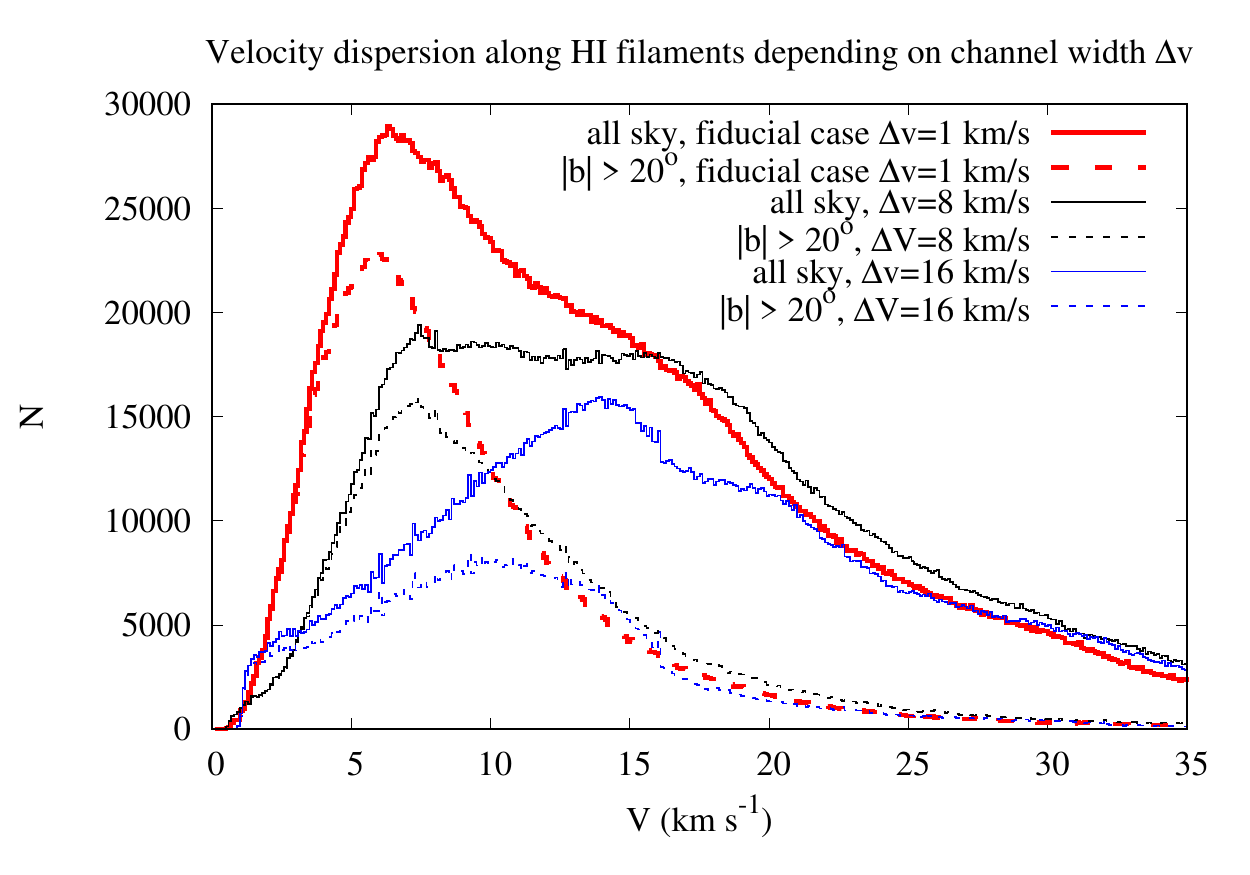}
   \caption{Distribution of \hi\ velocity dispersions $\mathcal{V}$
     along FIR filaments at channel widths of 1, 8, and 16 \kms. }
   \label{Fig_V}
\end{figure}

\begin{figure}[th] 
   \centering
   \includegraphics[width=9cm]{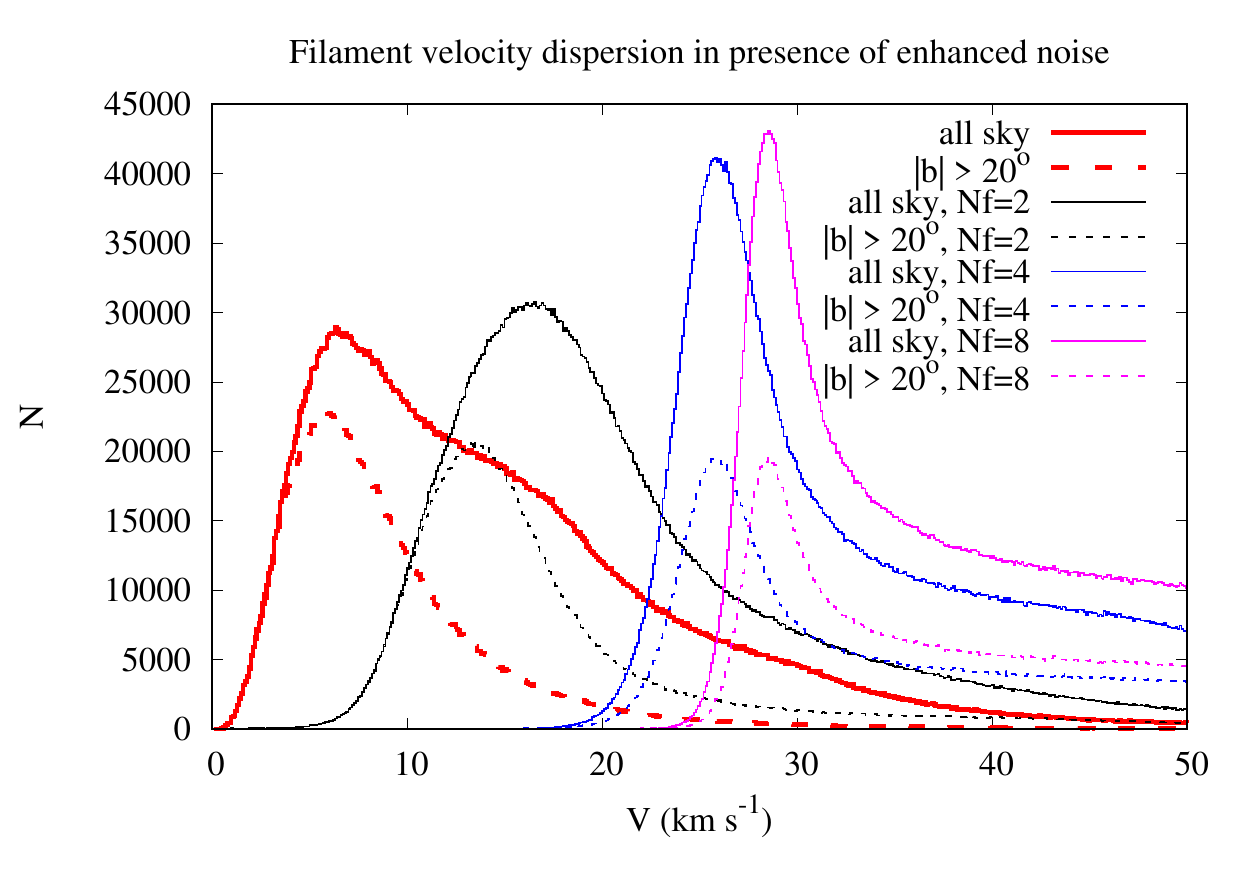}
   \caption{\hi\ velocity dispersions $\mathcal{V}$
     along FIR filaments at a channel width of 1 \kms\ depending on
     simulated noise enhancements $N_f = 2, 4,$ and 8 of the
     typical observed noise level. }
   \label{Fig_Noise}
\end{figure}

\subsection{Velocity dispersion along \hi\ filaments}
\label{Vel_disp}

In this subsection we consider the problem of whether the correlation
  between {\it Planck} and \hi\ data might be manufactured by the data
  processing. The correlation could be an artifact of selecting only the
  best of many noisy channels. In other words, the question is whether
  one could generate a map with a high correlation coefficient by using
  just N realizations of random noise given the number N is large
  enough. 

To investigate this question we consider the general case of a three-dimensional
distribution of \hi\ filaments that may be correlated with FIR
structures that are observable only in two dimensions. To establish a correlation,
we strictly demand that such \hi\ filaments are not only coincident and
aligned with the FIR in position but are also homogeneous in the third
dimension, thus uniform without irregularities in velocity. A
distribution is usually considered to be uniform if its parameters are
not arbitrary but lie between certain narrow bounds. Irregularities are
then measurable as deviations.

\begin{figure}[thp] 
   \centering
   \includegraphics[width=9cm]{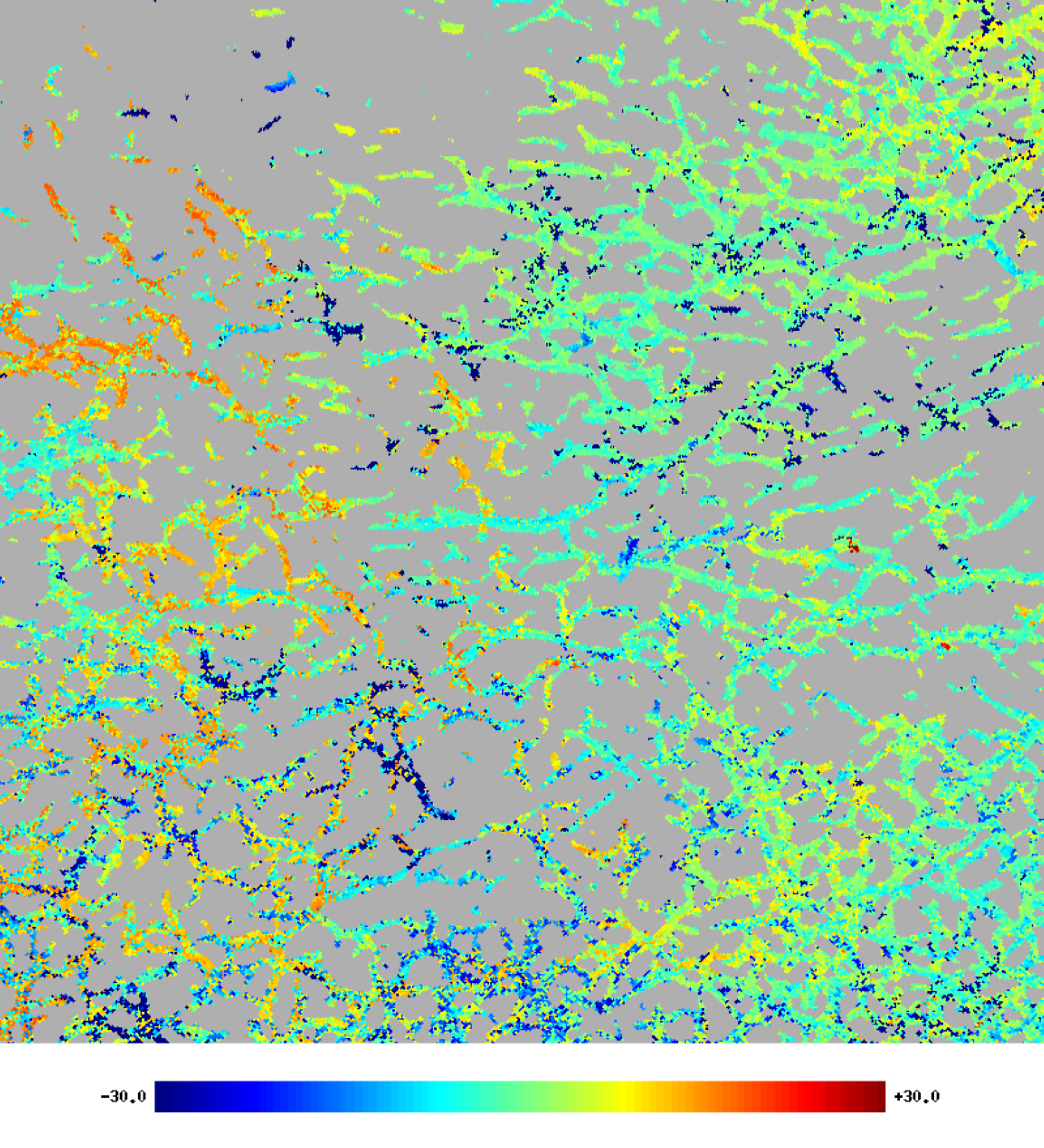}
   \includegraphics[width=9cm]{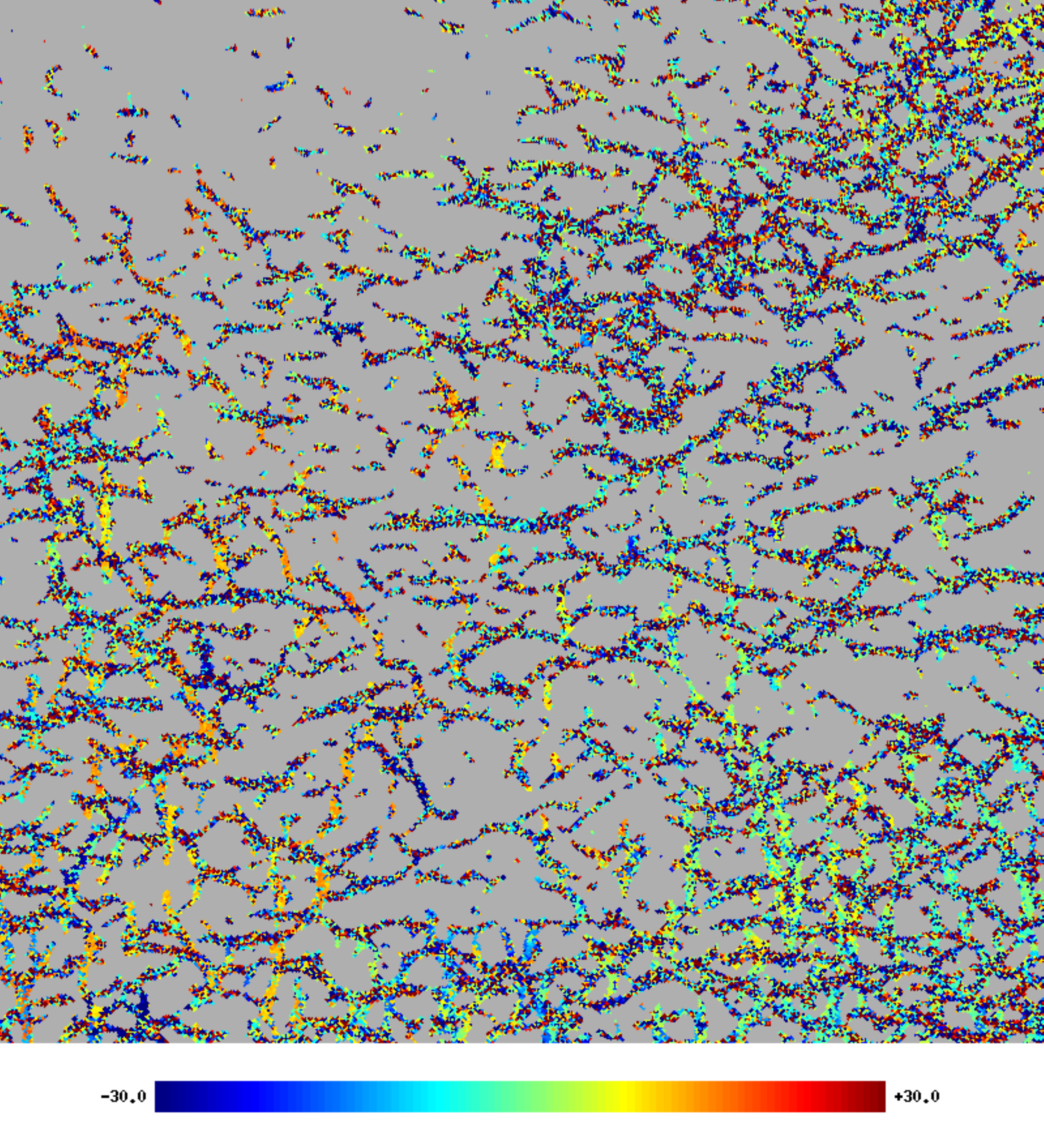}
   \caption{Velocity field in the case of \hi\ filaments derived for data
     with a 1 \kms\ channel width at an arbitrarily selected central
     position, $l = 160\degr$, $b = 30\degr$ in gnomonic projection. The
     field size is 27\fdg7; only velocities in the range $ -30 <
     v_{\mathrm{LSR}} < 30 $ \kms\ are displayed. Top: Velocities from
     original telescope data. Bottom: Results derived after simulating a
     noise amplification by a factor of $N_f = 4$. }
   \label{Fig_NoiseMaps}
\end{figure}

To determine irregularities in velocity we define a velocity dispersion
along the \hi\ filaments
\begin{equation}
\label{eq:defV}
\mathcal{V}\left(\boldsymbol{r},\delta\right)=\sqrt{\frac{1}{N}\sum_{i=1}^N\left[v_{\mathrm{LSR}}(\boldsymbol{r}+\boldsymbol{\delta}_i)-v_{\mathrm{LSR}}(\boldsymbol{r})\right]^2} \, .
\end{equation}
The sum extends over all pixels along the filament with positions $
(\boldsymbol{r}+\boldsymbol{\delta}_i) $ within an annulus centered on
$\boldsymbol{r}$ and having inner and outer radii
$\boldsymbol{\delta}_{\mathrm{inner}}$ and
$\boldsymbol{\delta}_{\mathrm{outer}}$, respectively.  According to our
definition of the Hessian operator with adapted Gaussian smoothing over
five pixels we select an inner radius
$\boldsymbol{\delta}_{\mathrm{inner}} = 9\arcmin$. For the outer radius
we chose $\boldsymbol{\delta}_{\mathrm{outer}} = 1 \degr$, a value
that appears to be appropriate as discussed below in
Sect. \ref{Envelope_pos}. We also tested smaller
$\boldsymbol{\delta}_{\mathrm{outer}}$ values and found no significant
biases when changing this parameter. We count only pixels along
filaments and determine thus the local velocity dispersion $\mathcal{V}$
only along the filaments.

Figure \ref{Fig_V} shows the distributions of velocity dispersions
$\mathcal{V}$ for three different settings with channel widths of 1, 8,
and 16 \kms.  For 1 \kms\ the filament velocities have a typical scatter
$ \mathcal{V} \sim 6 $ \kms\ all sky and $ \mathcal{V} \sim 5.5 $ at high
latitudes. Close to the Galactic plane there is evidence for some
confusion, causing the bump in the distribution function at $
\mathcal{V} \sim 15$ \kms\ (see also Fig. \ref{Fig_Vel_HI},
top). $\mathcal{V}$ includes systematical effects from linear velocity
gradients. The measured velocity dispersions contain also statistical
uncertainties in the determination of $v_{\mathrm{fil}}$ caused by the
limited velocity resolution of 1 \kms\ (Sect. \ref{v_fil}).
$\mathcal{V}$ can therefore only partly be caused by fluctuations in the
velocity field, the FIR and \hi\ filaments are rather well defined and
coherent in velocity space.

Numerical studies of the condensation of the WNM into CNM structures
under the effect of turbulence and thermal instability by
\citet{Saury2014} indicate that the velocity field of the CNM reflects
the velocity dispersion of the WNM. These authors estimate a CNM
cloud-to-cloud velocity dispersion of 5.9 \kms, in good agreement with
Fig. \ref{Fig_V} in the case of a 1 \kms\ channel width. The velocity
dispersion distributions for the 8 and 16 \kms\ integrated maps deviate
by construction significantly from these theoretical estimates.

It remains to be discussed whether the correlation between FIR and
  \hi\ filaments at 1 \kms\ channel width could be flawed because the
  S/N in individual \hi\ channel maps is much lower than in the
  integrated map. The Hessian analysis of individual channel maps may
  increase the uncertainties even more and with 101 opportunities a
  channel with a concordant angle can nearly always be found.

To verify this assertion we consider how far noise degrades the
\hi\ data \citep[][Sect. A2.3]{Kalberla2019}. The system noise $T_{
  \mathrm{sys} }$ contains several independent contributions. Most
important is the thermal noise from the receiver system and the
elevation-dependent ground radiation including spill-over.  These
components are variable, but we use here an average thermal contribution
$T_{ \mathrm{sys} } = 30$ K. The line signal $T_{ \mathrm{B}}
(v_{\mathrm{LSR}}) $ adds to the noise contribution $T_{ \mathrm{Noise}}
(v_{\mathrm{LSR}}) $ that can be approximated as \citep{Haud2000}
\begin{equation} 
  T_{ \mathrm{Noise}} (v_{\mathrm{LSR}}) = \sigma_{ \mathrm{av}} [T_{
      \mathrm{sys}}  + T_{ \mathrm{B}} (v_{\mathrm{LSR}}) ] / T_{\mathrm{sys}},
\label{eq:noise}
\end{equation}
where $\sigma_{ \mathrm{av}}$ is the average noise level in the
baseline, determined at velocities without \hi\ line emission. We use
this approximation of the radiometer equation to increase artificially
the noise for all observed \hi\ line profiles step by step by a factor
of two. For each of these cases all steps of the data processing
pipeline are then repeated. Figure \ref{Fig_Noise} shows the
results. The velocity dispersions increase significantly and $
\mathcal{V} $ is already unacceptable large for a noise enhancement by a
factor $N_f = 2$. Likewise we are unable to get lucky imaging with
higher additional noise injection.

Figures \ref{Fig_V} and \ref{Fig_Noise}, using the dispersions
$\mathcal{V}$, are somewhat abstract and it may be difficult to realize
the consequences. In Fig. \ref{Fig_NoiseMaps} we demonstrate therefore
that the derived coherent velocity field in the upper panel is degraded
significantly when enhancing the telescope noise level by a factor of
$N_f = 4$. In the lower panel of Fig. \ref{Fig_NoiseMaps} it is just
possible to recognize a few \hi\ filaments as coherent
structures. Considering a noise amplification by a factor $N_f = 8$ (not
shown), the image is completely decorrelated. Likewise it is not
possible to generate or improve coherent structures for larger values of
$N_f$. This result reflects a well known physical experience; the
additional noise that we consider here is incoherent (independent in
position and velocity) and by adding incoherent events it is not
possible to generate coherent structures with a high correlation
coefficient. Generating a coherent result in N random realizations as a
sum of incoherent events demands N to be virtually infinite. In other
words: the \hi\ structures that are derived from Hessians at 1
\kms\ resolution are not random but highly correlated, not only in
position but also in velocity.  Observers are working hard to get
significant data. The average noise term in Eq. (\ref{eq:noise}) scales
with the integration time $\tau$ for individual positions as $\sigma_{
  \mathrm{av}} \propto \tau^{-1/2}$.  The velocity field presented at
bottom of Fig. \ref{Fig_NoiseMaps} simulates observations that are
therefore unacceptable short in observing time by a factor of 16.

Observers common experience is also that it is necessary to resolve
structures to be analyzed. In the case of \hi\ lines, CNM structures with an
approximate FWHM width of 3 \kms\ are common and according to the
Nyquist sampling theorem \citep{Shannon1975} a spectral resolution of at
least 1/2 of this value is necessary, independent of any other
requirements concerning the noise level. The possibility that
  averaging over broader channels would increase the correlation by
  reducing noise violates the sampling theorem.  Applying such an
averaging causes a loss of information about the \hi\ structures and
does according to Fig. \ref{Fig_V} not improve the velocity dispersion
along the filaments. Also from Fig. \ref{Fig_Aligne_857} and Table
\ref{table:1} we find for none of the different alignment measures any
evidence for an improved correlation with increasing channel width.

This subsection on the velocity dispersion summarizes one of the most
important results from our investigations. Comparing local orientation
angles $\theta$ in the FIR at 857 GHz along filaments with
\hi\ orientation angles for 101 channels it is no surprise to find some
agreement at one of these velocity channels. For structures that are
unrelated this should result in a more or less random velocity
distribution. Along the filaments we find however a surprisingly close
agreement of the orientation angles with a well-defined coherence in
velocity space (see Figs. \ref{Fig_NoiseMaps} and \ref{Fig_Vel_HI}, top),
a necessary condition for alignment between FIR and \hi\ structures
along magnetic field lines as conjectured by \citet{Clark2018} and
\citet{Clark2019b}.

\begin{figure*}[thp] 
   \centering
   \includegraphics[width=14.5cm]{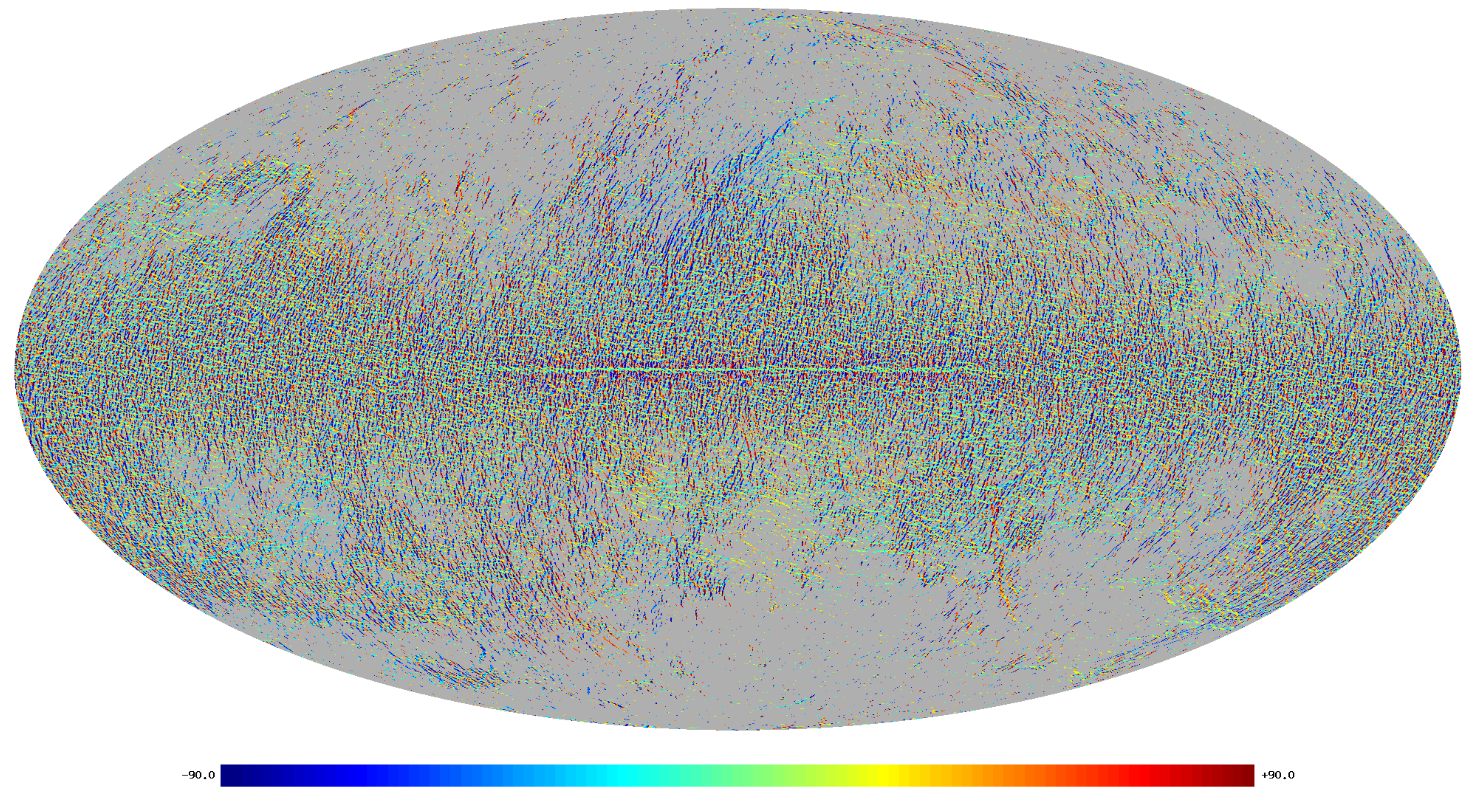}
   \includegraphics[width=14.5cm]{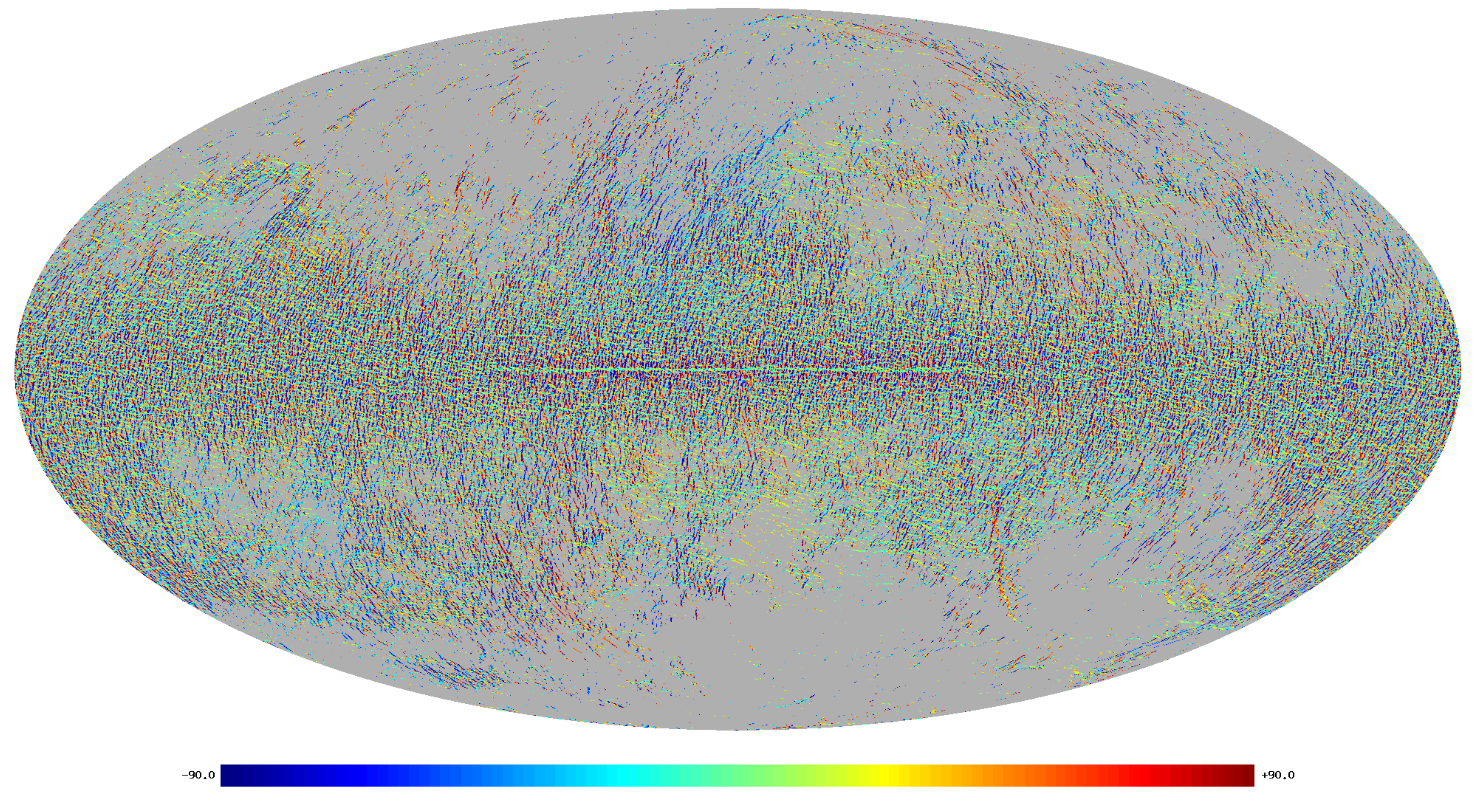}
   \includegraphics[width=14.5cm]{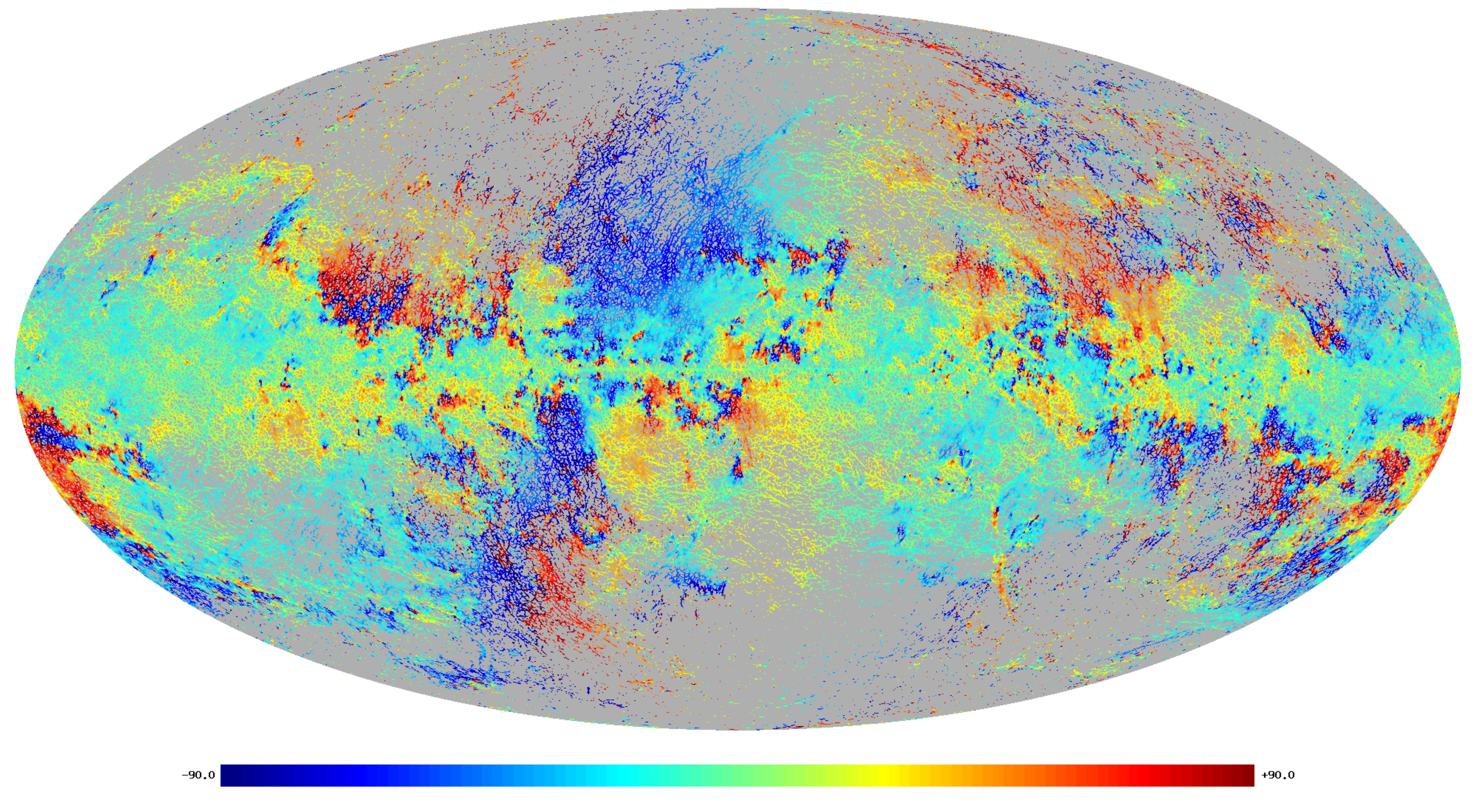}
   \caption{Orientation angles according to Eq. (\ref{eq:theta}). Top:
     From the FIR distribution at 857 GHz. Middle: From the best fit
     \hi\ distribution. Bottom: Distribution of orientation angles
     according to Eq. (\ref{eq:theta2}) calculated from Stokes parameters
     $U$ and $Q$ of the 353 GHz FIR distribution.}
   \label{Fig_Theta}
\end{figure*}

\subsection{FIR and \hi\ alignment conditions }

To clarify conditions that lead to a good alignment between FIR and
\hi\ filaments, we need to sort out several effects that influence the
data analysis.

First of all we need to consider data processing and observational
uncertainties. The application of the Hessian operator, thus the
calculation of first- and second-order derivatives, leads unambiguously to
an amplification of uncertainties. In addition the definition of the 5x5
Hessian matrix, applied to a HEALPix data cube introduces a spatial
filtering, depending on the dimension defined by the nside parameter.

To minimize the uncertainties, also to minimize the calculation efforts,
it is reasonable to use Gaussian derivatives for the calculation of the
Hessian matrix \citep[e.g.,][]{Soler2020}. We do not follow this track
but apply smoothing and calculation of the Hessian matrix independently,
allowing $ H(x,y) $ to be determined for smoothed FIR and unsmoothed
\hi\ data with the same software. We adapt a Gaussian smoothing to the
FIR data only, but with a FWHM smoothing kernel that matches the 5x5
pixel size of the Hessian operator. Assuming that the 857 and 545 GHz
{\it Planck} FIR data are essentially a representation of similar ISM
structures we reproduce the angular orientation of filamentary
structures in 857 and 545 GHz FIR data according to Eq. (\ref{eq:theta})
within $ \sigma_{\mathrm{Voigt}} \sim 4\fdg3 $ all sky and $
\sigma_{\mathrm{Voigt}} \sim 7\fdg5 $ toward high latitudes. The scatter
in $\delta \theta$ decreases by about a factor of four if we apply a
Gaussian smoothing. Applying the same alignment test to smoothed 857 and
353 GHz FIR data, we find that the dispersions $\sigma_{\mathrm{Voigt}}$
between both FIR data sets increase roughly by a factor of 3.5; we
explain this by sensitivity limitations, caused by the significant total
intensity difference of the CMB and CIB in both frequency bands
  (\citet{Odegard2019}, Fig. 3). Using angular orientations from
Stokes parameters at 353 GHz in comparison to those from 857 GHz we find
also an increased mismatch. As mentioned above, this mismatch is only
partly due to sensitivity limitations. The application of the Hessian
matrix causes a reduced sensitivity to low spatial frequencies but in
Sect. \ref{Curvature_dist} it will become clear that spatial filtering
does not affect our analysis severely.

Considering similarities between 857 GHz FIR and \hi\ we find some
alignment for column densities but the correlation is weak. Searching
for an angular alignment between FIR and \hi\ in single channels we
obtain a far better agreement but only for a velocity resolution of
$\delta v_{\mathrm{LSR}} = 1$ \kms\ the agreement is perfect. The
alignment measures are close to that of our test case at 857 and 545
GHz. Such an alignment can only be meaningful if the coherence between 
FIR and \hi\ can be verified also in velocity space. Continuity in velocity
along the filaments is discussed in Sect. \ref{Vel_disp} and later in Sect. \ref{spatial_dist}. For the rest of this
paper we consider only the best fit \hi\ filaments with a velocity
resolution of $\delta v_{\mathrm{LSR}} = 1$ \kms.

To demonstrate the small-scale structure of orientation angles $\theta$
in filaments we show in Fig. \ref{Fig_Theta} in the top and middle
panels the distributions of $\theta$ for FIR at 857 GHz and for the
\hi\ in narrow velocity intervals. Both distributions are almost
indistinguishable. At the bottom we display $\theta_S$ from Eq.
(\ref{eq:theta2}), calculated from 353 GHz $U$ and $Q$ Stokes maps.
Still, there is some correlation of this with maps in the upper panels
(see also Table \ref{table:1}) but it is obvious that parameters derived
from the Stokes maps differ from those derived by a Hessian analysis.
Large-scale structures are suppressed by the Hessian analysis despite
the fact that all maps in Fig. \ref{Fig_Theta} have a similar spatial
resolution. Some of the narrow structures resemble chromospheric fibrils
that tend to be aligned with the magnetic field
\citep{Asensio2017}. For the interpretation we need to be careful
with respect to spatial filtering caused by the Hessian operator
\citep{Aragon2007}. Only at the very end of our paper in
Sect. \ref{Curvature_dist} it will become clear that neither sensitivity
limitations nor biases caused by spatial filtering are affecting our
conclusions severely.

The focus of our discussions in this section went from data processing
to more physical influences like velocity widths and scale sizes. In
the following we intend to explore astrophysical issues.

\section{Coherence conditions }
\label{Coherence}

The close angular alignment between FIR emission and \hi\ filaments at
narrow velocity intervals implies that the FIR emission of such
filaments has to come from regions where gas and dust are well mixed and
in dynamical equilibrium. The best fit local \hi\ velocities
$v_{\mathrm{fil}}(l,b)$ are accordingly characteristic for the filamentary
FIR emission structures. In this section we explore conditions and
limitations for FIR and \hi\ coherence in the ISM.

\subsection{\hi\ Doppler temperatures }
\label{T_D}

\begin{figure}[th] 
   \centering
   \includegraphics[width=9cm]{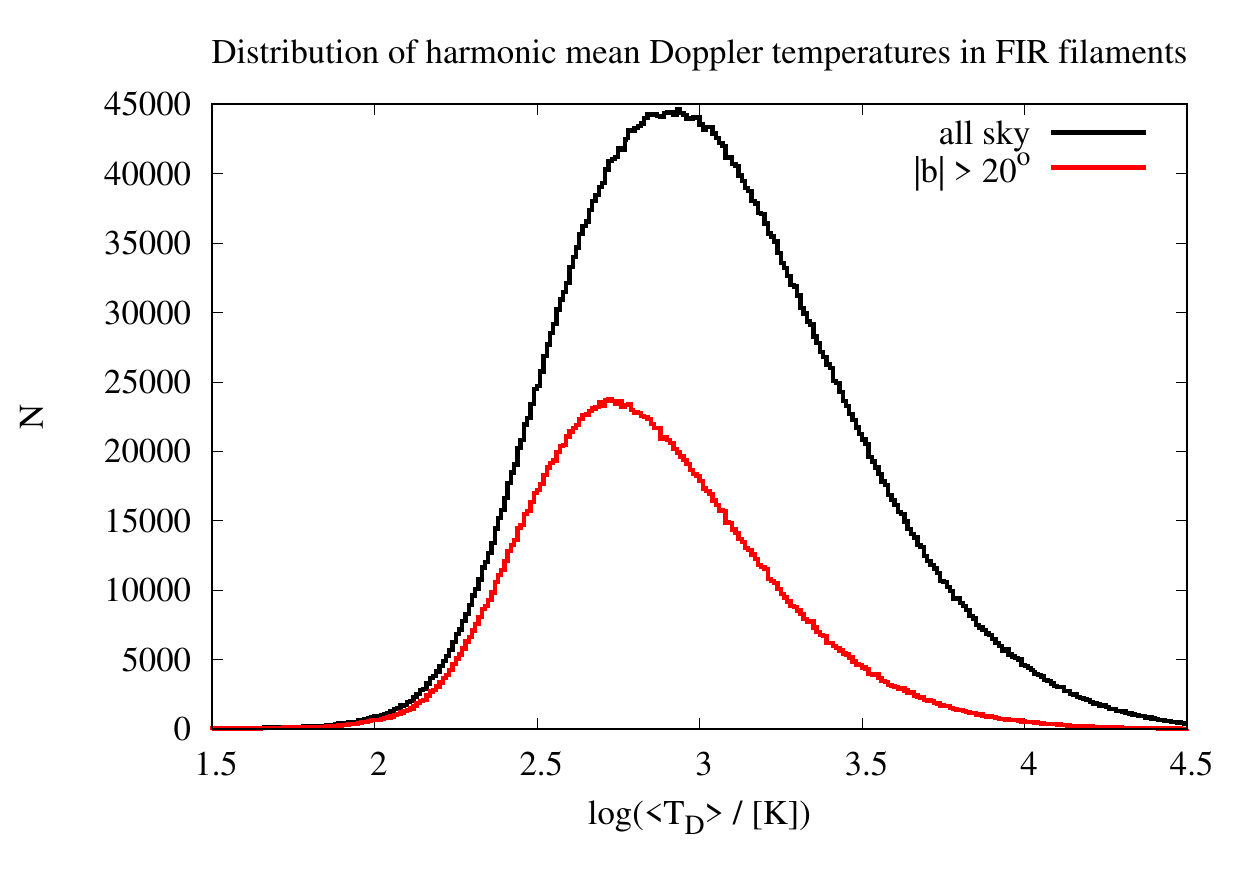}
   \caption{Distribution of harmonic mean \hi\ Doppler temperatures in
     coherent FIR and \hi\ filaments. }
   \label{Fig_histo_TD}
\end{figure}

\hi\ filaments are cold with median Doppler temperatures of 220 K
(e.g., \citet{Clark2014}, \citet{Kalberla2016}). For single dish
observations we cannot determine excitation temperatures, we can only
use a Gaussian decomposition to derive Doppler temperatures
$T_{\mathrm{D}}$ as upper limits of the true exitation temperatures of
individual \hi\ components. In the case of several such components, we need
to use harmonic mean Doppler temperatures defined as the brightness
temperature ($T_{\mathrm{B}}$) weighted average Doppler temperature for all Gaussian
components along the line of sight,
\begin{equation}
  \langle T_{\mathrm{D}}(v_{\mathrm{LSR}}) \rangle = \frac{ \int T_{\mathrm{B}}(v_{\mathrm{LSR}}) T_{\mathrm{D}}\ d v_{\mathrm{LSR}}}
        {\int T_{\mathrm{B}}(v_{\mathrm{LSR}})\  d v_{\mathrm{LSR}}}.
\label{eq:T_D}
\end{equation}
This $\langle T_{\mathrm{D}}(v_{\mathrm{LSR}}) \rangle$ definition is
analogous to the definition of optical depth weighted average spin
temperatures $\langle T_{\mathrm{S}}(v_{\mathrm{LSR}}) \rangle$
(e.g., \citet{Murray2020}). Figure \ref{Fig_histo_TD} shows the
log-normal distribution of $\langle T_{\mathrm{D}}(v_{\mathrm{fil}})
\rangle$ for \hi\ filaments at high latitudes $|b| > 20 \degr$ and all
sky. While the all sky distribution may suffer from confusion caused by
the Galactic plane, the high latitude data indicate that most of the gas
in the filaments belongs to the CNM with low Doppler temperatures.

\subsubsection{\hi\ phase composition in filaments }
\label{phase_fil}

\begin{figure}[th] 
   \centering
   \includegraphics[width=9cm]{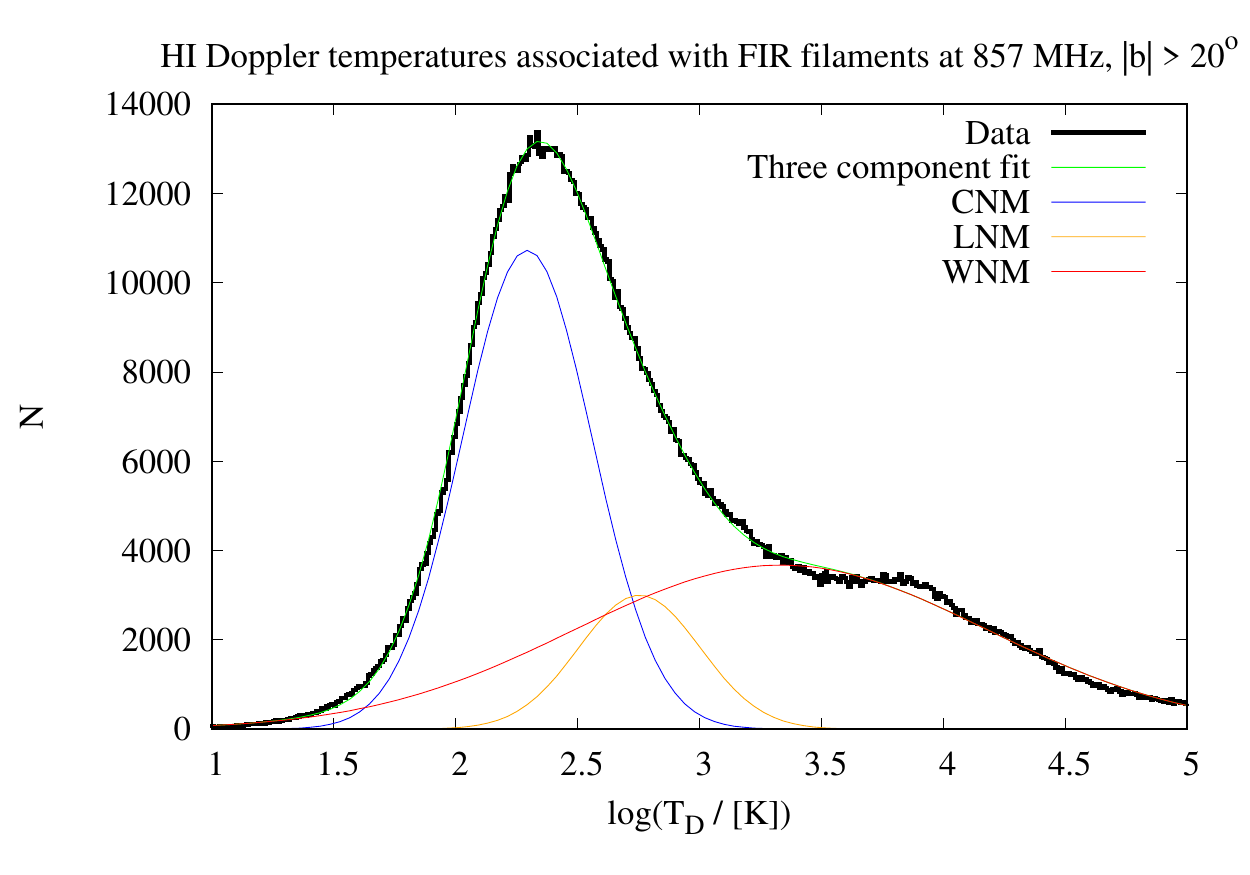}
   \includegraphics[width=9cm]{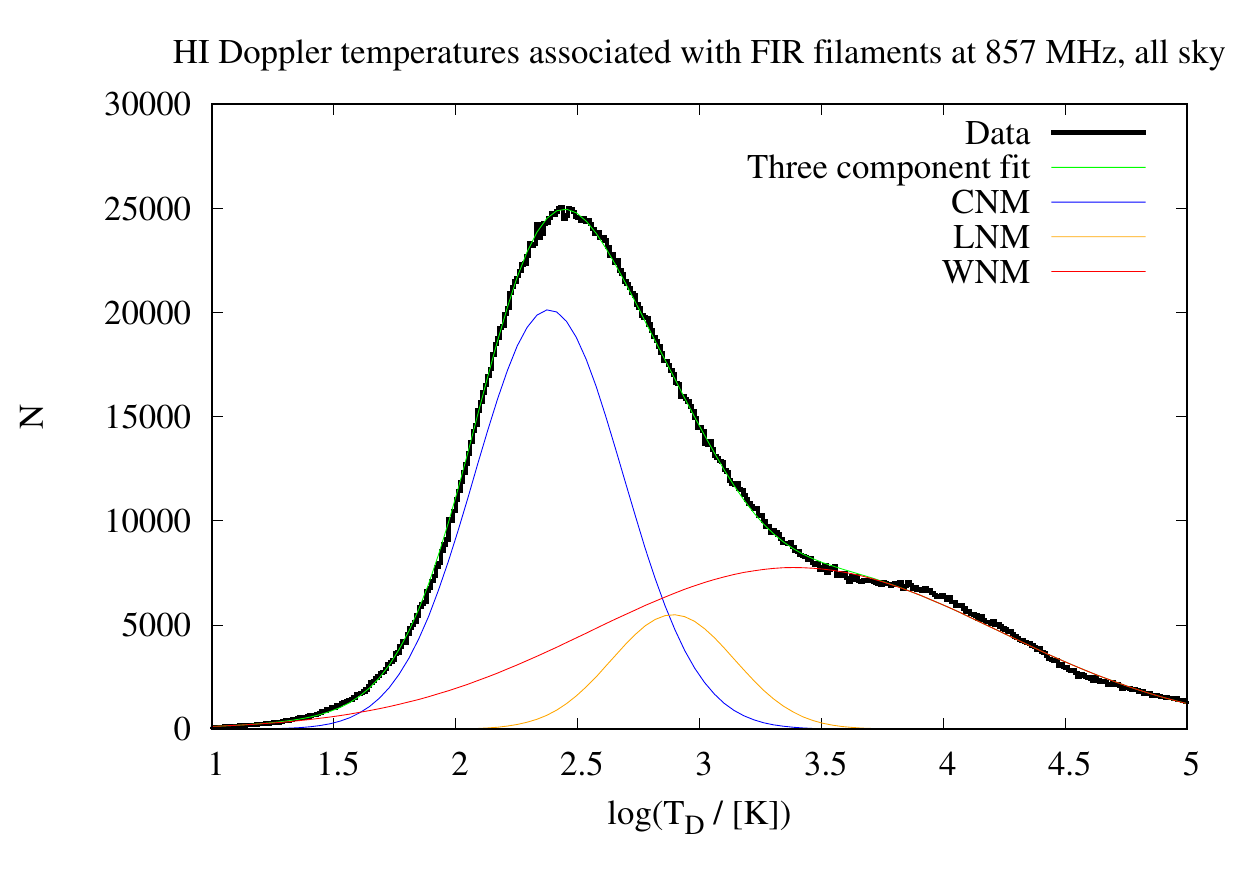}
   \caption{Distribution of \hi\ Doppler temperatures of coherent FIR
     and \hi\
     filaments compared with a Gaussian three-component fit. Top: High
     latitude data, $|b| > 20 \degr$. Bottom: All sky. }
   \label{Fig_histo_all}
\end{figure}

To determine the phase composition of the coherent FIR and \hi\ filaments we
search at each position along the filaments for the Gaussian component
with the center velocity $v_{\mathrm{Gauss}}$ with the least deviation $
|v_{\mathrm{fil}} - v_{\mathrm{Gauss}}| $ from the filament velocity. We
determine the Doppler temperature $ T_{\mathrm{D}} = 21.86\ \Delta v^2 $
of this component, here $\Delta v$ is the FWHM velocity width of this
Gaussian.

Figure \ref{Fig_histo_all} shows the derived log-normal distribution of
\hi\ Doppler temperatures in filaments. We also apply a three-component
fit to the $ T_{\mathrm{D}} $ distribution. At high Galactic latitudes
we find the CNM as the dominant component with a geometric mean $
T_{\mathrm{D, CNM}} = 196 $ K. We get for the unstable lukewarm neutral
medium (LNM) $ T_{\mathrm{D, LNM}} = 567 $ K and $
T_{\mathrm{D, WNM}} = 2156 $ K for the WNM.  These values can be
compared to characteristic values at high Galactic latitudes $
T_{\mathrm{D, CNM}} = 283 $ K, $ T_{\mathrm{D, LNM}} = 2014 $ K, and $
T_{\mathrm{D, WNM}} = 11879 $ K, as determined by \citet{Kalberla2018}
from a Gaussian analysis, unconstrained to filamentary structures. The
formal errors of these values are low, around 1\%. We conclude that
\hi\ in coherent FIR and \hi\ filaments is for all phases significantly
colder than the \hi\ outside such filaments.

\subsubsection{\hi\ phase composition in blobs }
\label{blobs}

The Hessian operator allows two
eigenvalues, representing filaments and local enhancements (blobs) along
the filaments,  to be distinguished with Eq. (\ref{eq:lambda}). Figure \ref{Fig_lam} shows that there are numerous
local enhancements at a resolution of 18\arcmin. Such structures are
well defined at the resolution of 14\farcm5 for the GASS and 10\farcm8
for the EBHIS. We repeat the analysis from Sect. \ref{phase_fil} and
determine the phase distribution for these \hi\ blobs.

\begin{figure}[th] 
   \centering \includegraphics[width=9cm]{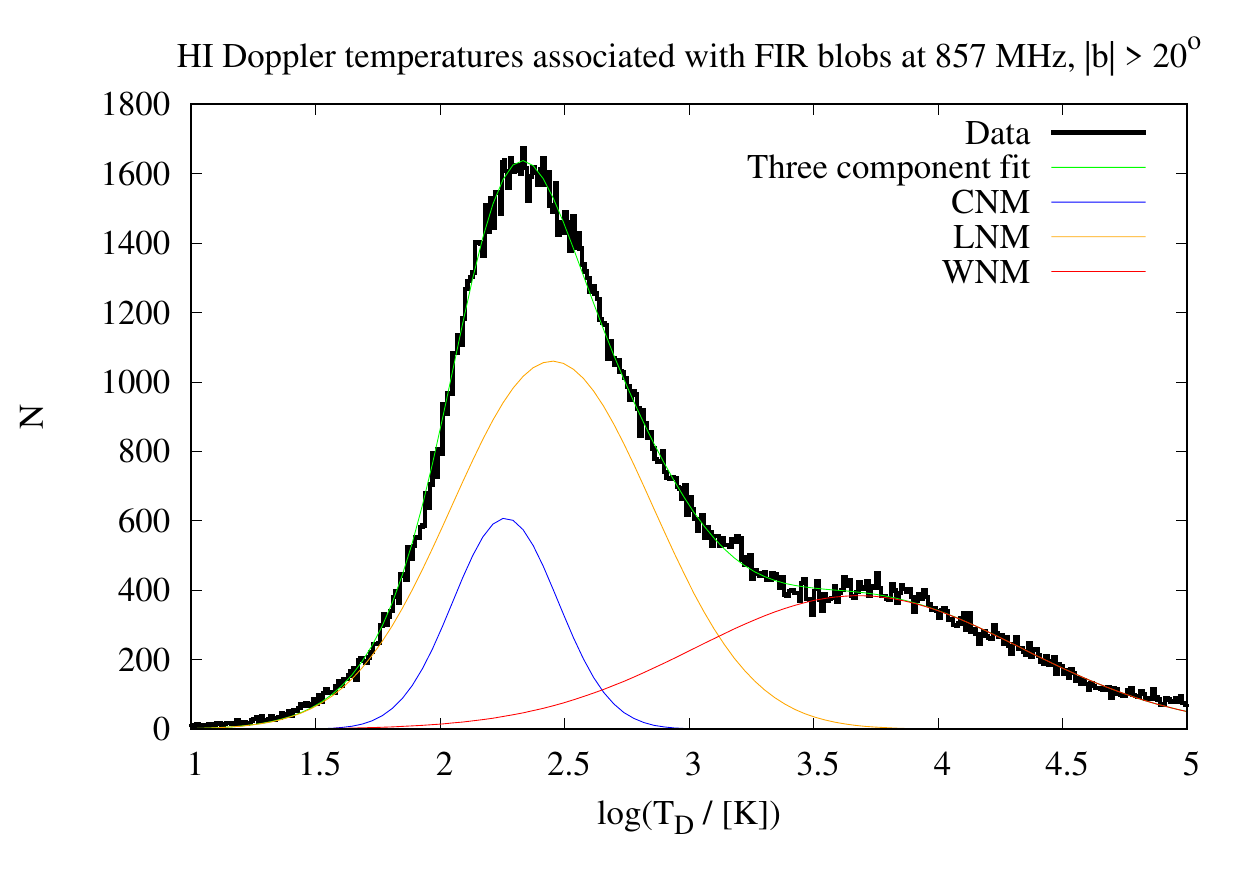}
   \includegraphics[width=9cm]{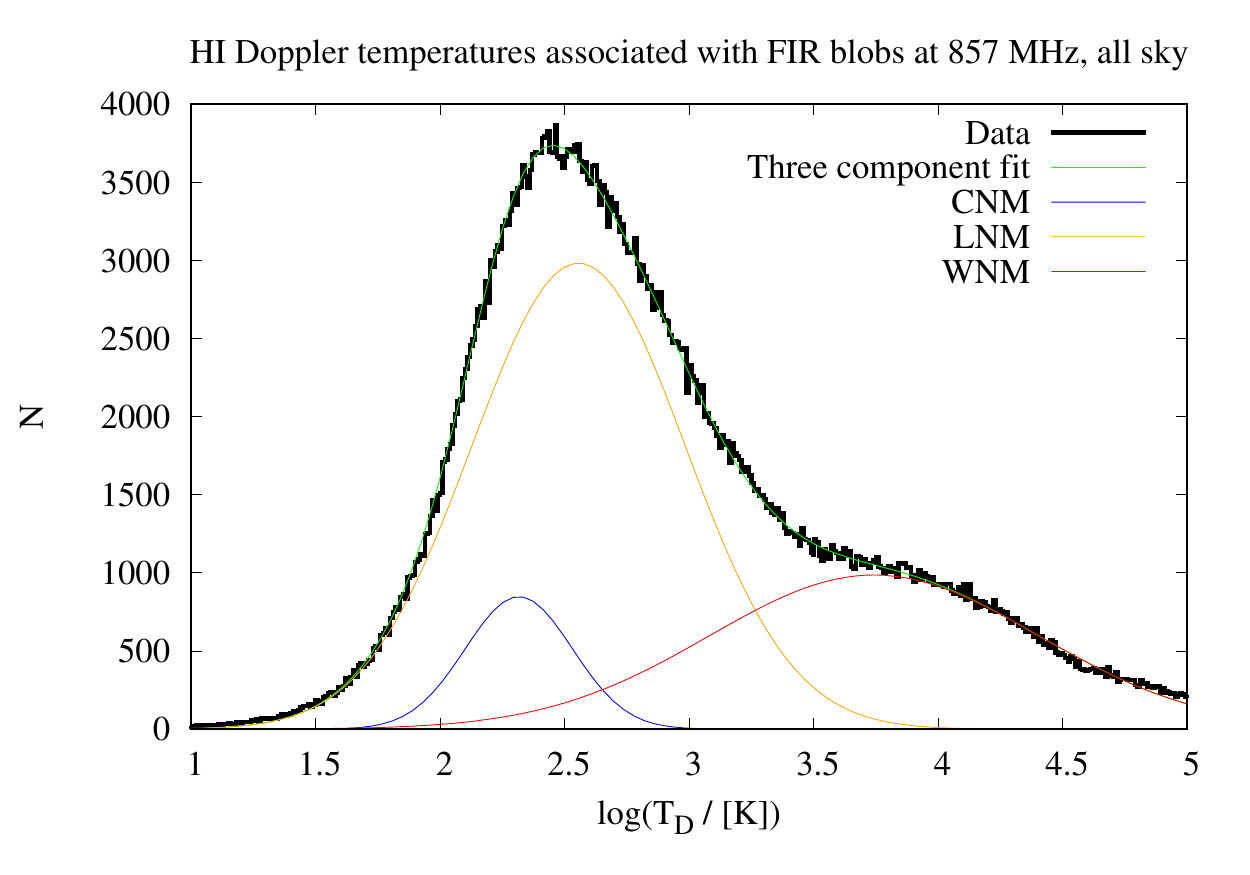}
   \caption{Distribution of \hi\ Doppler temperatures of local
     enhancements (blobs) along filaments and their
     approximations by Gaussian distributions (green). To distinguish
     the distributions from \hi\ phases, we separate the CNM (blue) from
     the warmer phase, LNM and WNM (red). }
   \label{Fig_histo_vel_blob}
\end{figure}

At high Galactic latitudes we find with Fig. \ref{Fig_histo_vel_blob} a
$ T_{\mathrm{D}} $ distribution that is in its shape similar to
Fig. \ref{Fig_histo_all} for the filaments. The CNM with a geometric
mean $ T_{\mathrm{D, CNM}} = 183 $ K is cold but the LNM with $
T_{\mathrm{D, LNM}} = 282 $ K dominates the low temperature wing. This
part is colder than the LNM of the filaments. Opposite for the WNM, here
we get with $ T_{\mathrm{D, WNM}} = 4766 $ K somewhat higher
temperatures in comparison to the filaments. All sky we obtain
$T_{\mathrm{D, CNM}} = 207 $ K, $ T_{\mathrm{D, LNM}} = 360 $ K and $
T_{\mathrm{D, WNM}} = 5528 $ K. In summary, local enhancements along
filaments contain also predominantly cold \hi\ gas.  The formal
parameters for the phase decompositions in Figs. \ref{Fig_histo_all} and
\ref{Fig_histo_vel_blob} may not be very well defined but we clearly
find in all phases significantly lower Doppler Temperatures than the
averages determined previously by \citet{Kalberla2018}. Comparing these
blobs with sources from the {\it Planck} catalog of Galactic cold
clumps \citep{Planck2016b} we find that only 2397 out of 13242 sources
listed in this catalog belong to the population of blobs. We conclude
that most of the blobs shown in Fig. \ref{Fig_lam} share the
properties of the filamentary structures; they are just condensations
along the filaments.

\subsection{Velocity distribution }
\label{vel_dist}

\begin{figure}[th] 
   \centering
   \includegraphics[width=9cm]{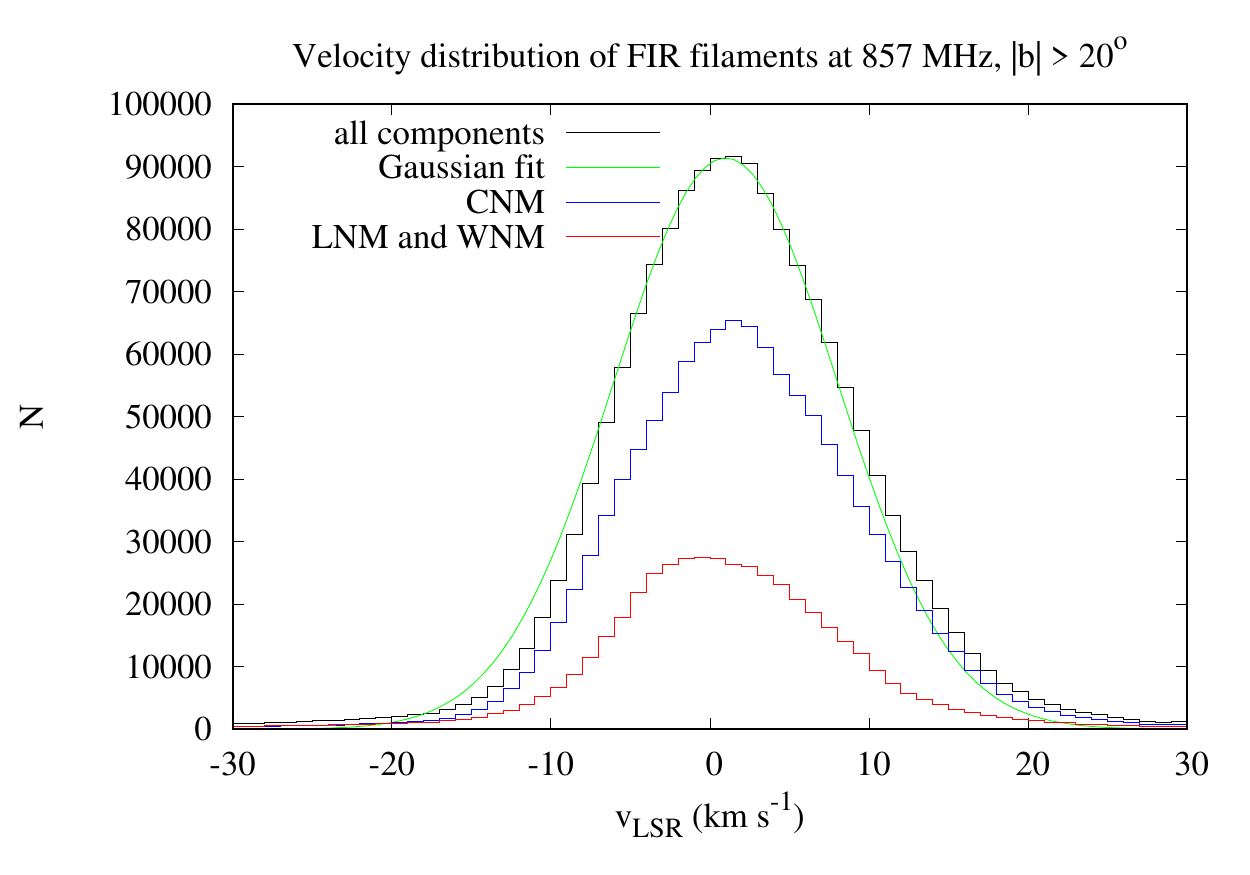}
   \includegraphics[width=9cm]{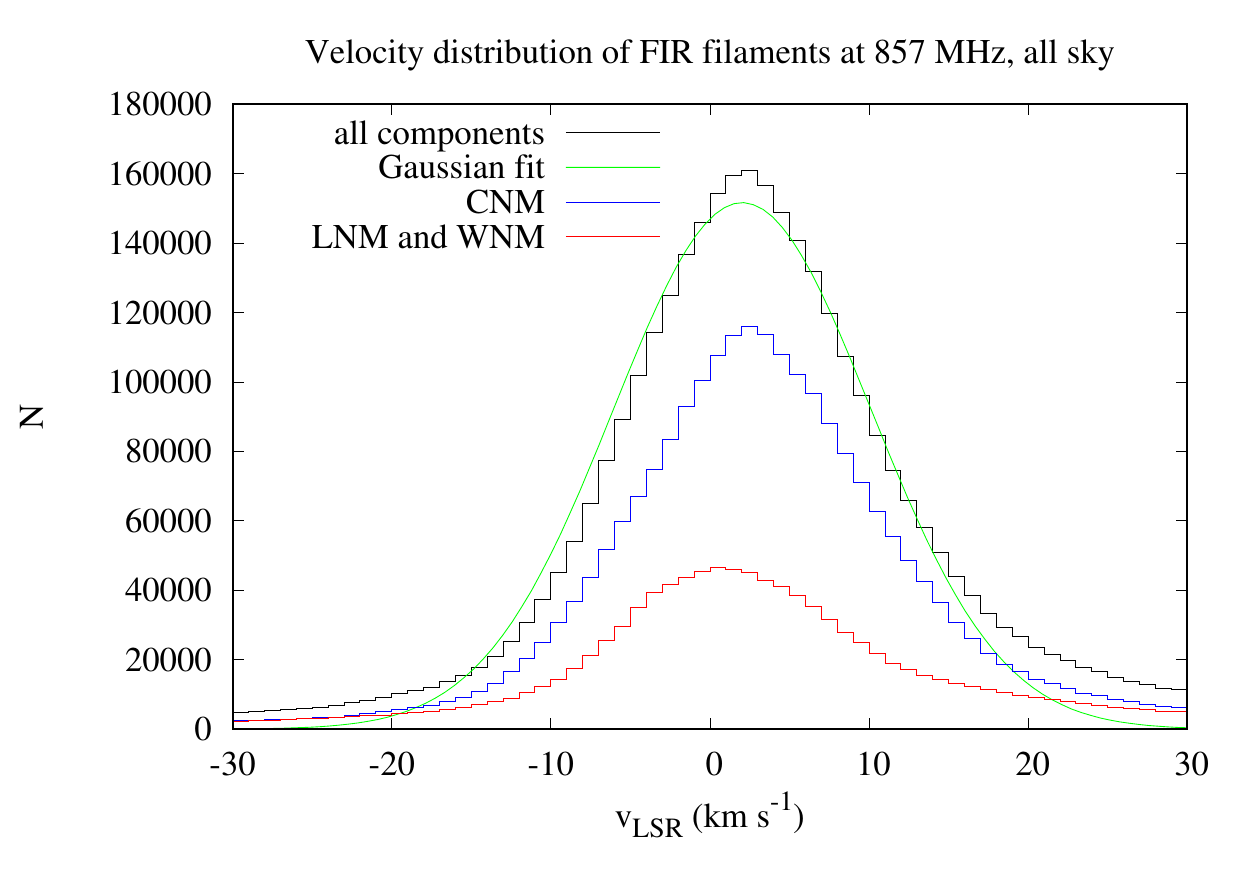}
   \caption{Histograms of the velocity distributions for filamentary
     structures and their approximations by Gaussian distributions
     (green). To distinguish the distributions from \hi\ phases, we
     separate the CNM (blue) from the warmer phase, LNM and WNM
     (red). }
   \label{Fig_histo_vel}
\end{figure}

Figure \ref{Fig_histo_vel} shows the velocity distributions for \hi\ in
coherent FIR and \hi\ filaments. Filaments are local phenomena, they do not
share high rotational velocities from the rotation curve in the Galactic
plane. Using unconstrained all sky data we find that filaments are
mostly in the range $ |v_{\mathrm{LSR}}| \la 30 $ \kms.  At high
Galactic latitudes we find barely filaments with $ | v_{\mathrm{LSR}}| >
20 $ \kms. Our analysis covers the range $ | v_{\mathrm{LSR}}| < 50 $
\kms\ but this limited range is complete and allows an unbiased
description of filamentary structures. A Gaussian fit to the velocity
distribution leads to a FWHM width of 16.6 \kms\ at high latitudes and to
19.1 \kms\ for the all sky case and confirms the narrow velocity
distribution of small-scale filamentary structures derived previously
from USM by \citet{Kalberla2016}.

To check whether filament velocities may depend on the phase composition
we distinguish velocity distributions according to contributions from
the CNM and the warmer phases. Figure \ref{Fig_histo_vel} shows that the
warmer contributions tend to be asymmetric, indicating dynamical
interactions.

\subsection{Polarization: \hi\ coherence in velocity }
\label{HI_pol}

\citet{Clark2018} investigated the alignment of \hi\ features with the
plane-of-sky magnetic field orientation. To quantify the correlation
between \hi\ structures and the magnetic field as characterized by the
polarization fraction of 353 GHz dust emission, she proposed characterizing \hi\ structures by their \hi\ coherence or
\hi\ polarization. This metric is defined for the degree of coherence of
the \hi\ orientation as a function of velocity by Stokes-like parameters
$U_\mathrm{HI}$ and $Q_\mathrm{HI}$. \citet{Clark2014} and
\citet{Clark2019b} use the orientation of \hi\ structures with angles
derived from a RHT.  Here we use angles $\theta$
according to Eq. (\ref{eq:theta}), derived after application of the
Hessian operator. These are weighted by the local \hi\ brightness
temperature $T_{\mathrm{B}}$ and integrated over the line of sight velocity
$v_{\mathrm{LSR}}$,
\begin{equation}
U_\mathrm{HI} = \int T_B(v_{\mathrm{LSR}})\ \mathrm{cos}( 2 \theta (v_{\mathrm{LSR}}) )\ dv_{\mathrm{LSR}}
\label{eq:U_HI}
\end{equation}
and
\begin{equation}
Q_\mathrm{HI} = \int T_B(v_{\mathrm{LSR}})\ \mathrm{sin}( 2 \theta (v_{\mathrm{LSR}}) )\ dv_{\mathrm{LSR}}
\label{eq:Q_HI}
,\end{equation}
and the \hi\ coherence \citep{Clark2014} or \hi\ polarization
\citep{Clark2019b} is defined as
\begin{equation}
p_\mathrm{HI} = \frac{ \sqrt{U_\mathrm{HI}^2 + Q_\mathrm{HI}^2}} {\int
  T_B(v_{\mathrm{LSR}})\ dv_{\mathrm{LSR}} }.
\label{eq:t_HI}
\end{equation}
\begin{figure}[th] 
   \centering
   \includegraphics[width=9cm]{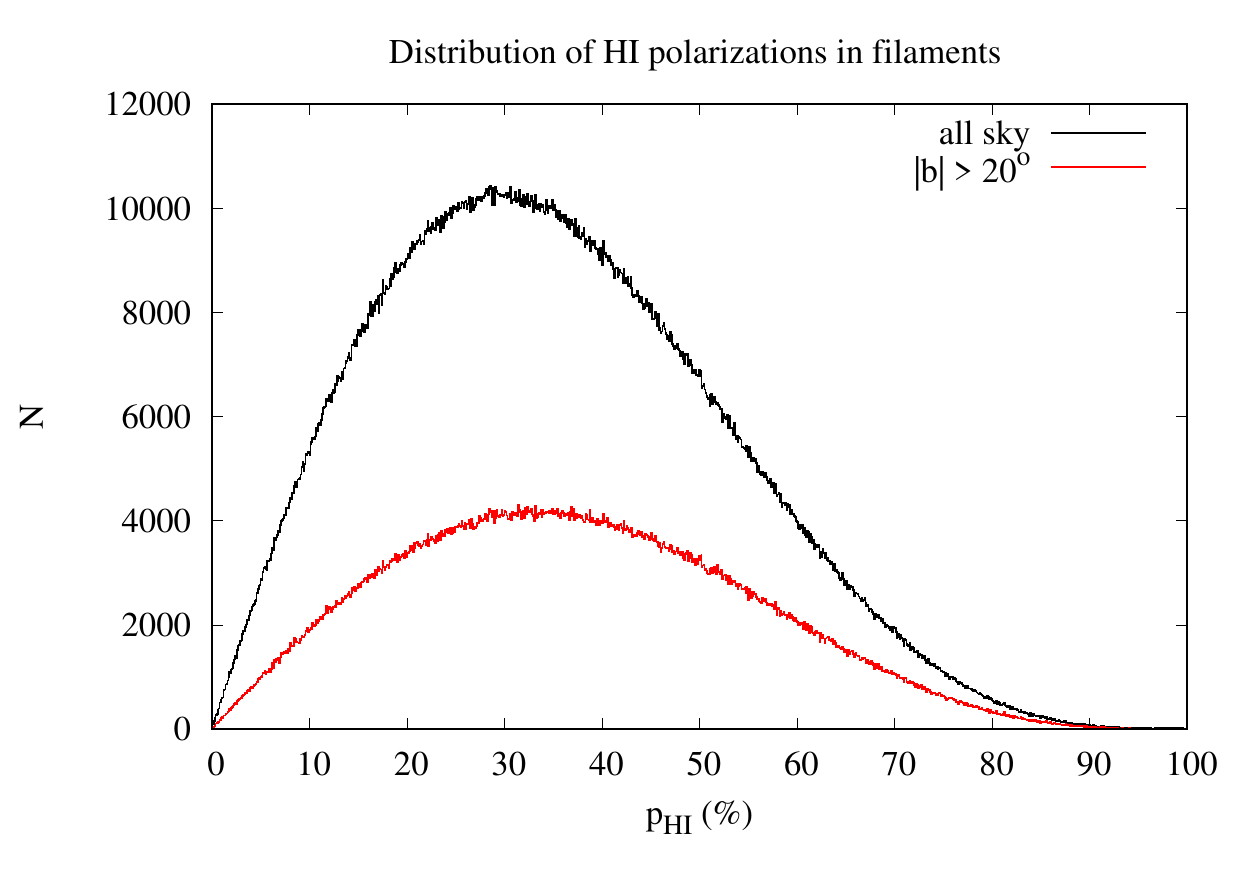}
   \caption{Distribution of the \hi\ polarization $p_\mathrm{HI}$ according
     to Eq. (\ref{eq:t_HI}), all sky and for high latitudes at $|b| > 20
     \degr$. }
   \label{Fig_HI_coherence}
\end{figure}

\citet{Clark2019b} applied this concept to map magnetically coherent
regions of space. They have proven the existence of \hi\ regions that
are highly correlated with the 353 GHz Q and U maps of polarized dust
emission observed by {\it Planck} on scales of 80\arcmin. Comparing 857
GHz data on scales of 18\arcmin\ with \hi\ emission from GASS and EBHIS
we use this concept for our high resolution analysis and display the
distribution of derived \hi\ polarization data $p_\mathrm{HI}$ in
Fig. \ref{Fig_HI_coherence}. These distributions are close to Gaussians
with peaks at $p_\mathrm{HI} \sim 0.3 $ but with extended wings up to
$p_\mathrm{HI} \sim 0.8 $. As discussed below in
  Sect. \ref{pol_disp} in more detail, the dust polarization fraction is
  limited and $p_\mathrm{HI}$ is not representative for the dust.

The derived distributions for the \hi\ polarization $p_\mathrm{HI}$ in
Fig. \ref{Fig_HI_coherence} are remarkably well defined and the peaks
appear only little affected by confusion effects in the Galactic
plane. On scales of 18\arcmin\ $p_\mathrm{HI}$ is significantly enhanced
in comparison to previous low resolution
investigations. \citet{Planck2016} determine from 353 GHz data on
multipole scales $30 \la l \la 300$ a mean polarization fraction of
magnetized filaments of 11\%. \citet{Clark2018} and \citet{Clark2019b}
report \hi\ polarizations around 15\% on scales of 90\arcmin\ and
80\arcmin,\ respectively, with a maximum near 30\%. From 353 GHz {\it
  Planck} data at 1\degr resolution \citet{Planck2015} determine a
maximum polarization of 19.8\% and \citet{Planck2020a} derive a maximum
of 22\% at 80\arcmin\ resolution.

For all these publications the spatial filters were chosen in such a way
that they highlight all the bright filaments. Our aim is to compare FIR
and \hi\ filaments at the highest possible resolution with the best
sensitivity. Our data are matched to the 18\arcmin\ resolution of the
Hessian operator and can only be consistent with previous results if
there are significant small-scale effects along the filaments and within
the analyzed beam.  The CNM as the coldest and densest part of the
\hi\ distribution dominates the structure of the filaments. The densest
portions are in the very centers of the filaments and the magnetic field
strengths correlates with volume density. These cold and dense parts
cause the local bending for both, FIR and \hi\ filaments with
fluctuations in $\theta$. Such a tangling must cause a de-correlation of
the polarization observed at larger scales.

\begin{figure}[th] 
   \centering
   \includegraphics[width=9cm]{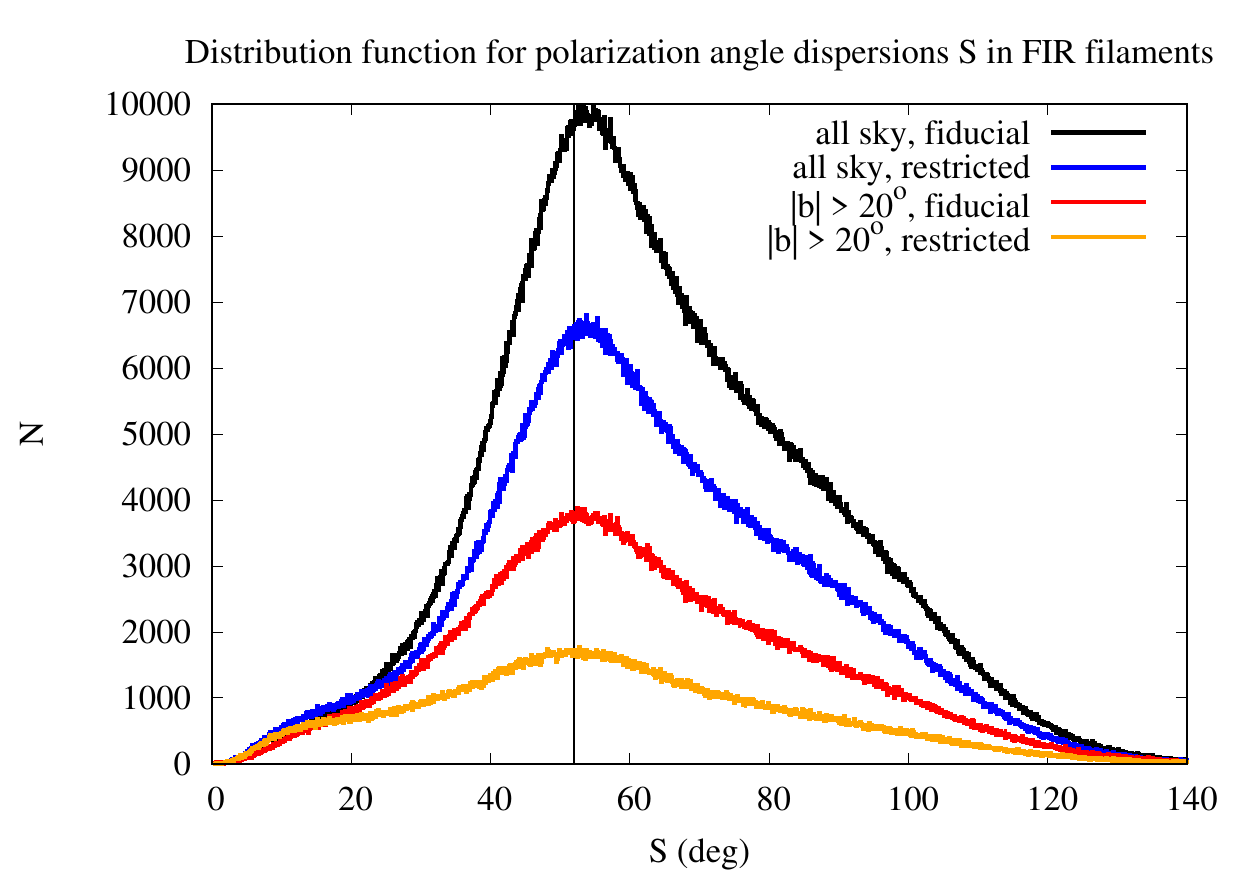}
   \caption{Distribution of the polarization angle dispersion
     $\mathcal{S}$ along FIR filaments at a spatial resolution of
     18\arcmin. The peak at $\mathcal{S} = 52\degr$, annotated with a
     vertical line, indicates that many positions have significant
     deviations of polarization angles along the filaments. The black and
     red lines mark filaments with eigenvalues $\lambda_- <
     -1.5\ \mathrm{K/deg}^{-2}$, and blue and orange lines represent 
     filaments limited to $\lambda_- < -3\ \mathrm{K/deg}^{-2}$.  }
   \label{Fig_S}
\end{figure}

\subsection{Polarization angle dispersion}
\label{Pol_ang_disp}

To derive the mean scatter of polarization angles along the FIR
filaments at 857 GHz we use the polarization angle dispersion function
$\mathcal{S}$, introduced in \citet{Planck2015}
\begin{equation}
\label{eq:defS}
\mathcal{S}\left(\boldsymbol{r},\delta\right)=\sqrt{\frac{1}{N}\sum_{i=1}^N\left[\theta(\boldsymbol{r}+\boldsymbol{\delta}_i)-\theta(\boldsymbol{r})\right]^2} \, .\end{equation}
The sum extends over all pixels along the filament with positions $
(\boldsymbol{r}+\boldsymbol{\delta}_i) $ within an annulus centered on
$\boldsymbol{r}$ and having inner and outer radii
$\boldsymbol{\delta}/2$ and $3\boldsymbol{\delta}/2$, respectively.
According to our definition of the Hessian operator with adopted
Gaussian smoothing over five pixels we select a lag of
$\boldsymbol{\delta} = 18\arcmin$. Contrary to the previous usage of
$\mathcal{S}$ we count only pixels along filaments with eigenvalues
$\lambda_- < -1.5\ \mathrm{K/deg}^{-2}$.

\begin{figure}[thp] 
   \centering
   \includegraphics[width=9cm]{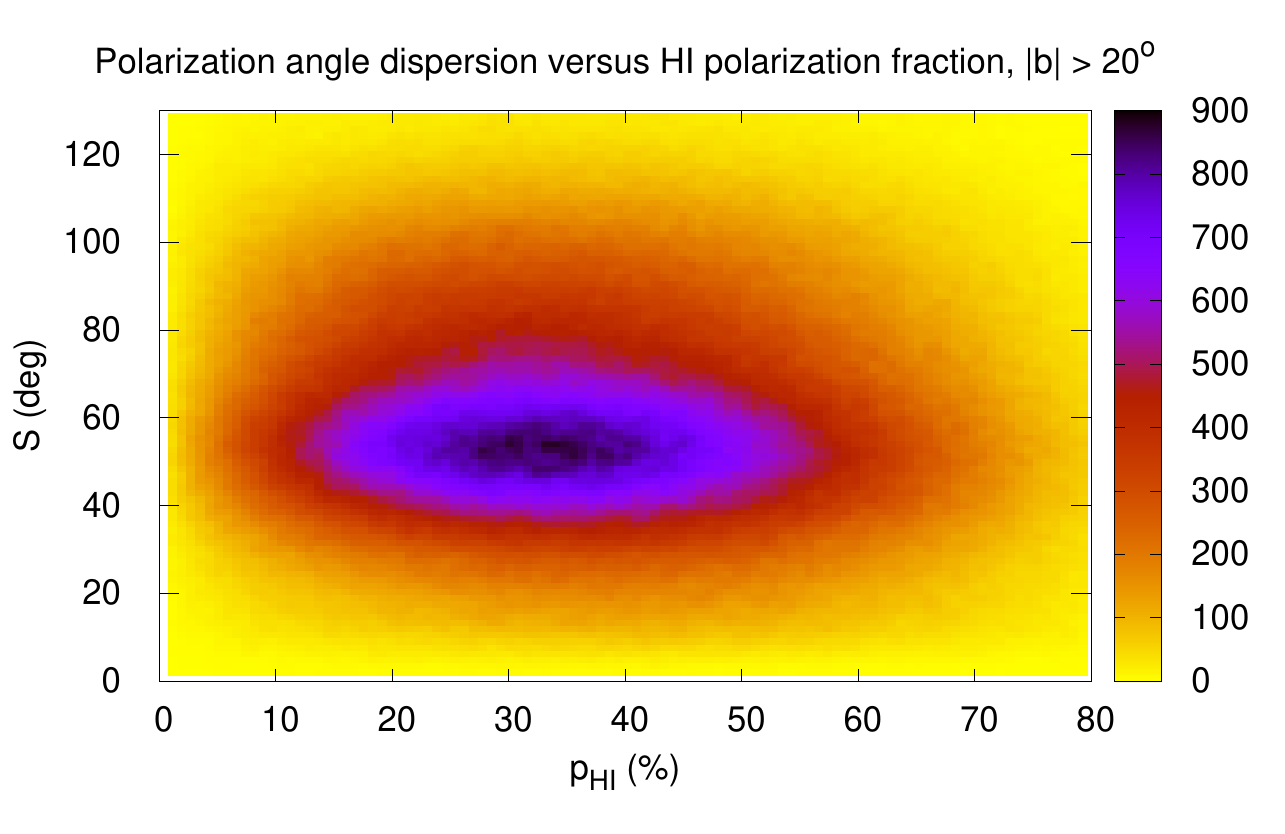}
   \includegraphics[width=9cm]{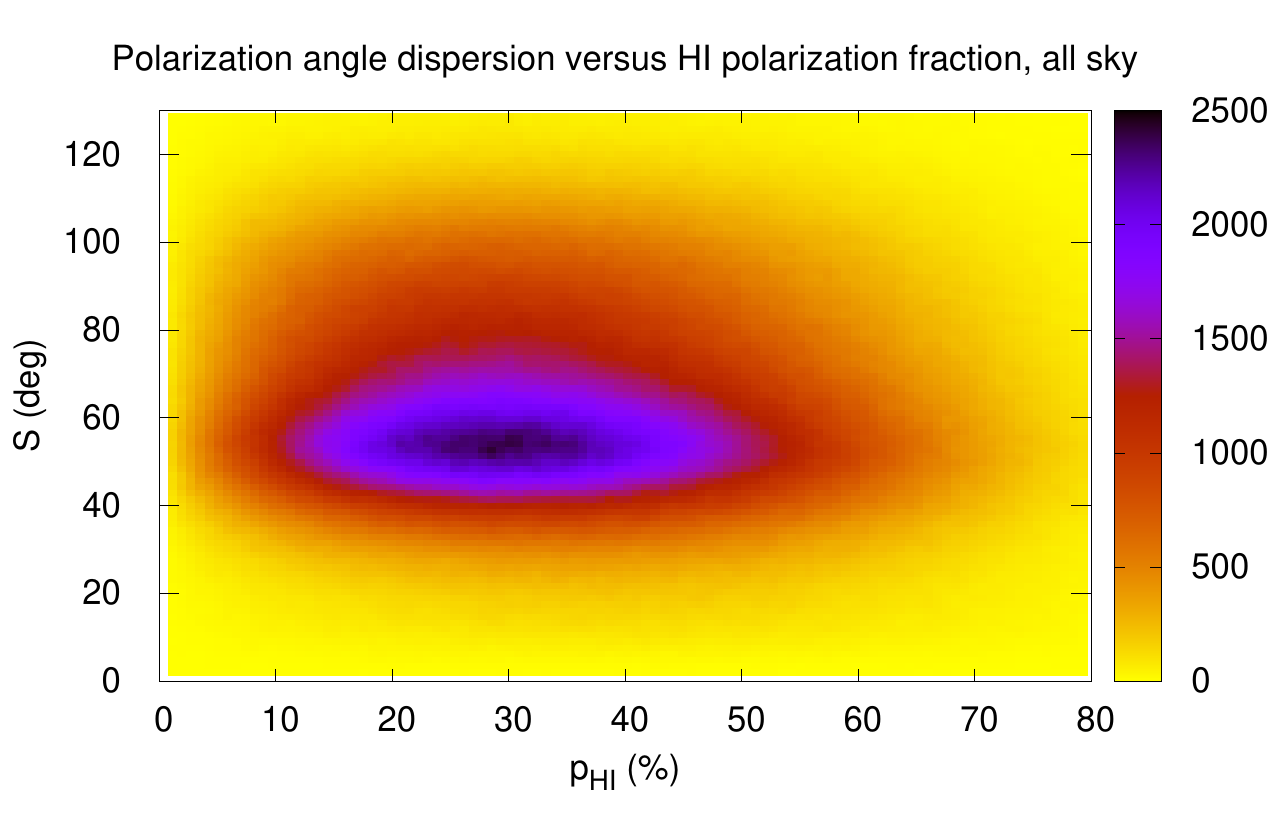}
   \caption{Two-dimensional density distribution functions for the polarization
     angle dispersion $\mathcal{S}$ along FIR filaments at a spatial
     resolution of 18\arcmin\ as a function of the \hi\ polarization
     fraction. Top: High Galactic latitudes. Bottom: All sky.  }
   \label{Fig_S_pol}
\end{figure}

Figure \ref{Fig_S} shows that polarization angles along filaments have
considerable dispersions. We consider here for the moment only the
  black and red lines in that figure.  The peaks of these distribution
functions are close to 52\degr. \citet{Planck2015} and
\citet{Planck2020a} use $\mathcal{S}$ to quantify the regularity of the
magnetic field.  They conclude that a distribution peaking around
$\mathcal{S} = \pi/\sqrt{12}\ (\sim 52\degr)$ is characteristic for a
chaotic and spatially completely uncorrelated distribution of
polarization angles
$\theta(\boldsymbol{r}+\boldsymbol{\delta}_i)-\theta(\boldsymbol{r})$.
Opposite to this conclusion we demonstrate in Sect. \ref{Curvature}
that the \hi\ polarization angles follow a distinct nonrandom pattern
if one takes the relation between orientation angles $\theta$ at
neighboring positions into account. The observed orientation angels for
neighboring positions are correlated due to systematic field curvatures
with increased tangling on small scales. The definition for
$\mathcal{S}$ (Eq. \ref{eq:defS}) disregards such effects. As shown in
Sect. \ref{Curvature_dist}, the distribution of curvatures is highly
nonrandom, indicating that the filaments follow a systematical bending
of the magnetic field lines. The peak at $\mathcal{S} \sim 52\degr$ is
consistent with an uniform distribution of orientation angles without
preferred orientation \citep[e.g.,][Appendix A]{Naghizadeh1993} but may
not be mistaken as a proof for a purely random distribution of filament
structures. Here we consider only the question whether the extended
wings for $\mathcal{S} \ga 90\degr$ could be caused by blending of
unrelated filaments. To check for effects from such a blending we
excluded in Eq. (\ref{eq:defS}) positions with velocity deviations
exceeding 10 \kms. We find no significant change in the histograms
compared to the unbiased distribution shown in Fig. \ref{Fig_S}. In
presence of the narrow velocity distribution derived in
Sect. \ref{vel_dist} blending may be unavoidable. At high Galactic
latitudes on average 2.5 to 3 \hi\ clouds are expected along the line of
sight \citep{Panopoulou2020}. In \citet{Planck2020a} it is assumed that
the polarization angle dispersion measure is characteristic for a
turbulence in layered structure of the magnetic field along the line of
sight.

The unexpected strong peak of the polarization angle dispersion at
  $\mathcal{S} \sim 52\degr$ in Fig.  \ref{Fig_S} is an intrinsic
  property of the small-scale structures we observe along the filaments
  as a response to the Hessian operator. To check the dependence of
  $\mathcal{S}$ on the chosen confidence level $\lambda_- <
  -1.5\ \mathrm{K/deg}^{-2}$ that is used throughout this paper, we
  exacerbate this constraint by a factor of two. In response only the
  most prominent filamentary structures are considered. Toward high
  Galactic latitudes this constraint affects 50\% of all filaments and
  all-sky 30\% are discarded. However, the peak of the polarization angle
  dispersion distribution in Fig.  \ref{Fig_S} remains at $\mathcal{S}
  \sim 52\degr$ (blue and orange lines). On average about 109 positions
  along the filaments contribute to the measurement of $\mathcal{S,}$
  while in the case of $\lambda_- < -3\ \mathrm{K/deg}^{-2}$ only 103
  positions are used.

\begin{figure}[thp] 
   \centering
   \includegraphics[width=9cm]{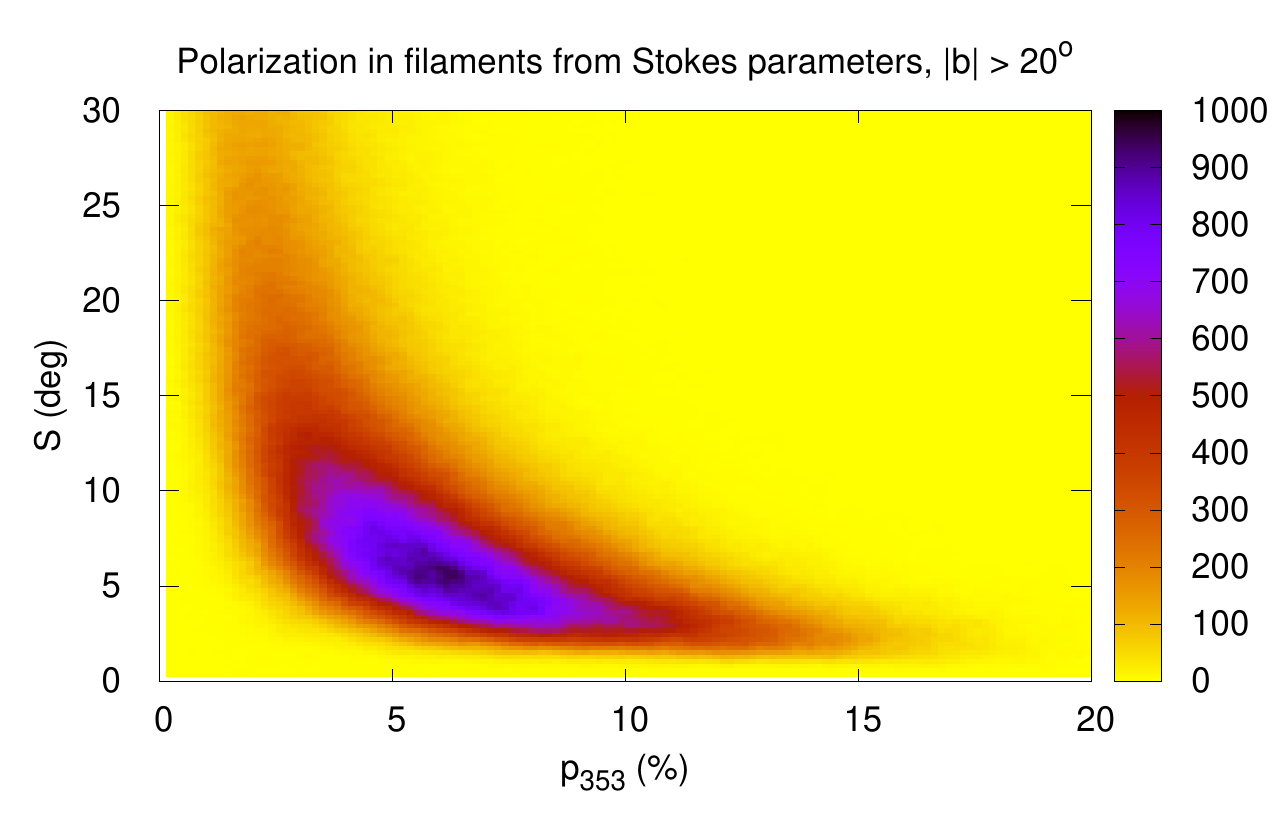}
   \includegraphics[width=9cm]{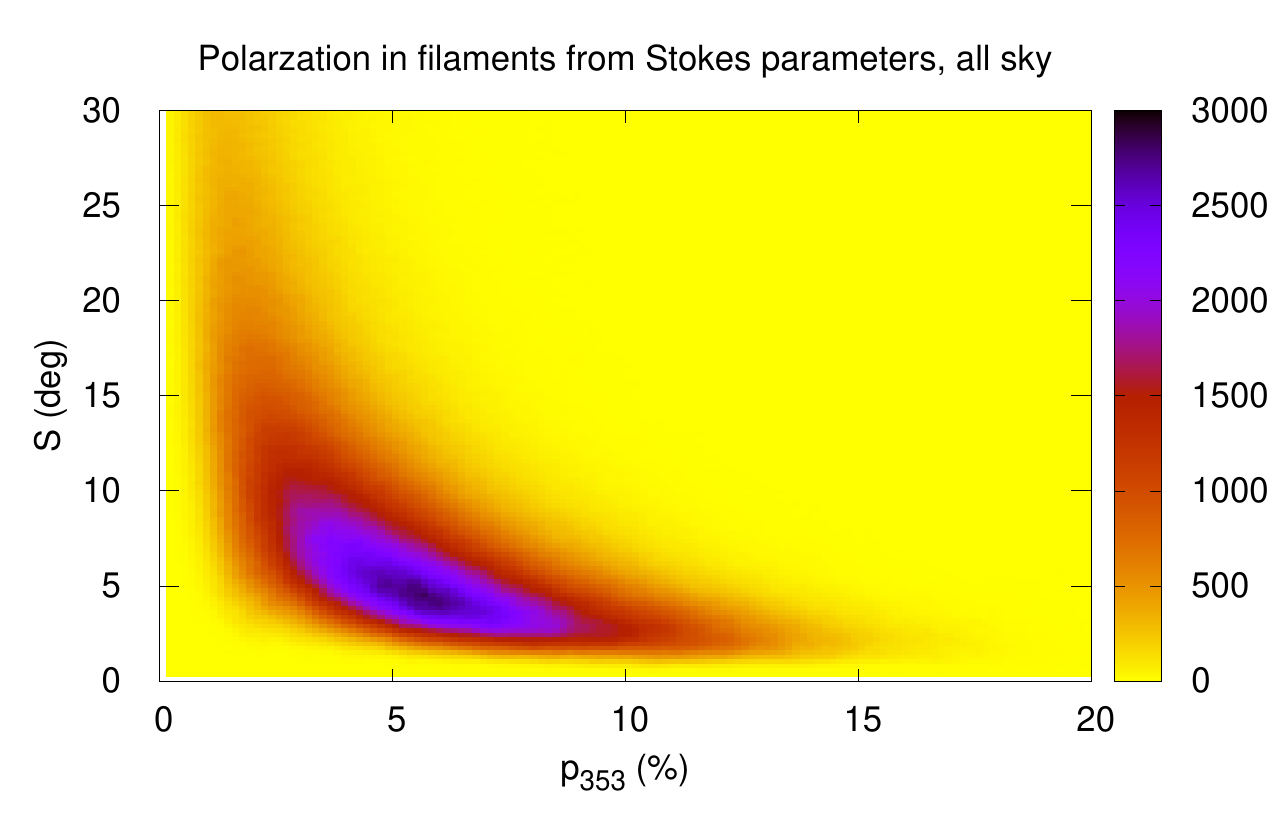}
   \caption{Two-dimensional density distribution functions for the polarization angle
     dispersion $\mathcal{S}$ along FIR filaments at a spatial
     resolution of 18\arcmin\ as a function of the polarization fraction
     derived from Stokes parameters at 353 GHz. Top: High Galactic
     latitudes. Bottom: All sky.  }
   \label{Fig_S_pol_353}
\end{figure}

\subsection{Polarization angle dispersion versus polarization fraction }
\label{pol_disp}

\citet{Planck2015} report that at 353 GHz $\mathcal{S}$ is in general
anticorrelated to the observed polarization fraction. This result was
confirmed by \citet{Planck2020a} and \citet{Clark2019b} report a similar
relation also for \hi\ polarization with a resolution of 160\arcmin. In
the following we focus on the polarization angle dispersion along
individual filaments at considerable smaller scales of 18\arcmin.

The
$\mathcal{S}$ from Eq. (\ref{eq:defS}) essentially measures fluctuations
of the polarization angle perpendicular to the line of sight on scales
of $3\boldsymbol{\delta}$, three times the selected resolution and in
our case on a scale of $\sim 54\arcmin$. Tangling causes fluctuations in
the plane-of-sky magnetic field orientation. Figure \ref{Fig_S_pol}
shows that,  in this case,  $\mathcal{S}$ does not depend on the
\hi\ polarization $p_\mathrm{HI}$ according to Eq. (\ref{eq:t_HI}). The
\hi\ polarization $p_\mathrm{HI}$ is defined as a measure of coherence
along the line of sight, hence across the \hi\ fibers. The dispersion
$\mathcal{S}$ is along the filament and both are apparently not
correlated. Our results clearly contradict \citet[][Fig. 11]{Clark2019b},
who observe an anticorrelation between $\mathcal{S}$ and $p_\mathrm{HI}$
similar to the anticorrelation in the case of polarization fractions
$p_\mathrm{353}$ from Stokes parameters. \hi\ polarization fractions and
polarization angle dispersions on scales of 160\arcmin\ and 18\arcmin
are not compatible. The scale differs by a factor of $\sim 9$ and we
conclude that just on angular scales of $\sim 18\arcmin $ we are able to
resolve details of the local magnetic field structure manifested in the
bending or tangling of fibers perpendicular to the line of sight.

\begin{figure}[thp] 
   \centering
   \includegraphics[width=9cm]{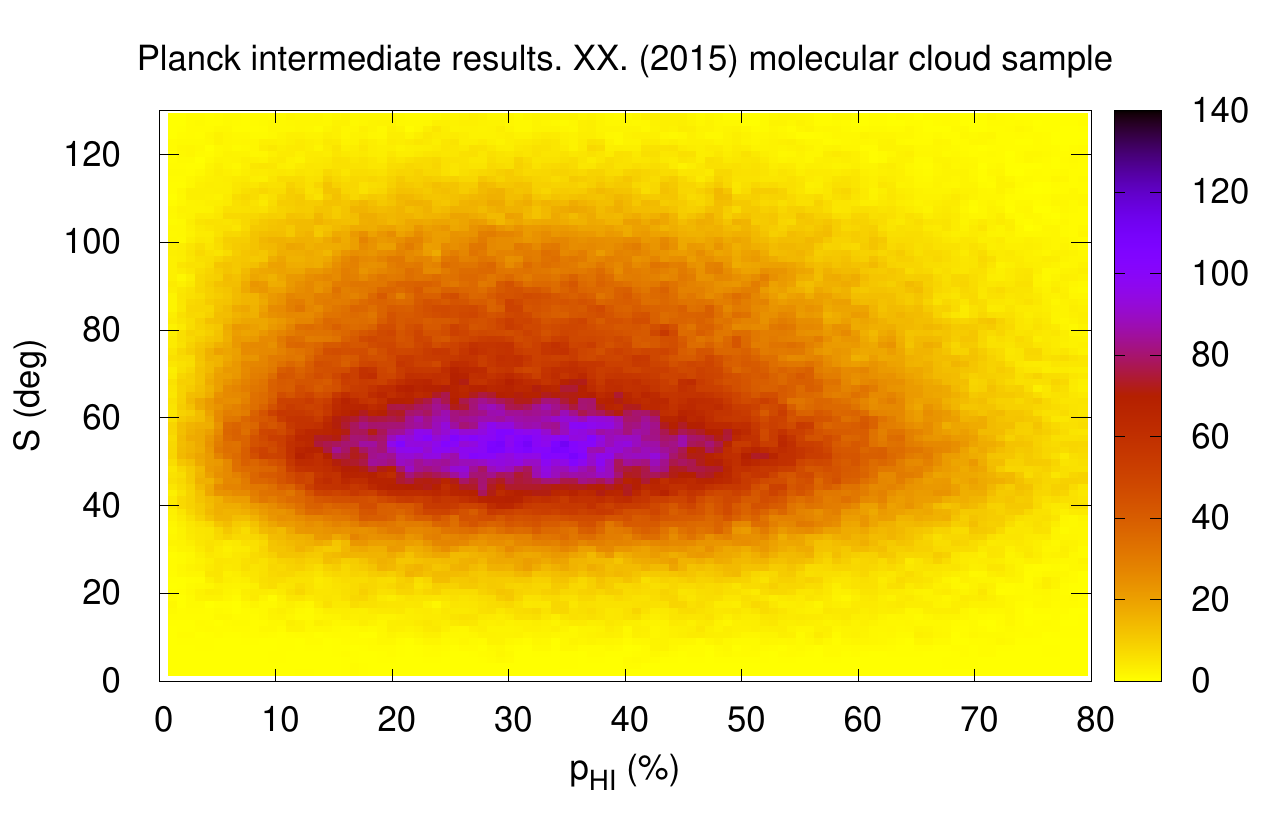}
   \includegraphics[width=9cm]{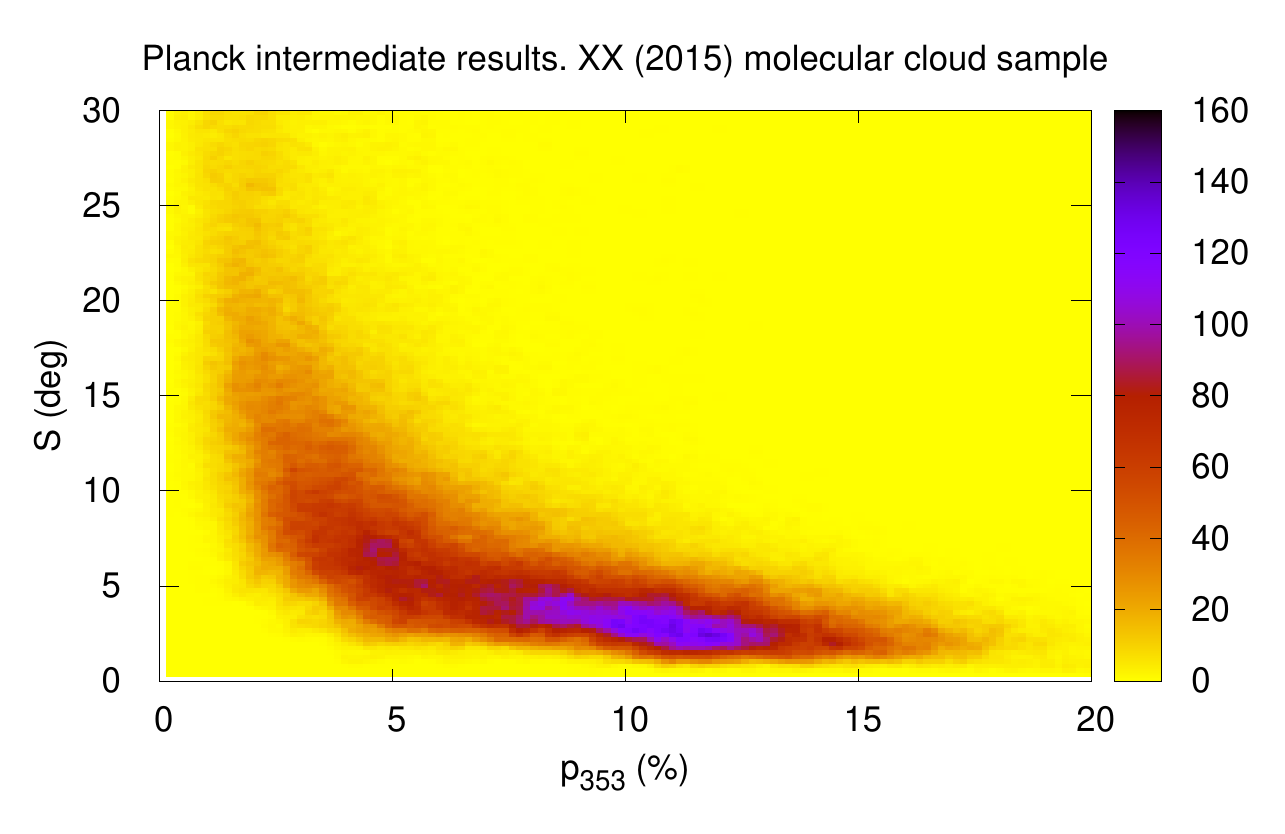}
   \caption{Two-dimensional density distribution for the polarization angle
     dispersion $\mathcal{S}$ along FIR filaments in molecular clouds at
     a spatial resolution of 18\arcmin\ as function of polarization
     fractions in \hi\ filaments (top) and at 353 GHz (bottom). We used
     the same molecular cloud sample as considered by
     \citet{Planck2015b}.  }
   \label{Fig_S_pol_mol}
\end{figure}

In Fig. \ref{Fig_S_pol_353} we display dependences between polarization
angle dispersion along the same filaments as in Fig. \ref{Fig_S_pol},
but now for polarization fractions determined from Stokes parameters at
353 GHz. These two-dimensional distributions confirm the previously  published 
anticorrelations between polarization angle dispersion and polarization
of FIR filaments. We also used the molecular cloud sample as considered
by \citet{Planck2015b} for an additional comparison between polarization
angle dispersions and polarization fractions. The result is shown in
Fig. \ref{Fig_S_pol_mol}. The only difference from the previously
  published two-dimensional distributions for the diffuse ISM is the larger
fractional polarization at 353 GHz for the denser molecular clouds. In
the case of the molecular clouds, a decrease in the maximum polarization
fraction with increasing column density is reported by
\citet{Planck2015b}. So we would expect for the molecular cloud sample a
trend to lower $p_\mathrm{353}$ values but we find the opposite.  Our
investigations in Sect. \ref{Comparison} show that there is in general
only a weak correlation for orientation angles $\theta$ between FIR
filaments and \hi\ structures in column densities. Accordingly we find
only a weak and probably insignificant correlation between \hi\ column
densities and polarization dispersions $\mathcal{S}$ and fractions
$p_\mathrm{HI}$.

Additionally, in simulations of MHD turbulence of polarized thermal emission from
Galactic dust by \citet{Planck2015b}, the polarization angle dispersion
$\mathcal{S}$ is found to anticorrelate with the polarization fraction.
These authors considered a lag of $\boldsymbol{\delta} = 16\arcmin$,
comparable to our resolution.  The model assumption is that there is a
large-scale anisotropic component of the magnetic field. Cold dense
filaments and clumps are condensing of the magnetized WNM.  In this case
a turbulent magnetic field component is linked to turbulent velocity
perturbations caused by a converging flow. The high \hi\ polarization
$p_\mathrm{HI}$, as well as the absence of a correlation with
$\mathcal{S}$ from our investigations are inconsistent with these
studies. Also, the case for a strong turbulence, advocated by
\citet[][Sect. F.6]{Planck2020a} on basis of an anticorrelation between
$\mathcal{S}$ an $p$ at 353 GHz, is not supported by our investigations
if we consider filaments with coherence in FIR and \hi.

\begin{figure}[thp] 
   \centering
   \includegraphics[width=9cm]{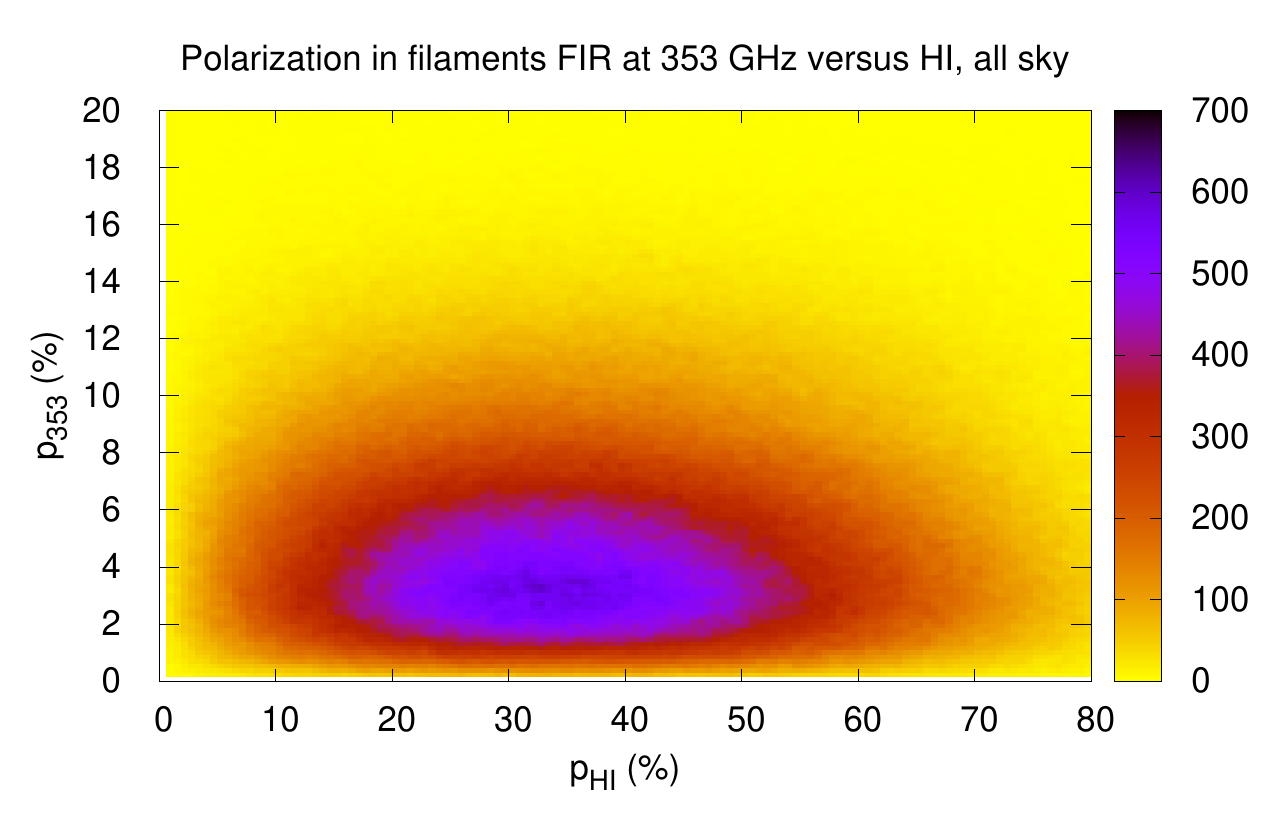}
   \caption{Two-dimensional density distribution showing the relation between
     polarization fractions of filaments in \hi\ and at 353 GHz. }
   \label{Fig_pp}
\end{figure}

\begin{figure*}[thp] 
   \centering
   \includegraphics[width=14.5cm]{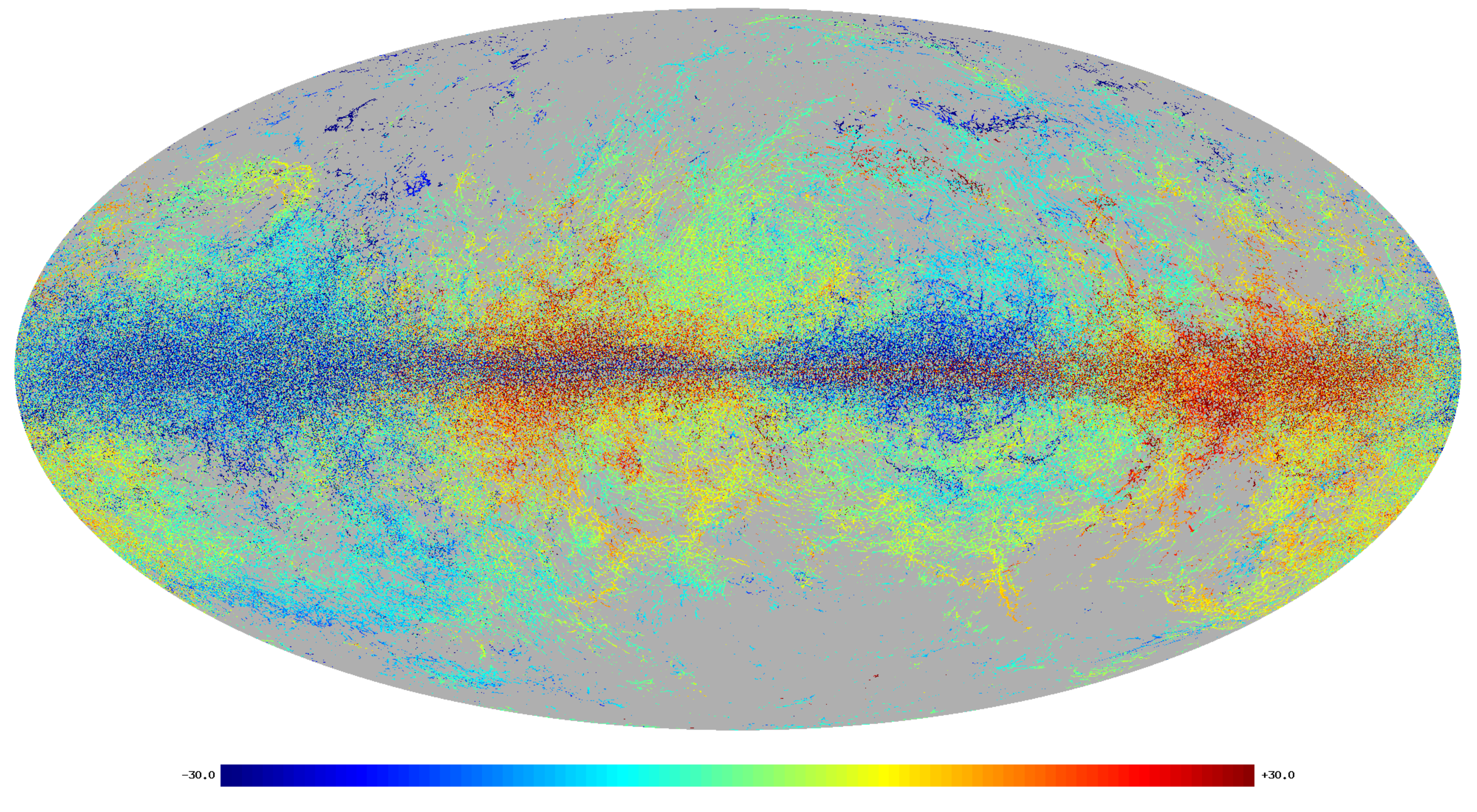}
   \includegraphics[width=14.5cm]{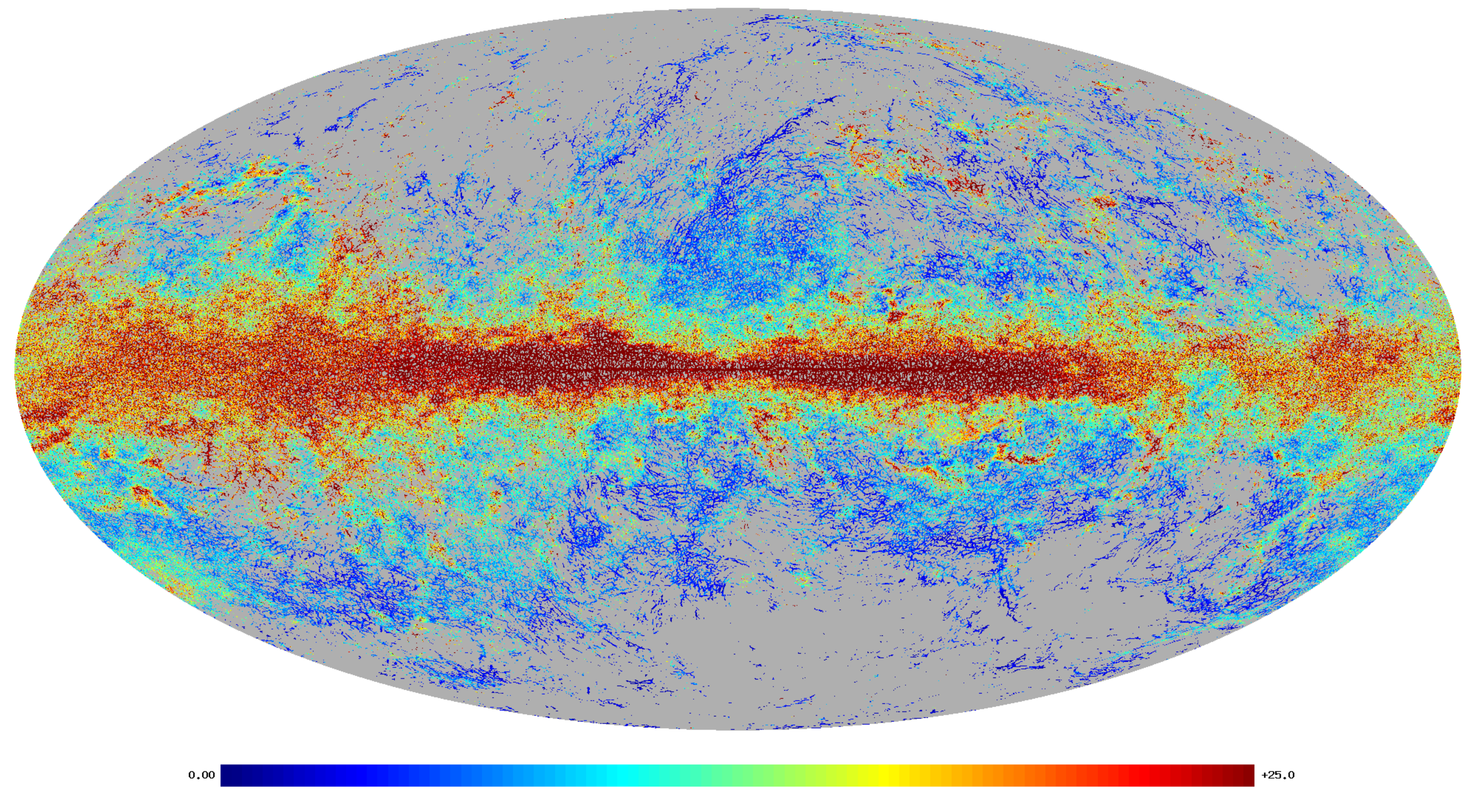}
   \includegraphics[width=14.5cm]{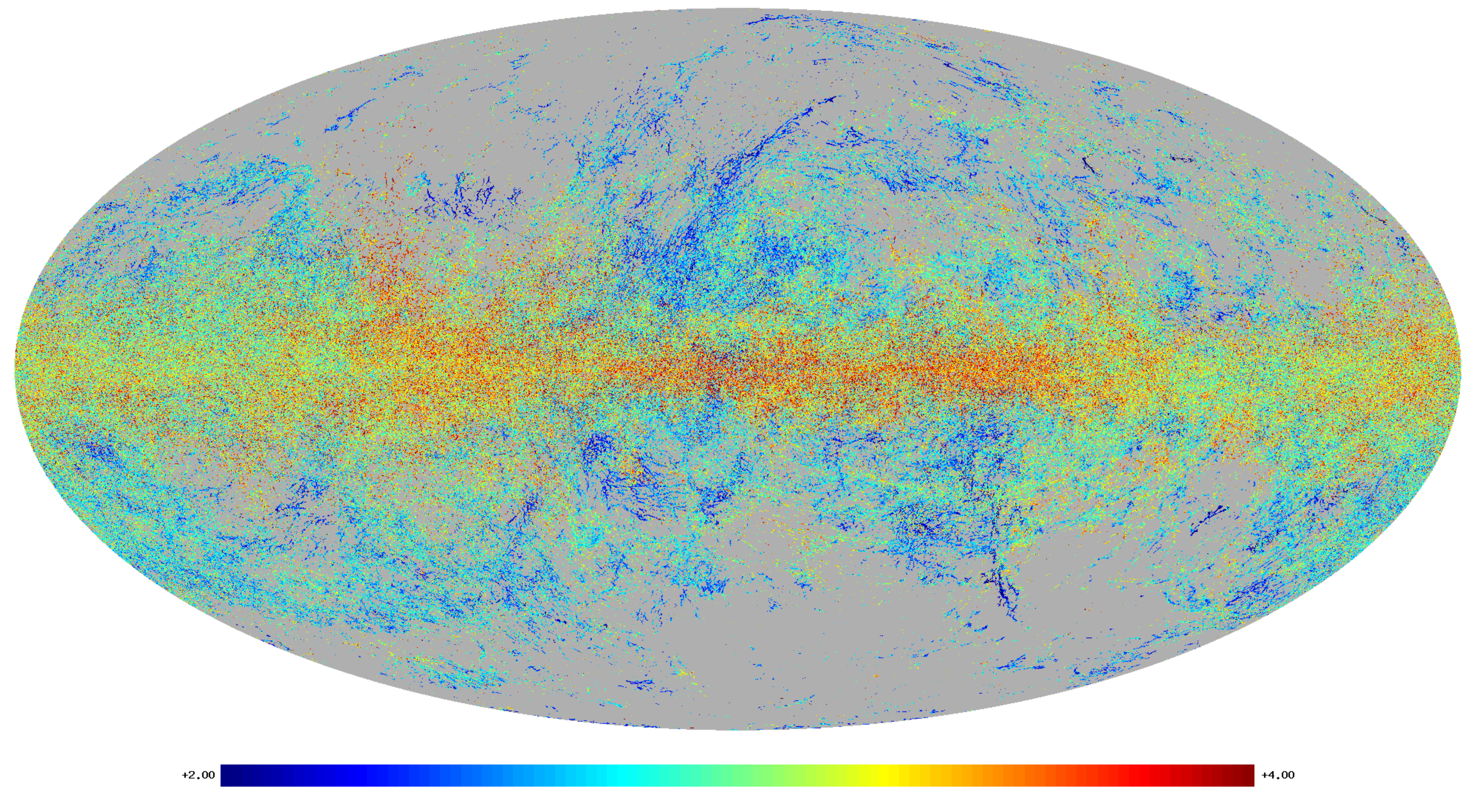}
   \caption{All-sky Mollweide displays of derived \hi\ parameters.  Top:
     Velocity distribution for coherent FIR and \hi\ filaments in the
     velocity range $ -30 < v_{\mathrm{LSR}} < 30 $ \kms.  Middle:
     Distribution of the velocity dispersions along filaments for $0 <
     \mathcal{V} < 25$ \kms.  Bottom: Spatial distribution of harmonic
     mean \hi\ Doppler temperatures for $ 2 <
     \mathrm{log}(T_{\mathrm{D}}) < 4 $.  }
   \label{Fig_Vel_HI}
\end{figure*}

The discrepancies between the polarization fractions derived from
  the \hi\ Hessians and the Stokes parameters at 353 GHz indicate
  systematical differences caused by the data processing. The derived
  polarization is certainly affected by the relative orientation of the
  local magnetic field with respect to the line of sight. When comparing
  gas and dust we assume that both are affected in a similar way. But we
  need to take into account that different parts of the ISM are
  sampled. Figure \ref{Fig_pp} shows the relation between polarization
  measures $p_\mathrm{353}$ and $p_\mathrm{HI}$. For the full sample of
  filamentary structures considered by us the polarization derived from
  Stokes parameters is on average only 10\% of the \hi\ polarization.
  As detailed by \citet[][Sect. 6.1]{Clark2019b}, the observed
  polarization fraction is an important constraint on the intrinsic
  polarizing efficiency. In the case of dust grains, the size distributions
  and alignment functions are important, as is the local orientation of
  the magnetic field \citep{Draine2009}. This leads to a limited
  observable polarization of $\sim 22$\% \citep{Planck2020a}. The
  \hi\ polarization $p_\mathrm{HI}$ reflects only the internal coherence
  of the gas distribution and not the intrinsic polarization efficiency
  of dust. \citet{Clark2019b} speculate that the most coherent
  sightlines may have $p_\mathrm{HI} \sim 1$. At a resolution of
  80\arcmin\ they find $p_\mathrm{HI} \le 0.26$, only slightly more than
  the observed maximum dust polarization of $\sim 22$\% at 353 GHz. Our
  analysis results in observed dust polarizations that are comparable to
  previous investigations; however, we get far stronger
  \hi\ polarizations, up to 80\% and on average $\sim 30$\%. It is not
  possible to explain high \hi\ polarization fractions with noise or
  systematic effects. We interpret therefore the lower \hi\ polarization
  fractions observed by \citet{Clark2019b} with beam depolarization due
  to tangling within their larger beam. In a similar way the high
  polarization angle dispersions $\mathcal{S}$, derived by us from the
  Hessians, can be explained. The \hi\ filaments are well resolved
  structures in narrow velocity intervals of 1 \kms, with a previously
  unreachable spatial resolution. Our smaller beam allows
  fluctuations in orientation angles to be observed on small scales. We emphasize that
  the Hessian analysis is most sensitive on such small scales.

\subsection{Spatial distribution of coherent structures}
\label{spatial_dist}

Figure \ref{Fig_Vel_HI} shows on top the spatial distribution of the
velocity field that we derive for the coherent FIR and \hi\ filaments for the
dominant velocities $ -50 < v_{\mathrm{LSR}} < 50 $ \kms, but to improve
the presentation we display only $ -30 < v_{\mathrm{LSR}} < 30 $
\kms. Contributions outside this velocity range are statistically
unimportant (Fig. \ref{Fig_histo_vel}). In the middle we present the
distribution of the velocity dispersions $\mathcal{V}$ along filaments
and below the harmonic mean Doppler temperatures in logarithmic scaling.

\hi\ filaments are local structures that are embedded in the diffuse
\hi\ distribution. Derived harmonic mean Doppler temperatures according
to Eq. (\ref{eq:T_D}) are therefore upper limits to Doppler temperatures
for the \hi\ gas that is directly associated with the FIR
filaments. Figure \ref{Fig_Vel_HI} bottom shows a general trend that
coherent FIR and \hi\ filaments are cold, at least for high latitudes. Many
structures close to the Galactic plane may be severely affected by
confusion. The velocity dispersions $\mathcal{V}$ for the most prominent
filaments are low. In some regions we find filaments with significant
velocity gradients, indicating either dynamical interactions or overlap
of unrelated filaments from different velocity regimes.

\section{Filament curvature diagnostics}
\label{Curvature}

In Sect. \ref{v_fil} we determined FIR and \hi\ filaments as coherent
structures with nearly perfect angular alignment. This concordance is
confined to the narrowest velocity interval and the continuity of the
velocity field (Figs. \ref{Fig_NoiseMaps} and \ref{Fig_Vel_HI}) implies spatial coherence for FIR
and \hi\ structures in the plane of the sky. We adopt in the following
the paradigm by \citet{Clark2014} and \citet{Clark2019b} that such
coherent \hi\ structures can be used to probe the properties of the
associated magnetic field, in particular the field tangling. We find
that the FIR and \hi\ coherence is confined to particular cold gas and is
best defined for small-scale structures. The key for our understanding
of the relations between gas, dust and magnetic fields is therefore the
small-scale structure in narrow velocity intervals on arcmin scales
despite the fact that such structures are part of larger and more
prominent filaments \citep{Kalberla2016} that appear to be more
interesting upon visual inspection. \citet{Clark2014} and
\citet{Clark2019b} analyzed filamentary structures with a resolution of
60\arcmin\ to 80\arcmin\ but we focus here on scales of 18\arcmin.

\subsection{The filament curvature distribution}
\label{Curvature_dist}

The small-scale structure of MHD turbulence is characterized by a folded
structure of the fields, observable as transverse spatial oscillation of
the field direction, while the field lines remain mainly unbent on large
scales.  \citet{Clark2014} and \citet{Clark2019b} analyzed the
line of sight field tangling. Our high resolution analysis allows us to
focus on tangling as observed in projection perpendicular to the
line of sight, hence changes of the orientation angle from $\theta_0$
at the beam center to $\theta_1$ at an offset position. 

\begin{figure}[th] 
   \centering
   \includegraphics[width=9cm]{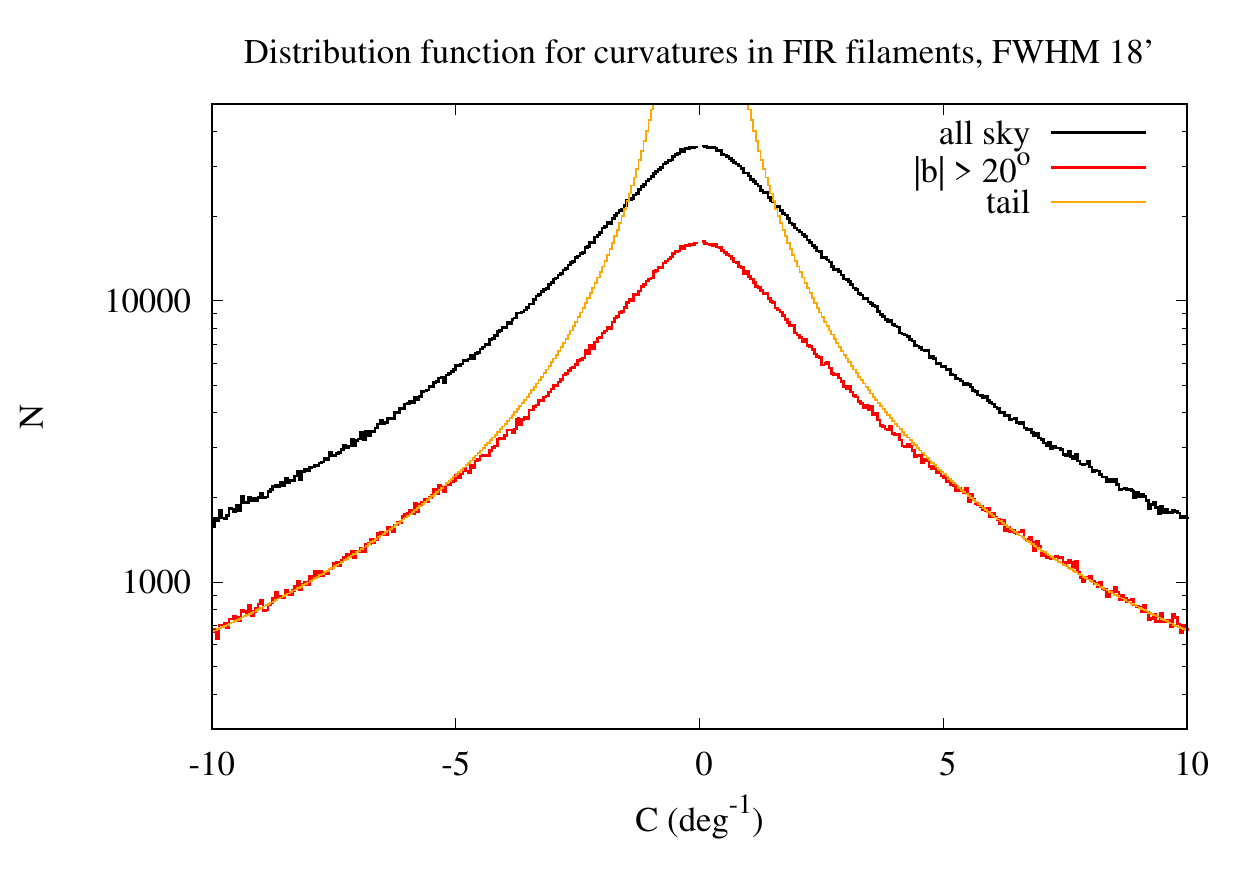}
   \caption{Distribution of local filament curvatures $C$ from
     \hi\ filaments and FIR at 857 GHz compared with a
     power tail $\propto C^{-13/7}$ according to
     \citet{Schekochihin2002}.  }
   \label{Fig_C}
\end{figure}

To parameterize the bending we need to consider that the magnetic field
points in the direction of the flux tube’s tangent vector while a
curvature is the response to a local perturbation perpendicular to the
field line. We determine for each position with orientation angle
$\theta_0$ along the filaments the local curvature $C = 1/R$ for a
structure with a radius determined from $R = \boldsymbol{\delta}/ (2
\sin ((\theta_0-\theta_1)/2) ) $. We select offsets $\boldsymbol{\delta}
= 9\arcmin$ with orientation angle $\theta_1$ on both sides of the
central position and average the radii. We verified that our results do
not depend critically on the particular choice of $\boldsymbol{\delta}$.
Figure \ref{Fig_C} displays the distribution of curvatures calculated
from FIR filaments derived at 857 GHz. Narrow channel \hi\ filaments
(see Fig. \ref{Fig_Theta} top and middle for comparison) lead within the
uncertainties to indistinguishable results.  Our observational findings
are quantitatively consistent with a power tail $\propto C^{-13/7}$
predicted by \citet{Schekochihin2002} for curvatures of field line
structures generated by a small-scale turbulent dynamo. According to
this model the small-scale magnetic turbulence is caused by folded
structures of the magnetic field $\mathbf{B}$ and the geometry of the
tangled magnetic fields can be studied in terms of the statistics of
their magnetic tension forces or curvatures $ \mathbf{K} =
\mathbf{\hat{b}} \cdot \nabla \mathbf{\hat{b}}$ with $ \mathbf{\hat{b}}
= \mathbf{B} / B $.

\begin{figure}[thp] 
   \centering
   \includegraphics[width=9cm]{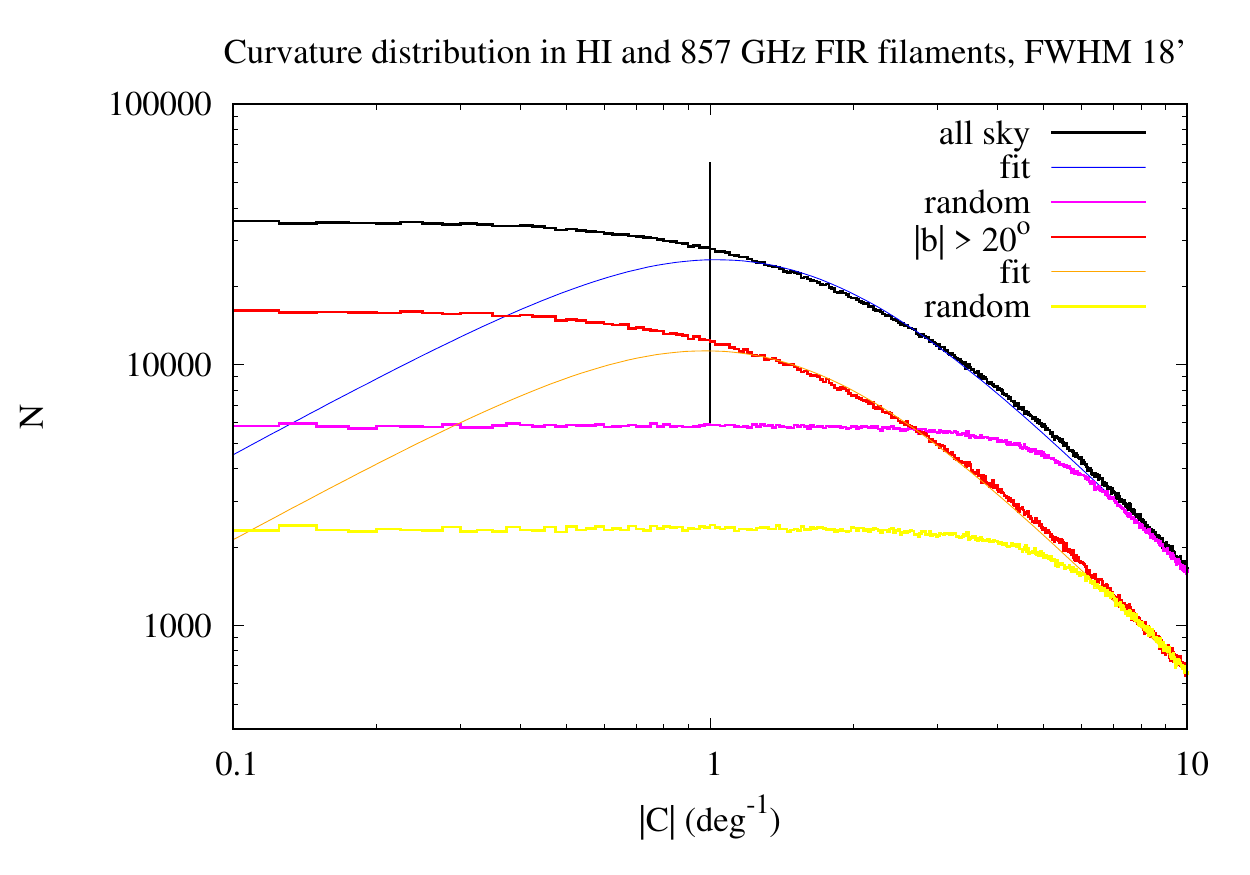}
   \includegraphics[width=9cm]{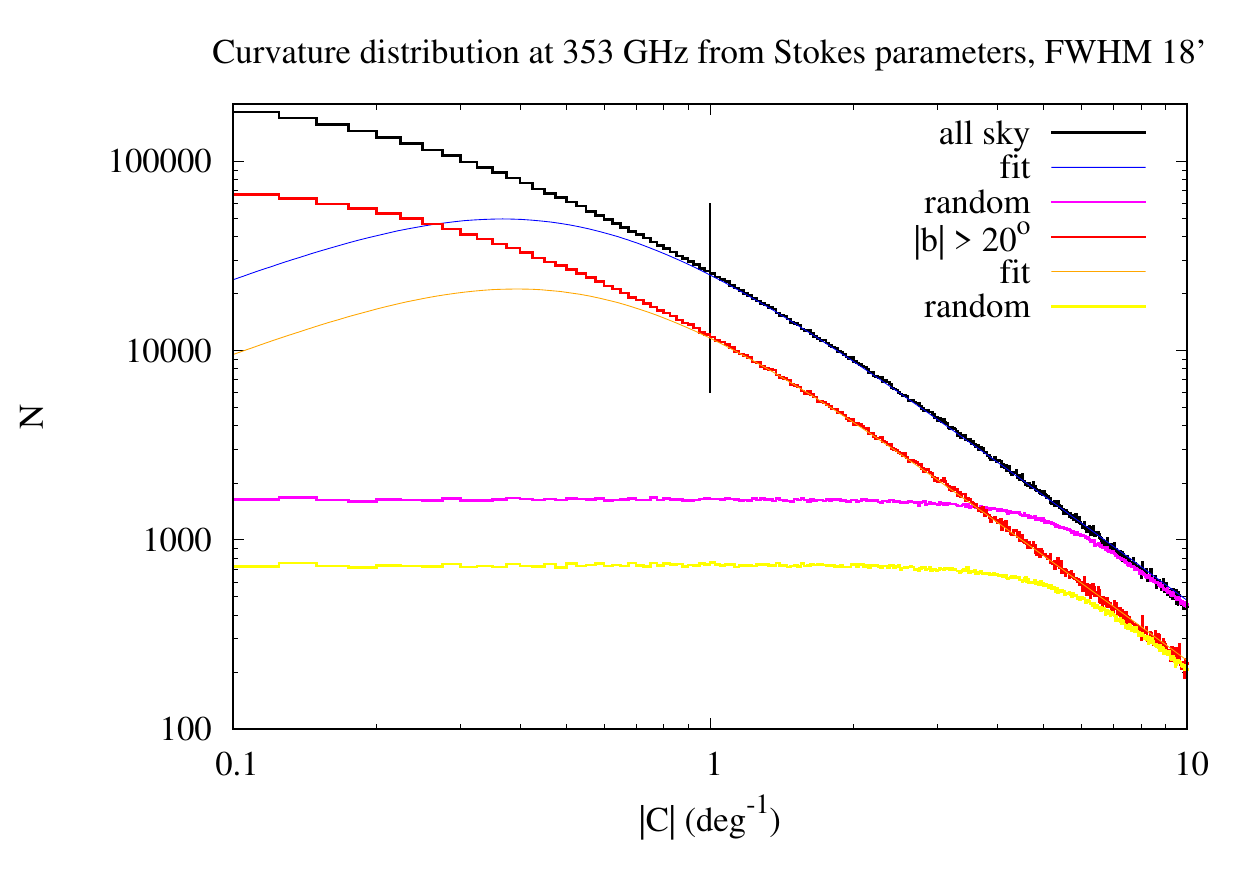}
   \caption{Distribution of observed local filament curvatures $|C|$ at
     a resolution of 18\arcmin\ in comparison to curvatures $|K|$ from
     Eq. (\ref{eq:kin}) (Eq. (9) of \citealt{Schekochihin2002}). Top:
     Curvatures for FIR filaments at 857 GHz and in \hi\ derived from
     Eq. (\ref{eq:theta}). Bottom: Curvatures from 353 GHz Stokes
     parameters $Q$ and $U$, using Eq. (\ref{eq:theta2}).  As indicated by
     the vertical line, the fit was only applied to $|C| >
     1\ \mathrm{deg}^{-1}$.  }
   \label{Fig_C2}
\end{figure}

\citet{Schekochihin2002} considered several model variants, the power
tail $\propto C^{-13/7}$ remains valid in all cases for large
curvatures. After an initially exponential growth the dynamo saturates
and the fully developed, forced, isotropic MHD turbulence is expected to
be the final saturated state of the small-scale dynamo. These authors
derive with their Eq. (9) a curvature distribution
\begin{equation}
  \label{eq:kin}
  P(K)  \propto \frac{K } {(1 + K^2)^{10/7} }
\end{equation}
with two asymptotes, $P(K) \propto K^1 $ for low curvatures and $P(K) \propto
K^{-13/7} $ for high curvatures.

In Fig. \ref{Fig_C2} we compare our data after fitting the scaling
parameter with this relation\footnote[3]{These results remain, in general,
correct if we also consider the back-reaction model, given in Eq. (24)
  of \citet{Schekochihin2002}, if we use in this relation a fixed
  parameter $\alpha = 1$.}. On top we display the curvature
distributions at 857 GHz derived after application of the Hessian
operator, using Eq. (\ref{eq:theta}). Below we plot the curvature
distributions calculated from 353 GHz Stokes parameters $U$ and $Q$
along the same filament positions using Eq. (\ref{eq:theta2}) to calculate
orientation angles.
Figure \ref{Fig_C2} shows systematical differences in the curvature
distributions that are related to different analysis methods.  Inherent
to the application of the Hessian operator is a low sensitivity to
structures larger than the applied pixel matrix, here 5x5 pixels or
1\fdg5x1\fdg5. The processing according to Eqs. \ref{eq:hessI}
to \ref{eq:theta} does however not bias significantly the calculation of
the orientation angels compared to the straight forward calculation from
Stokes parameters using Eq. (\ref{eq:theta2}). The shapes of the curvature
distributions remain consistent with the Eq. (\ref{eq:kin}). 

To estimate limitations from observational uncertainties we repeated the
calculations of the curvature distribution. To test the hypothesis that
the $\mathcal{S}$ map could be noise dominated, we replace the
orientation angles derived from Eqs. \ref{eq:theta} and \ref{eq:theta2}
with a cyclic uniform distribution. Repeating the complete data analysis
results in a flat curvature distribution. For a random uniform
distribution of orientation angles without any correlation of angles
between neighboring positions all curvatures have the same
probability. This is clearly inconsistent with observations. Our
simulations include beam effects, resulting in a decreased sensitivity
at high spatial frequencies. Accordingly all slopes in Fig. \ref{Fig_C2}
at high curvatures are affected by sensitivity limitations. The
predicted tail $\propto C^{-13/7}$ shown in Fig. \ref{Fig_C} is not
useful to constrain the model in this range. Comparing the
353\,GHz and 857\,GHz curvature distributions, we do not find any
obvious noise biases. This is on the first glance surprising because the
353\,GHz data are affected by noise much more severely than the 857\,GHz
ones but we show below that the S/N for the selected 353\,GHz filaments
is high enough.

We like to point out here first that the observed orientation angles for
neighboring positions are correlated due to systematic field
curvatures. The definition for $\mathcal{S}$ (Eq. \ref{eq:defS}) does
not account for such a systematic behavior. The distribution of
curvatures is highly nonrandom, indicating that the filaments follow a
systematical bending of the magnetic field lines. The peak at
$\mathcal{S} \sim 52\degr$ in Fig. \ref{Fig_S} is consistent with
strongly fluctuating orientation angles on small scales without
directional preferences as expected in the case of a tangled magnetic
field. This peak does however not imply a completely random and uncorrelated
distribution of orientation angles. Such a case requires the
curvatures (derived from orientation angels at adjacent positions along
the filaments) to also have a flat distribution.

We emphasize further that the derived distribution of observed
  local \hi\ filament curvatures is far from noise dominated. Using
Eq. (\ref{eq:noise}) we determine the S/N at each of the selected filament
positions for all \hi\ channels with $\delta v_{\mathrm{LSR}} = 1$
\kms\ considered by us throughout this paper as related to the
FIR. These data exceed an average S/N level of 87 all sky and 68 at high
latitudes. The Hessian operator is selective and most sensitive to
  \hi\ data with local S/N maxima (constant multiple rule). These are
  positions that are dominated by the CNM. The structures
  disclosed this way are predominantly observed close to zero
  velocity. For typical applications of a Hessian analysis a S/N of
better than five is considered to be sufficient
(e.g., \citet{Polychroni2013} or \citet{Soler2020}, we refer also to the
discussion in Sect. 2 of \citet{Schisano2014}).

In a similar way we
estimate the S/N for filaments at 353 GHz from the smoothed {\it Planck}
intensity map.  Following \citet{Planck2016b} we use a 12\fdg5 x 12\fdg5
field at $ l = 90\degr, b = -80\degr $ to determine the background
noise. Accordingly the 353 GHz FIR filaments used in our analysis exceed
an average S/N level of 30 all sky and 18 at high latitudes. This is
only about 1/4 of the S/N in \hi\ but still high enough that we do not
need to be worried about hidden noise biases in Fig. \ref{Fig_C2} and
other parts of our analysis. \citet{Montier2015} consider uncertainties
on polarization fraction and angle measurements of {\it Planck}
polarization data and conclude that these measurements are little
affected by uncertainties for an intensity S/N level above
10. \citet[][]{Skalidis2019} determined the dependences between beam
depolarization and smoothing radius of 353 GHz {\it Planck} data. They
describe in their Appendix A that a smoothing radius of
20\arcmin\ results in the best compromise to mitigate the beam
depolarization and work with sufficiently high quality data.

Previous investigations of filamentary structures by \citet{Planck2015},
\citet{Planck2016}, \citet{Clark2018}, \citet{Clark2019b}, and
\citet{Planck2020a} were focused mainly to angular resolutions between
80\arcmin\ and 160\arcmin. On these scales only the very tip $|C| \la 1
\ \mathrm{deg}^{-1}$ of the curvature distribution discussed in the
previous subsections is accessible but we extended this range by an
order of magnitude.  The agreement between our data and relation
\ref{eq:kin} is in general excellent for $|C| \ga 1\ \mathrm{deg}^{-1}$.
We select only the range $|C| > 1\ \mathrm{deg}^{-1}$ to fit the
observed curvatures to the distribution Eq. (\ref{eq:kin}). The
small-scale turbulent dynamo considered by \citet{Schekochihin2002} and
\citet{Schekochihin2004} dominates the curvature distribution only 
for such scales.

For Kolmogorov turbulence the energy that feeds the turbulence is
injected at an outer scale of $\sim 100$ pc \citep{Kalberla2019}. The
energy cascades then down to the viscous dissipation scale (called
Kolmogorov inner scale in purely hydrodynamic systems). It is important
to realize that the small-scale turbulent dynamo can feed energy in the
opposite way.  It acts initially below the viscous scale of $\sim
10^{-2}$ pc \citep{Schekochihin2004} but perturbations in the magnetic
field can propagate to larger scales. The dynamo leads initially to an
exponential magnetic energy growth.  A sketch of scale ranges and energy
spectra is given in Sect. 1.2 of \citet{Schekochihin2004} with their
Fig. 1. Folding of the magnetic fields causes magnetic tension forces
with a back reaction on the turbulent flow. The system becomes
increasingly nonlinear. Geometrically, one observes that stretched structures
and flux sheets are converted to flux ribbons. In this so-called
saturated stage, low curvatures are more dominant and the peak of the
curvature distribution has shifted up and to small curvatures (see
Fig. 25 of \citealt{Schekochihin2004}). For 353 GHz Stokes parameters
(unaffected by biases from spatial filtering) the peak is at $|C| \sim
0.4\ \mathrm{deg}^{-1} $, corresponding to radii $R \ga 2.5\degr$. The
kinetic energy on larger scales has to be fed in by external sources. It
is expected that the dynamo saturates with an equilibrium between
magnetic and kinetic energy. According to
\citet[][Sect. 12]{2020Schekochihin} details about the final state of
the saturated small-scale dynamo remain however an unsolved problem,
both numerically (due to lack of resolution) and theoretically (due to
lack of theoreticians). \citet[][his Fig. 1]{Beresnyak2012} argues that
most of the magnetic energy may be concentrated at the scale
corresponding to the peak in the curvature distribution.  Supernovae are
the dominant sources of turbulent energy in the ISM \citep{MacLow2004}
and may play a major role in this case.  The size distribution of
\hi\ shells in the outer Galaxy as potential sources for diameters
larger than 1\fdg5 was studied by \citet{Ehlerova2013} and is in good
agreement with the observed distribution of filaments at low and
intermediate latitudes.
 
According to \citet{Schekochihin2004} the timescale for the growth of
the small-scale magnetic energy is characterized by the viscous eddy
turnover time in the order of $10^5$ years. The mean-field dynamo theory
predicts a large-scale galactic field exponentiating timescale in a
rotation time of $\sim 10^8$ years. Thus the small-scale dynamo is very
fast. These timescales may be compared to the typical cooling time of
$\sim 1.4 \ 10^5$ years for the CNM and $\sim 5 \ 10^7$ years for the
WNM. Shocks due to supernovae are expected to occur every $\sim 5 \ 10^6
$ years \citep{Kalberla2009}. In this situation the WNM can hardly be in
equilibrium while phase transitions to the CNM can easily be triggered
by a small-scale dynamo. In consequence, cold structures in the diffuse
ISM that are driven by a small-scale dynamo, must evolve from small to
large scales.  \citet{Balsara2005} studied the magnetic field
amplification in the ISM by supernova-driven turbulence.
They find a field amplification that increases with volume density as
$|B| \propto \rho^{0.386}$ and conclude that the field amplification
takes place more vigorously in the lower temperature, denser gas.

Several authors have studied the role of the magnetic field in molecular
cloud formation and evolution. These investigations have been reviewed
by \citet{Hennebelle2019}. From MHD simulations the consensus is that
the magnetic field is strongly shaping the interstellar gas by
generating a lot of filaments. For the diffuse \hi\ distribution with
column densities below $N_H = 10^{21.7}\,{\rm cm^{-2}}$ the filaments
are aligned with the magnetic field. Recent investigations of phase
transitions by \citet{Falle2020} emphasize that cooling behind a shock
is affected by the magnetic field. Even a small initial magnetic field
can lead to a magnetically dominated state on the unstable part of the
equilibrium curve. These authors conclude that the magnetic field must
dominate in the final state of shocks even for an implausibly small
initial magnetic field strength and derive filament widths of $\sim
0.52$ pc (Figs. 11 and 12 of \citealt{Falle2020}).

The coherence between cold CNM and magnetized filamentary structures on
small scales is explainable in the framework of these
investigations. For the case of a magnetic pressure confinement of CNM
filaments \citet{Kalberla2016} derived a median filament thickness of
0.09 pc assuming that the distance to the filaments is on average 100
pc. This estimate is consistent with \citet{Clark2014} who use better
resolved Arecibo telescope observations and find largely unresolved
structures, corresponding to a scale of 0.12 pc or below.  Such
structures have so far not been related to the small-scale
dynamo. \citet{Shukurov2007} and \citet{Rincon2019} consider the
turbulent magnetic field as presumably confined to the WNM. If we,
however, think about the simplified case of a flux tube compression in
ideal MHD, we find that, in the case of phase transitions, the gas density and
the magnetic field strength in such tubes should be amplified
\citep[][Sect. 4.4]{Heiles2005}. Phase transitions coupled to flux tubes
may eventually lead to a balance between magnetic and gaseous pressure
with typical parameters as used by \citet{Kalberla2016} in their
Sect. 5.12 (see also \citealt{HeilesT2005}, Sect. 7.2, for discussion).
For a magnetically confined CNM it is necessary to have a sufficient
large magnetic field outside. 
Observational evidence for a flux tube scenario was reported by
\citet[][Sect. 3.3]{Kalberla2020}.
In this context it may be worth to reconsider the
equilibrium arguments given by \citet{Beresnyak2012} by including local
interactions from phase transitions.

Power spectra for the local CNM distribution show a remarkable power
excess at multipoles $l \ga 100 $, corresponding to scales below $ \sim
180 \degr / 100 = 1.8\degr$ \citep{Kalberla2019}. This is the range that
we find to be occupied by the small-scale dynamo in
Fig. \ref{Fig_C2}. Characteristic for this excess is a notable shift to
higher values of $l$ when decreasing the width of the analyzed velocity
slices or decreasing the Doppler temperature of the \hi\ distribution
under consideration (Fig. 15 of \citealt{Kalberla2020}). The power excess
for the coldest structures is at multipoles $ l \ga 500$, corresponding
to scales of $ 180 \degr / 500 \sim 20\arcmin$ (see Fig. 15 of
\citealt{Kalberla2020})

The so far unexplained excess in turbulent CNM power spectra
\citep{Kalberla2020} is compatible with the transition between large and
small-scale turbulent driving from our current analysis. On multipoles
$l \ga 100$ the small-scale dynamo is dominant. Last but not least we
need to take into account that $\lambda_+$, the second eigenvalue of the
Hessian operator $H$ from Eq. (\ref{eq:lambda}), discloses numerous
unresolved structures along filaments at similar scales (see
Fig. \ref{Fig_lam}). These sources are also cold (Sect. \ref{blobs})
and belong therefore to the excess part of the exponential power
distribution for the CNM.  The coherence between FIR and \hi\ filaments
exists only for cold \hi\ (Sect. \ref{phase_fil}) and turbulence is in
this case affected by the small-scale dynamo. The injection of turbulent
energy on small scales implies additional power at large multipoles and
correspondingly a flattening of the CNM power spectra in comparison to
multiphase \hi\ power spectra. Changes of the spectral index depending
on the phase composition of the \hi\ \citep{Kalberla2019} are expected.

\begin{figure}[thp] 
   \centering
   \includegraphics[width=9cm]{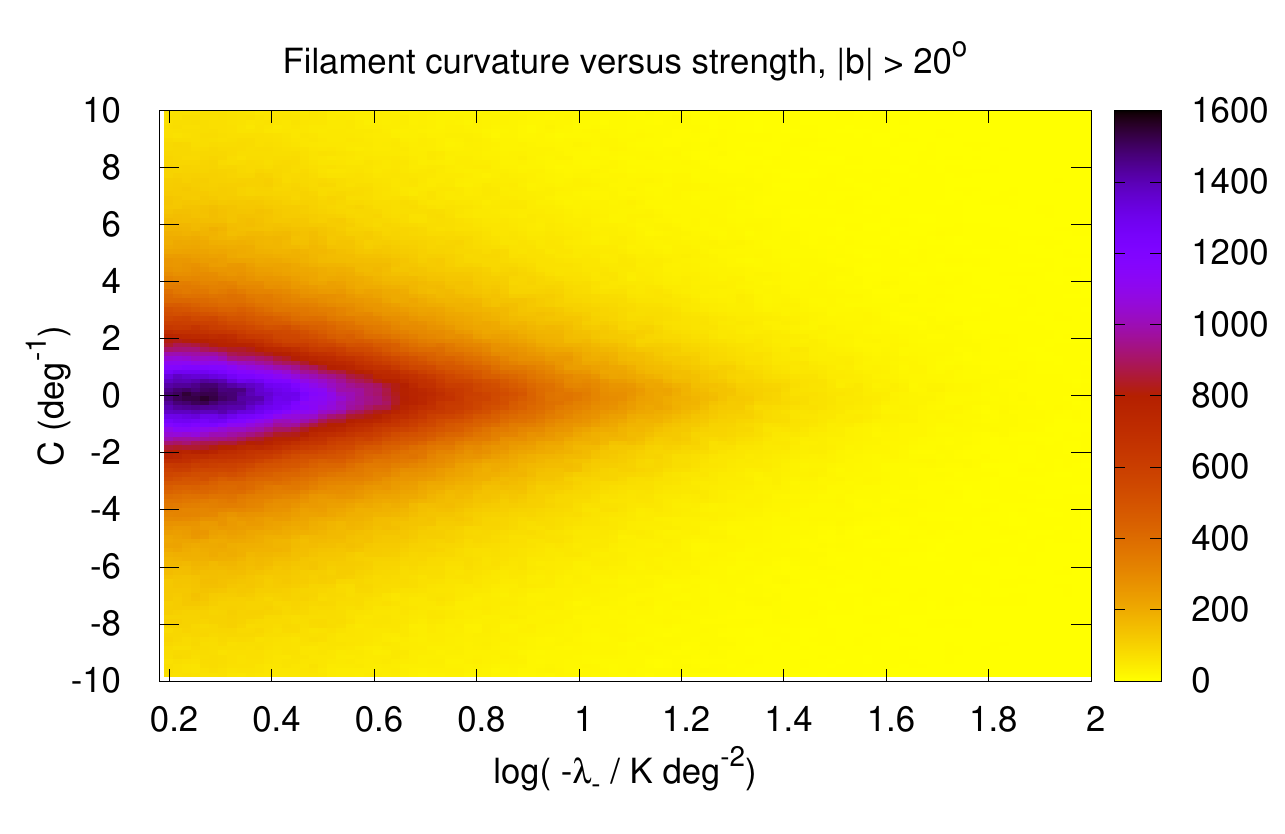}
   \includegraphics[width=9cm]{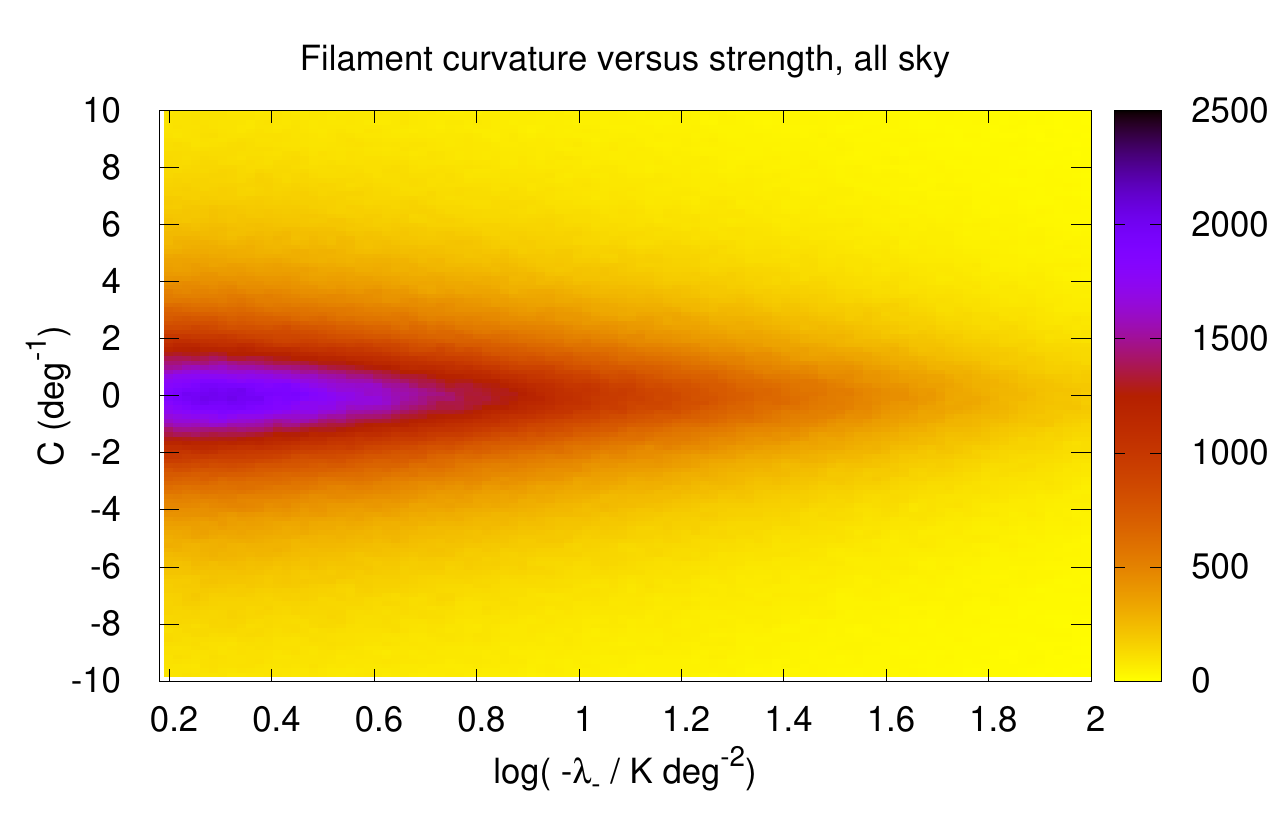}
   \caption{Two-dimensional density distribution of observed local filament
     curvatures $C$ at 857 GHz in comparison to eigenvalues $\lambda_- $
     from the Hessian analysis as a measure of the filament strength. }
   \label{Fig_SClam}
\end{figure}

\subsection{Curvature versus filament strength }
\label{Curve_strength}

In the saturated state of the small-scale dynamo the magnetic field
strength and the curvature are anticorrelated, $|B| \propto C^{-1/2}$
(Fig. 17 of \citealt{Schekochihin2004}, Figs. 5, 15, and 17 of
\citealt{Schekochihin2004}, and Fig. 5 of \citealt{St-Onge2018}).  Sharply
curved fields (high $C$ values or small radii) imply a high field
tension and the field strength is reduced. The anticorrelation between
field strength and curvature implies that the most prominent straight
filaments popping up in FIR or \hi\ maps must be indicating high values
for the field strength and polarization fraction (little depolarization
by tangling). For a prominent example we refer to a display of the
Riegel–Crutcher cloud, Figs. 10 and 12 of \citet{Clark2014}, as well as
\citet{McClure2006}.

\begin{figure}[ht] 
   \centering
   \includegraphics[width=9cm]{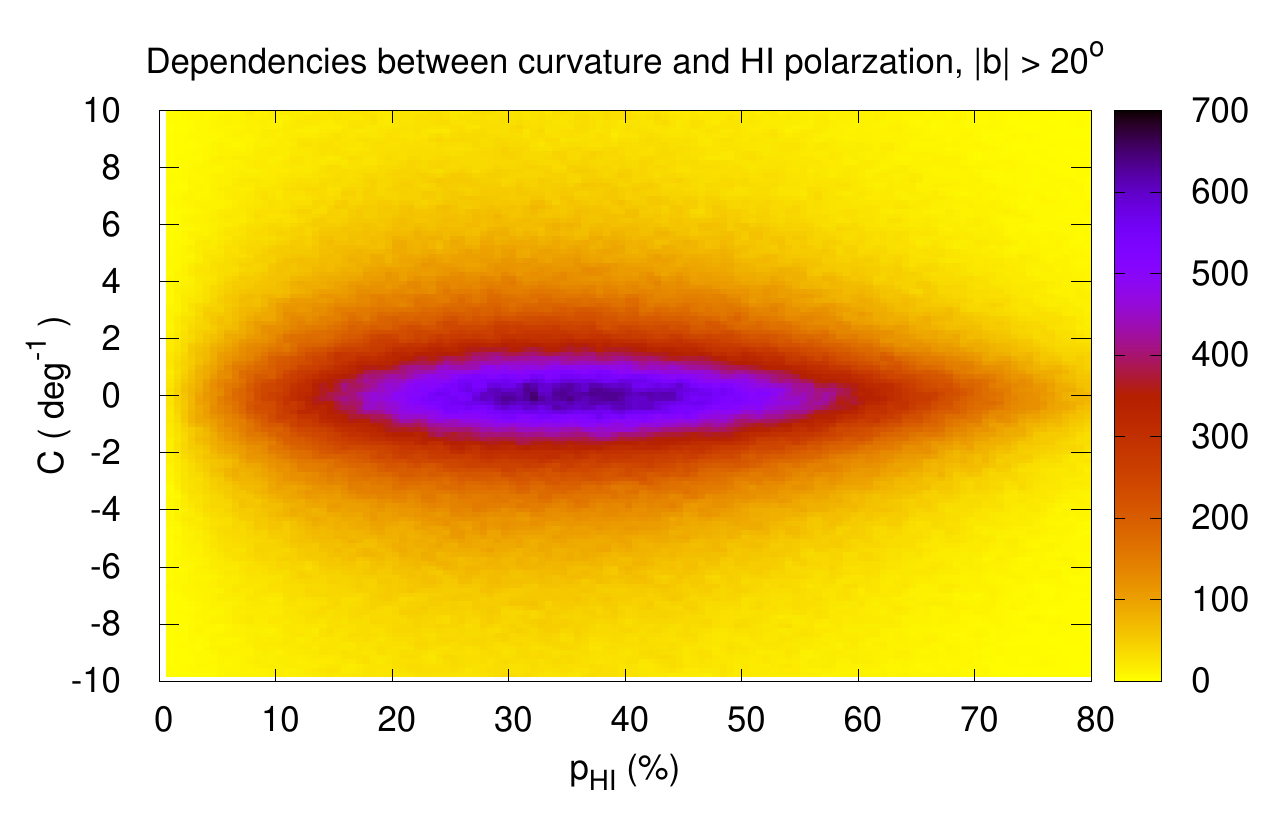}
   \includegraphics[width=9cm]{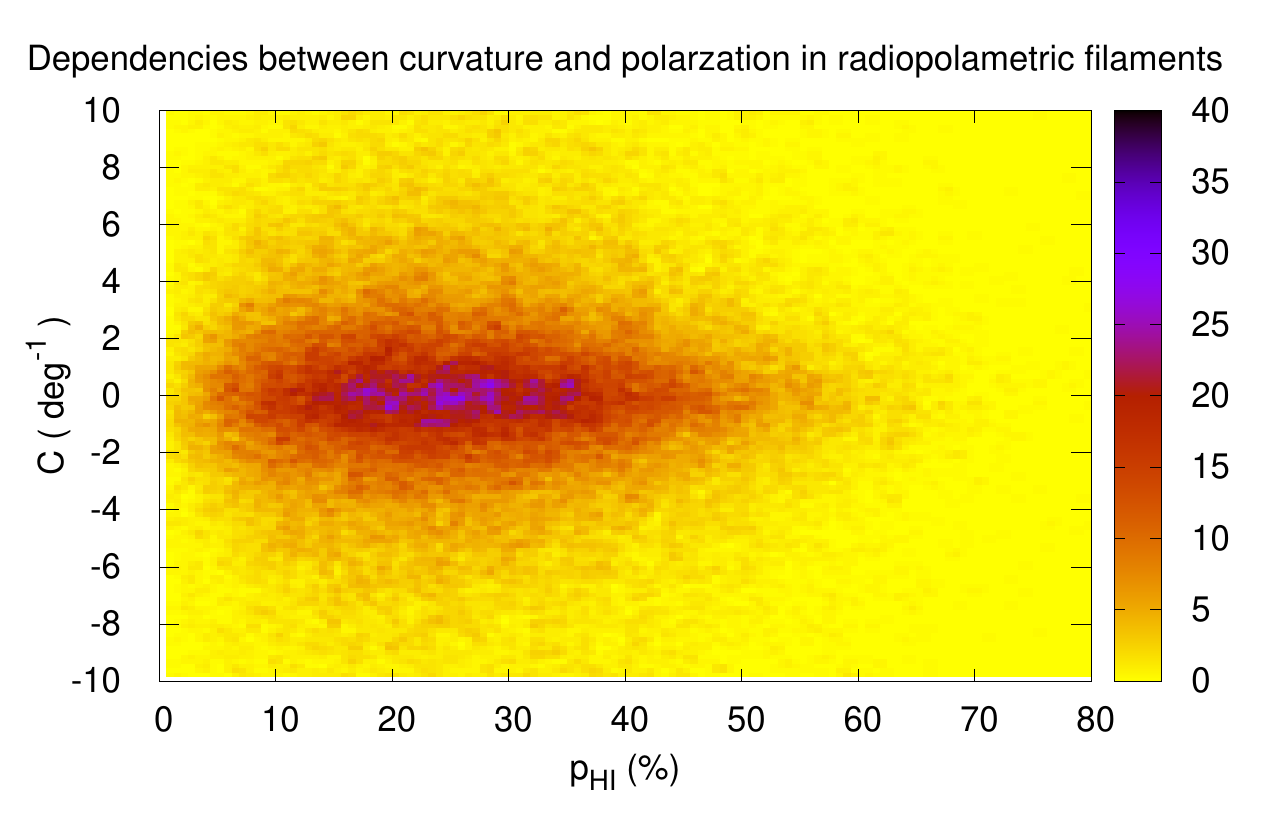}
   \includegraphics[width=9cm]{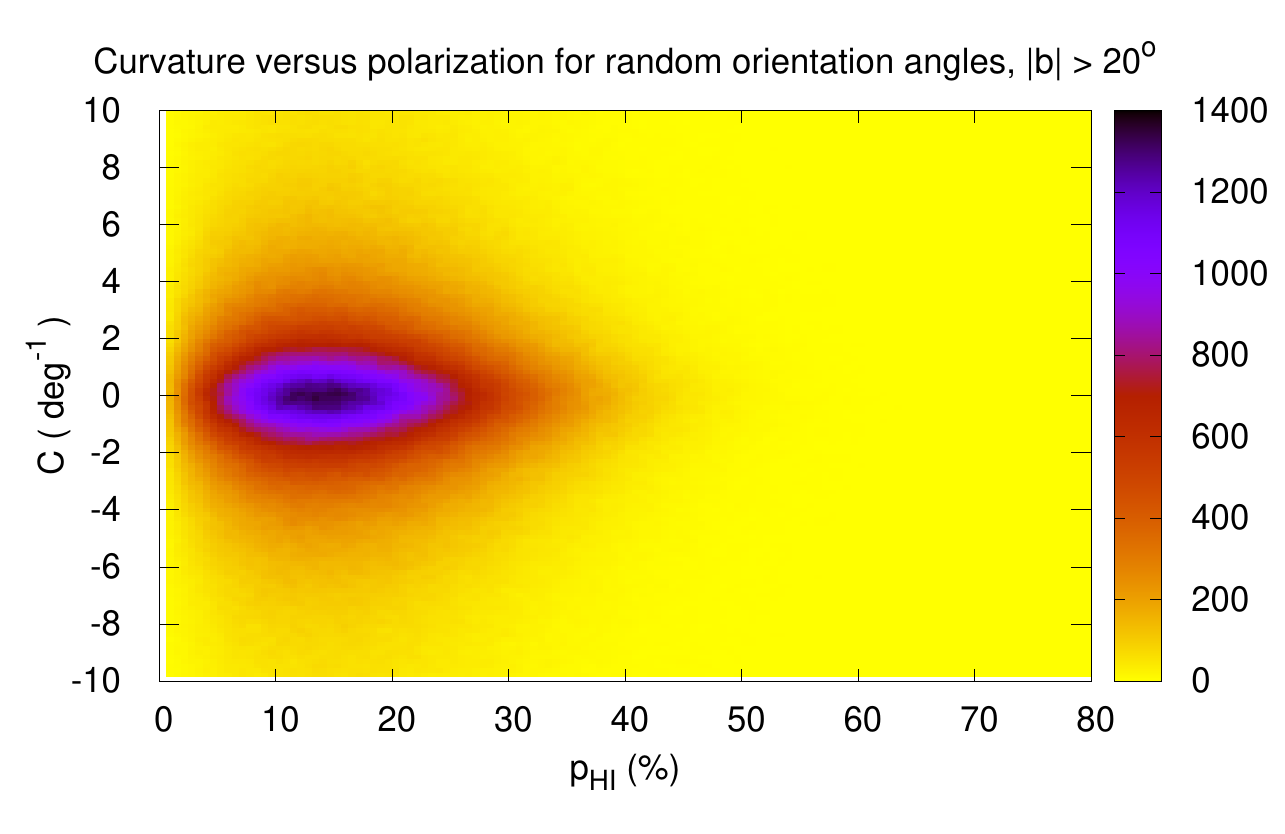}
   \caption{Two-dimensional density distribution of observed local filament
     curvatures $C$ in comparison to \hi\ polarization
     fraction. Top: High latitude sky. Middle: Selected fields with
     strong \hi\ anisotropies and associated radio-polarimetric filaments
     (\citealt{Kalberla2016b} and \citealt{Kalberla2017}). Bottom: Two-dimensional density
     distribution of spurious \hi\ polarization derived by modeling a
     random distribution of orientation angles along the line of sight.}
   \label{Fig_Sc_pol}
\end{figure}

To allow a general comparison between curvatures $C$ and strength of
filamentary structures we use the eigenvalues $\lambda_- $ as proxies
for the strength of the magnetic field. The relation between $\lambda_-$
and the intensity of a filamentary structure, hence the strength of the
magnetic field, is nonlinear.  $\lambda_-$ depends on the second-order
partial derivatives but we expect that the prominence of filaments is
related to $\lambda_-$ (see Fig. 5 of \citealt{Kalberla2016}). Figure
\ref{Fig_SClam} displays the two-dimensional distribution of observed local filament
curvatures $C$ and eigenvalues $-\lambda_-$ on a logarithmic scale,
verifying the expected relation. The most prominent filaments have low
curvatures, leading to their description as fibers. Tangled worm-like
structures have low intensities, are less prominent and have therefore
previously mostly been discarded when discussing filamentary structures
\citep[e.g., by][]{Kalberla2016}. However, these less prominent worms
belong definitively to the structures caused by a small-scale Galactic
dynamo and should not be excluded from the scientific discussion of the
ISM.

\begin{figure*}[htp] 
   \centering
   \includegraphics[width=7cm]{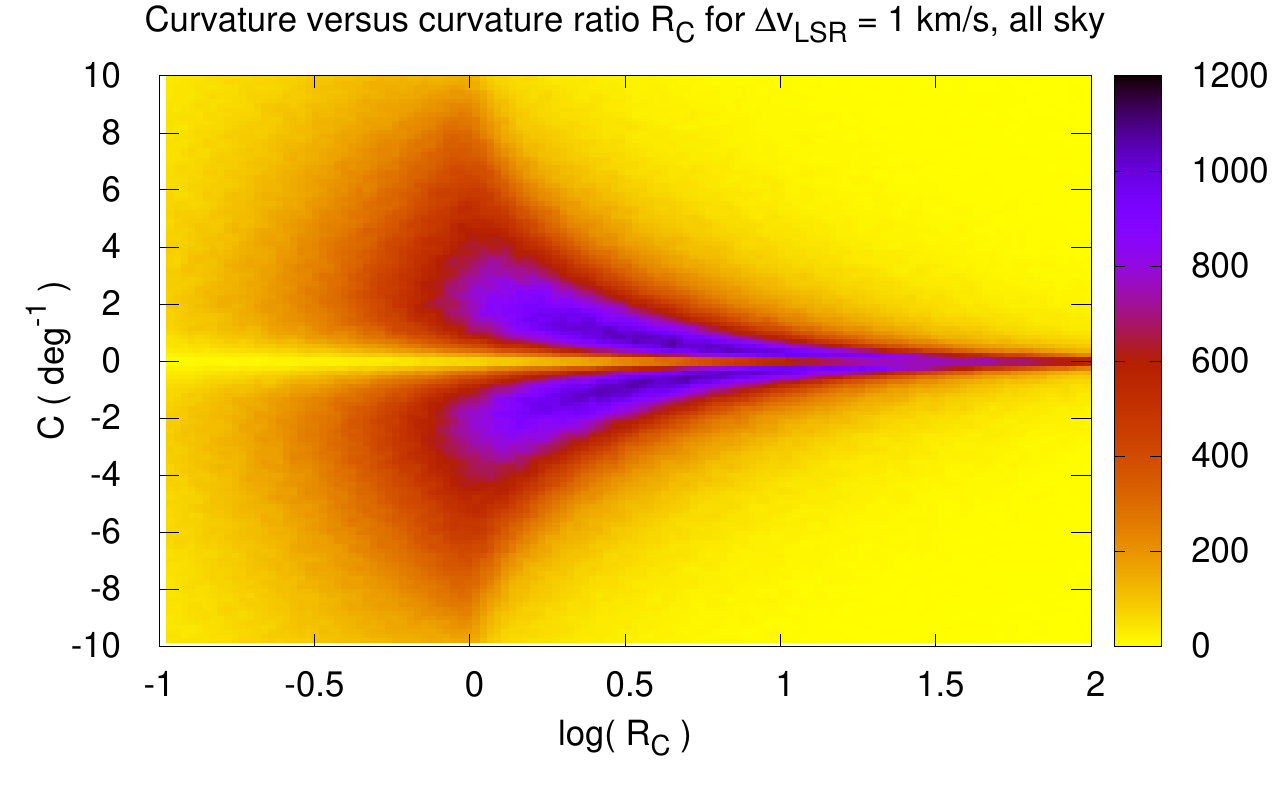}
   \includegraphics[width=7cm]{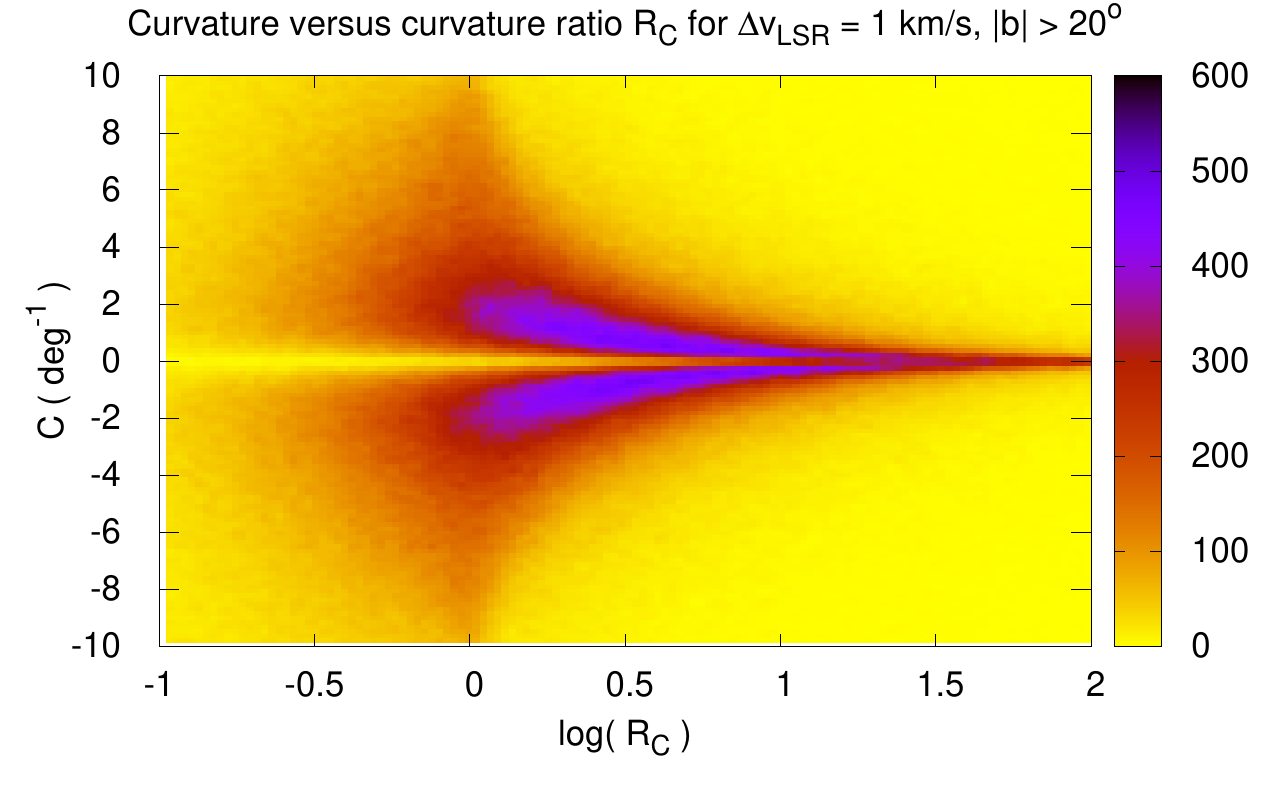}
   \includegraphics[width=7cm]{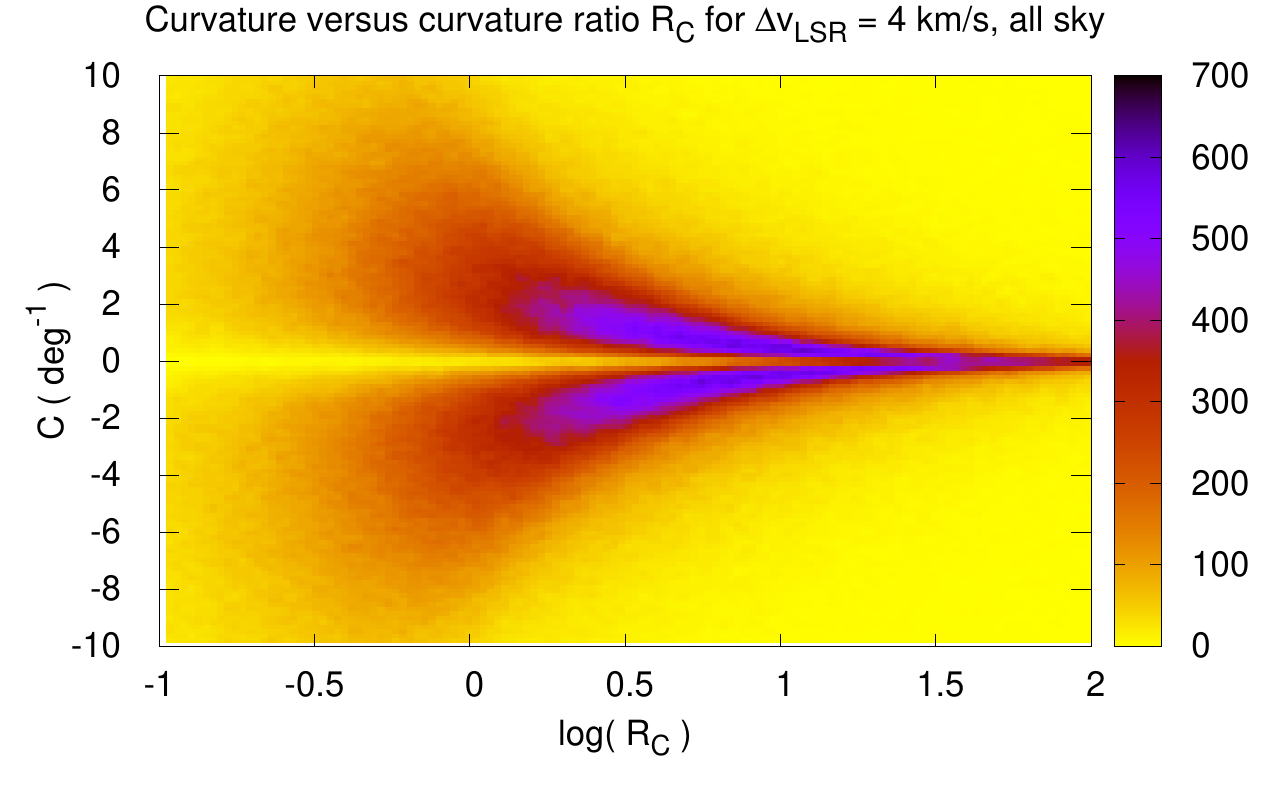}
   \includegraphics[width=7cm]{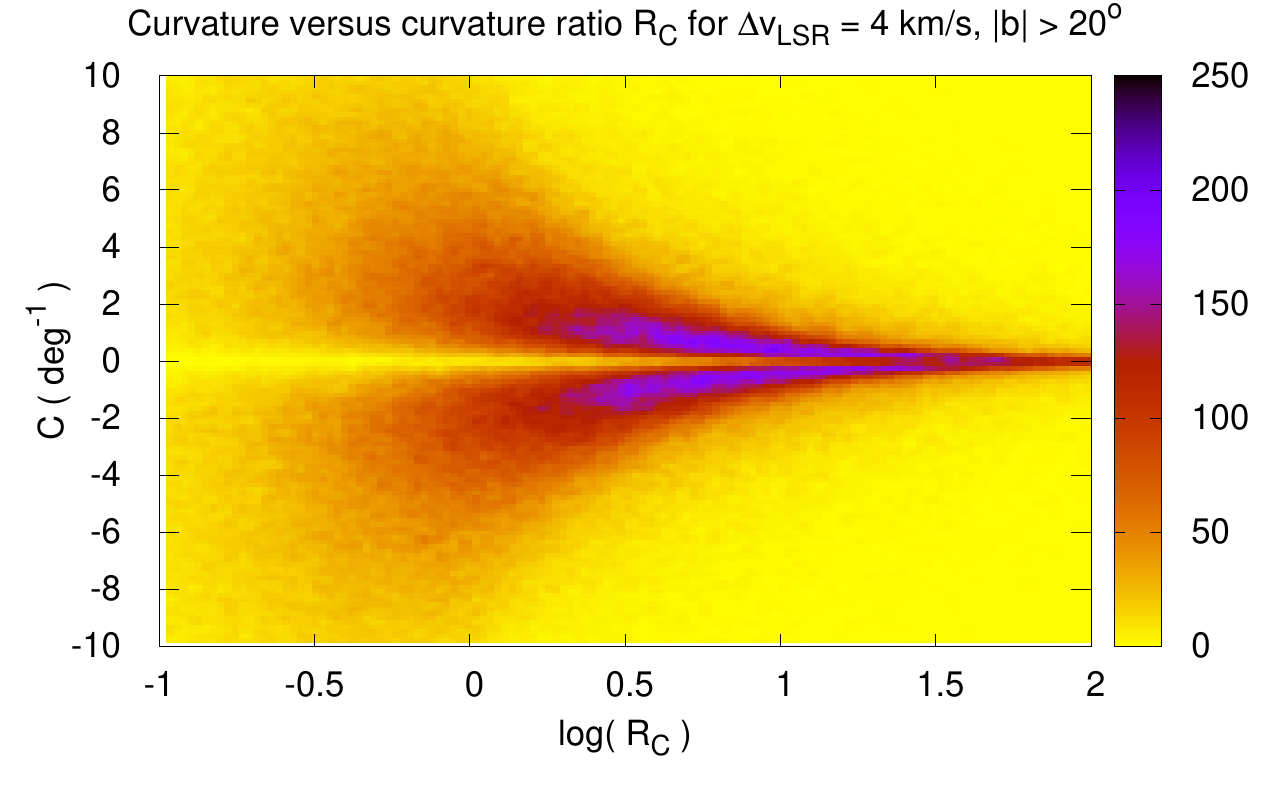}
   \includegraphics[width=7cm]{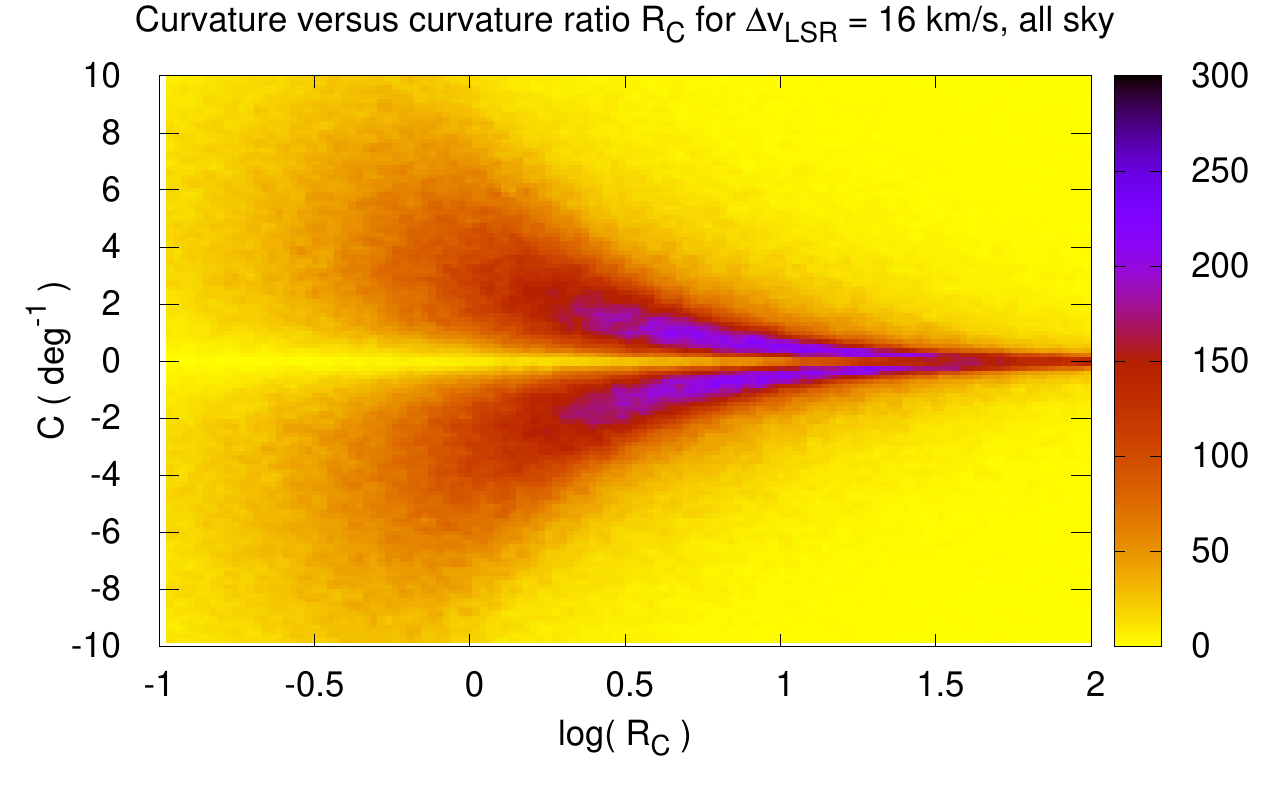}
   \includegraphics[width=7cm]{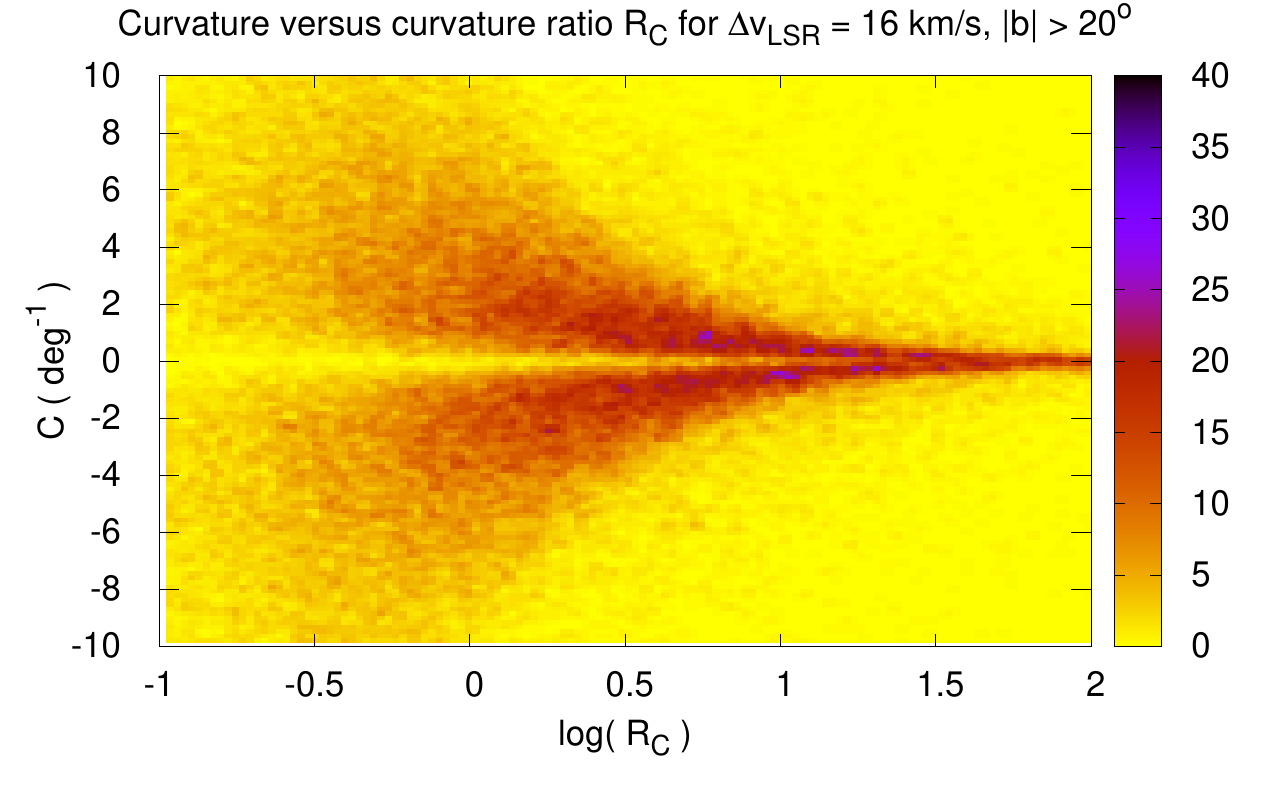}
    \caption{Two-dimensional histograms of filament curvatures $C$ in central parts
      of the filaments versus changes in curvature radius, characterized
      by the ratio $R_C$ for channels offset from $v_{\mathrm{fil}}$ by
      1, 4, and 16 \kms\ (top to bottom). Displayed are the all
      sky data (left) and at the high latitude data (right). }
   \label{Fig_Curve_depth_vel}
\end{figure*}

\subsection{Curvature versus \hi\ polarization fraction }
\label{Curve_pol}

A general relation between the curvature $|C|$ and the strength of the
magnetic field implies that also the \hi\ polarization fraction
$p_\mathrm{HI}$ must be related to $C$. Here we must take geometrical
effects into account. The curvature $|C|$ is derived from bending of
filaments in the plane of the sky while the \hi\ polarization fraction
$p_\mathrm{HI}$ is defined along the line of sight. Small curvatures
imply high magnetic field strengths with high internal coherence and
\hi\ polarization fractions. Figure \ref{Fig_Sc_pol} shows on top the
observed relation between $C$ and $p_\mathrm{HI}$ at latitudes $|b| >
20\degr$; all sky results are similar but not displayed. The highest
polarization fractions are reached for filaments with the lowest
curvatures as expected for a small-scale dynamo.

We repeat the
calculations by modeling as in Sect. \ref{Curvature_dist} a random
distribution of orientation angles. We consider the same sample of
positions along the filaments except that now the angles $\theta$ from
Eq. (\ref{eq:theta}) are replaced by a random distribution in
velocity. The result is shown in the bottom panel of
Fig. \ref{Fig_Sc_pol}.  We find nowhere an indication that a spurious
polarization signal of this kind could affect our results, the
orientation angles $\theta$ in our analysis are well defined

The FIR and \hi\ coherence is not the only case that hints to relations
between \hi\ filaments and magnetism.  Three fields that allow a
detailed comparison of filamentary structures observable with the Low
Frequency Array (LOFAR) as polarimetric filaments in close connection to
cold \hi\ filaments have been studied previously by
\citet{Kalberla2016b} and \citet{Kalberla2017}. The LOFAR observations
revealed strikingly linear coherent structures in Faraday depth,
including some prominent filaments several degrees in length. The
orientations of the LOFAR structures are affected by magnetic fields in
the warm ionized medium. The observed alignment between magneto-ionic
structures and \hi\ filaments in these fields implies that the
\hi\ filaments are shaped by the magnetic field.  The data from these
polarimetric studies are displayed in the middle of
Fig. \ref{Fig_Sc_pol} for comparison. This plot shows statistical
uncertainties from the restricted number of analyzed filaments but there
are no indications for any noise biases as documented in the lower
panel. The two-dimensional density distribution of $C$ against
$p_\mathrm{HI}$ shows a similar trend as the distribution from
FIR and \hi\ structures at high latitudes, except that the polarization
fractions in the case of the polarimetric filaments are somewhat
lower. We question that the lower polarization is significant; a part of
the filaments are close to the Galactic plane and may suffer from
confusion.  The investigations on these radio-polarimetric structures
indicate anisotropies and systematic changes of turbulent power spectra
for those parts of the \hi\ velocity distribution that are associated
with the magneto-ionic medium, we refer to \citet{Kalberla2016b} and
\citet{Kalberla2017} for details.

\begin{figure*}[tp!] 
   \centering
   \includegraphics[width=7cm]{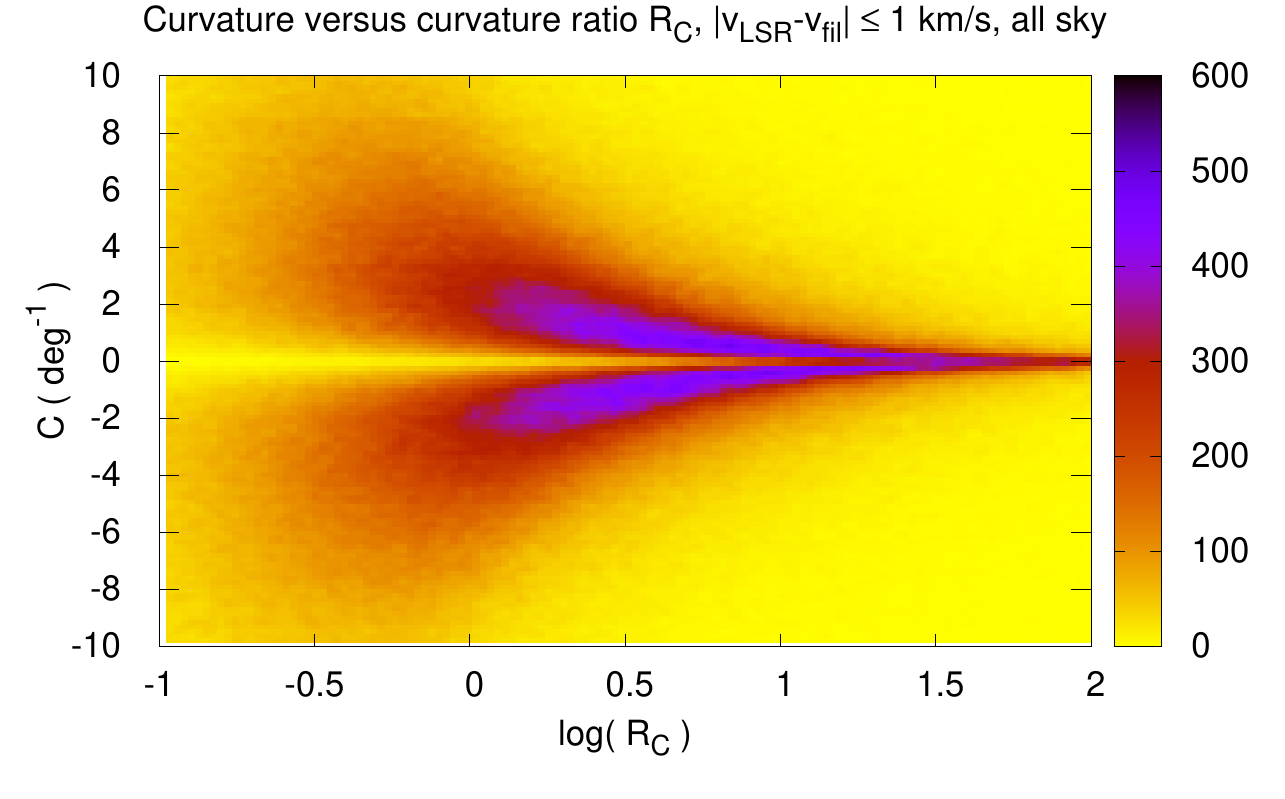}
   \includegraphics[width=7cm]{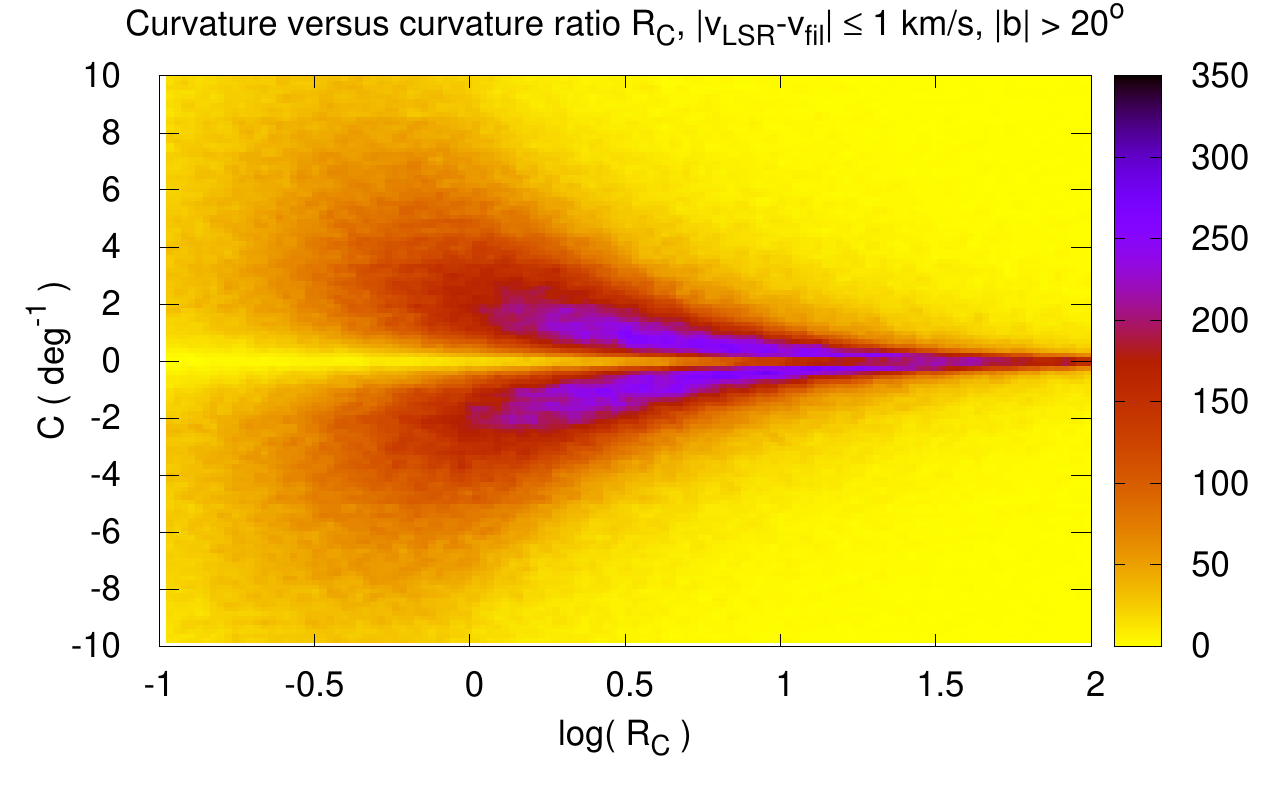}
   \includegraphics[width=7cm]{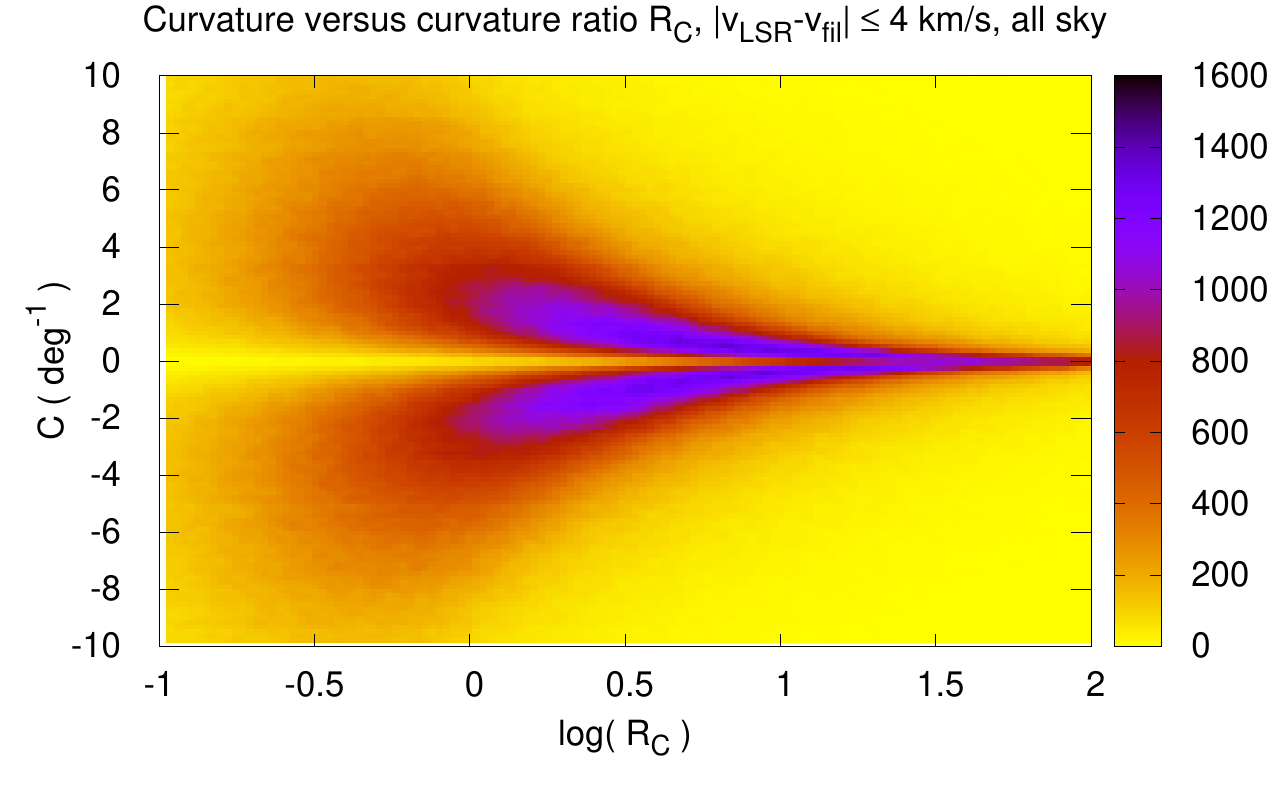}
   \includegraphics[width=7cm]{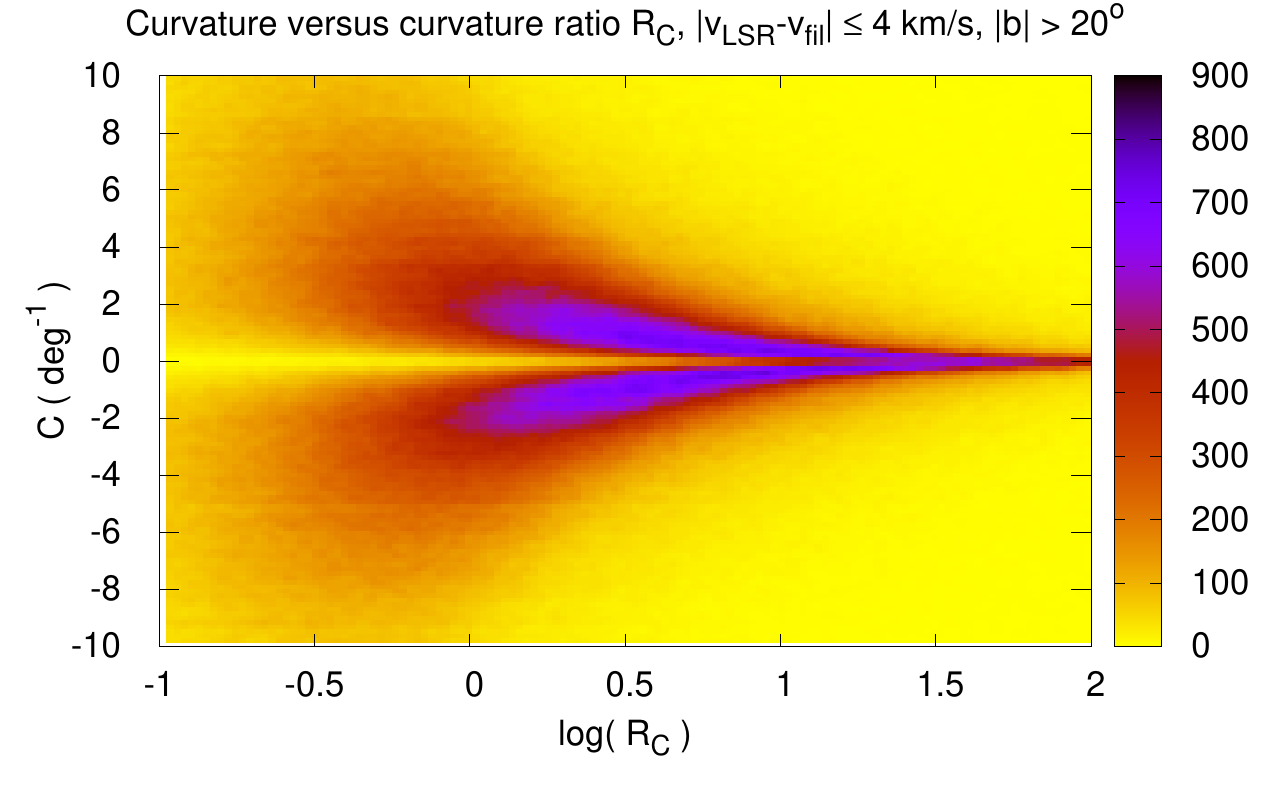}
   \includegraphics[width=7cm]{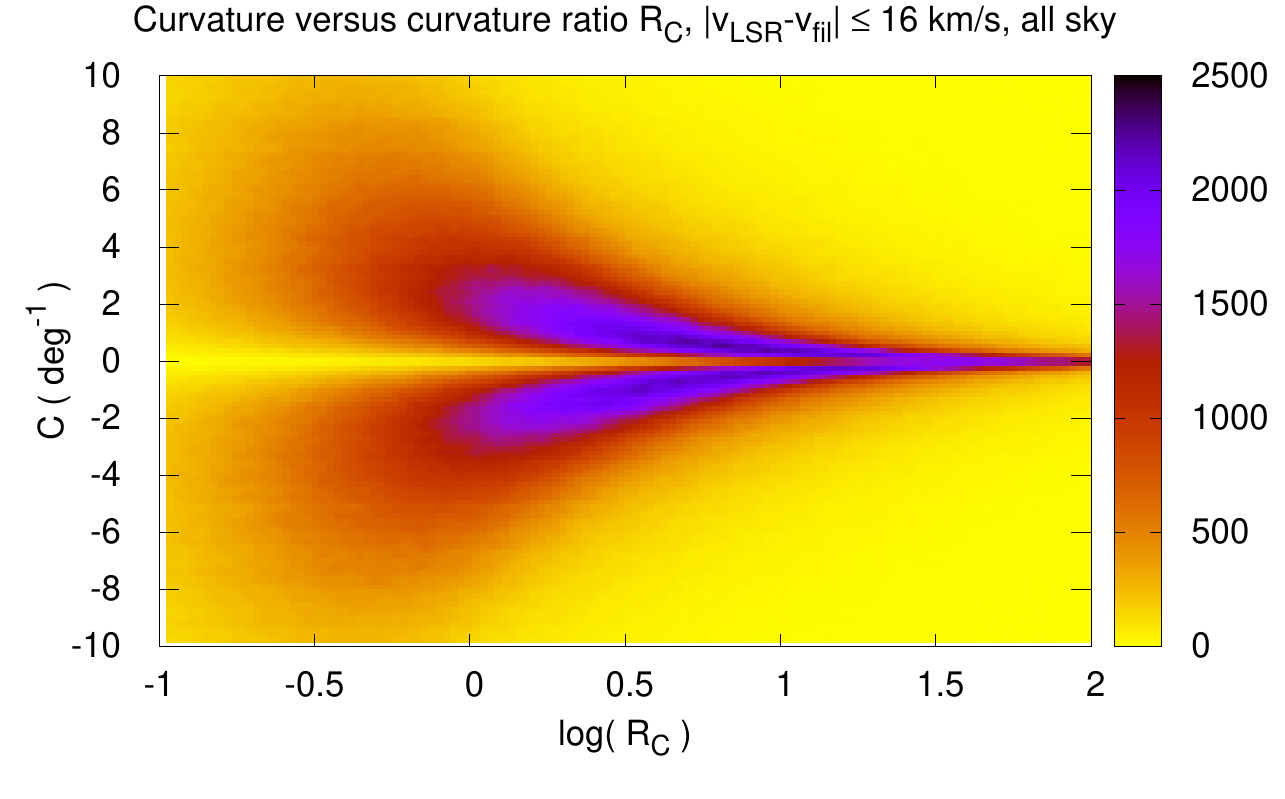}
   \includegraphics[width=7cm]{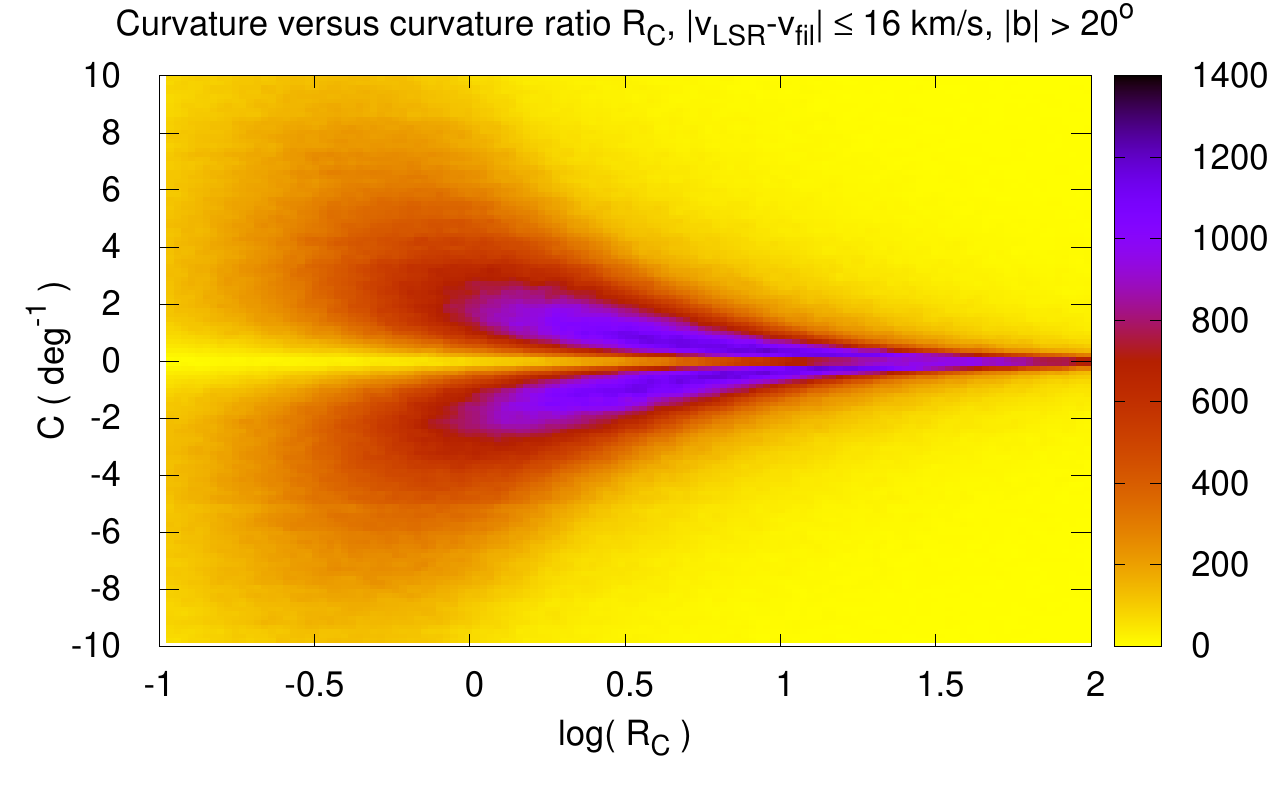}
    \caption{Two-dimensional histograms of filament curvatures $C$ in central parts
      of the filaments versus changes in curvature radius, characterized
      by the ratio $R_C$ for curvatures at the closest positions off the
      filaments. We integrate $R_C$ for velocities at channels
      $|v_{\mathrm{LSR}}-v_{\mathrm{fil}}|$ by 1, 4, and 16 \kms\ (top
      to bottom). Displayed are the all sky data (left) and the
      high latitude data (right). }
   \label{Fig_Curve_depth_pos}
\end{figure*}

\section{Curvature changes across flux tube envelopes }
\label{Curve_dev}

The determination of filament properties and curvatures in the previous
sections was limited either to FIR filaments or to \hi\ structures in
narrow channels with best fit velocities $v_{\mathrm{fil}}$ as
determined in Sect. \ref{v_fil}. We adopt here the working hypothesis
that these filaments stand for structures of magnetized flux tubes in
the diffuse ISM. Coherent FIR and \hi\ structures at positions with the best
fit velocities $v_{\mathrm{fil}}$ (Fig. \ref{Fig_Vel_HI}) represent
in such a model the central parts, the bones of the flux tubes.
We intend in this section to determine the properties of
filament envelopes along the flux tubes at positions deviating from the
central parts. We distinguish structures with positional offsets in the
plane of the sky but also entities along the line of sight with
velocities that differ from $v_{\mathrm{fil}}$.

\subsection{Structural changes at offset velocities }
\label{Envelope_vel}

Curvatures that we discussed in the previous sections depend on
positions and orientation angles that are defined only along the
filaments at velocities $v_{\mathrm{fil}}$. Considering deviating
velocities $\Delta v_{\mathrm{LSR}} = |v_{\mathrm{LSR}} -
v_{\mathrm{fil}}|$ we may expect that on average structures with
increasing offsets $\Delta v_{\mathrm{LSR}}$ are also offset
increasingly in distance. For a turbulent medium the characteristic
velocities $v_l$ and $v_0$ at scales $l$ and $l_0$ are related according
to $v_l / v_0 = (l/l_0)^{1/3}$
\citep[e.g.,][Sect. 7.6]{Frisch1996}. Considering \hi\ channels at
velocities offsets from $v_{\mathrm{fil}}$ would accordingly allow the flux tube properties far from the central parts to
be determined, though an
exact scaling between velocity and distance is missing.

\begin{figure*}[tp!] 
   \centering
   \includegraphics[width=7cm]{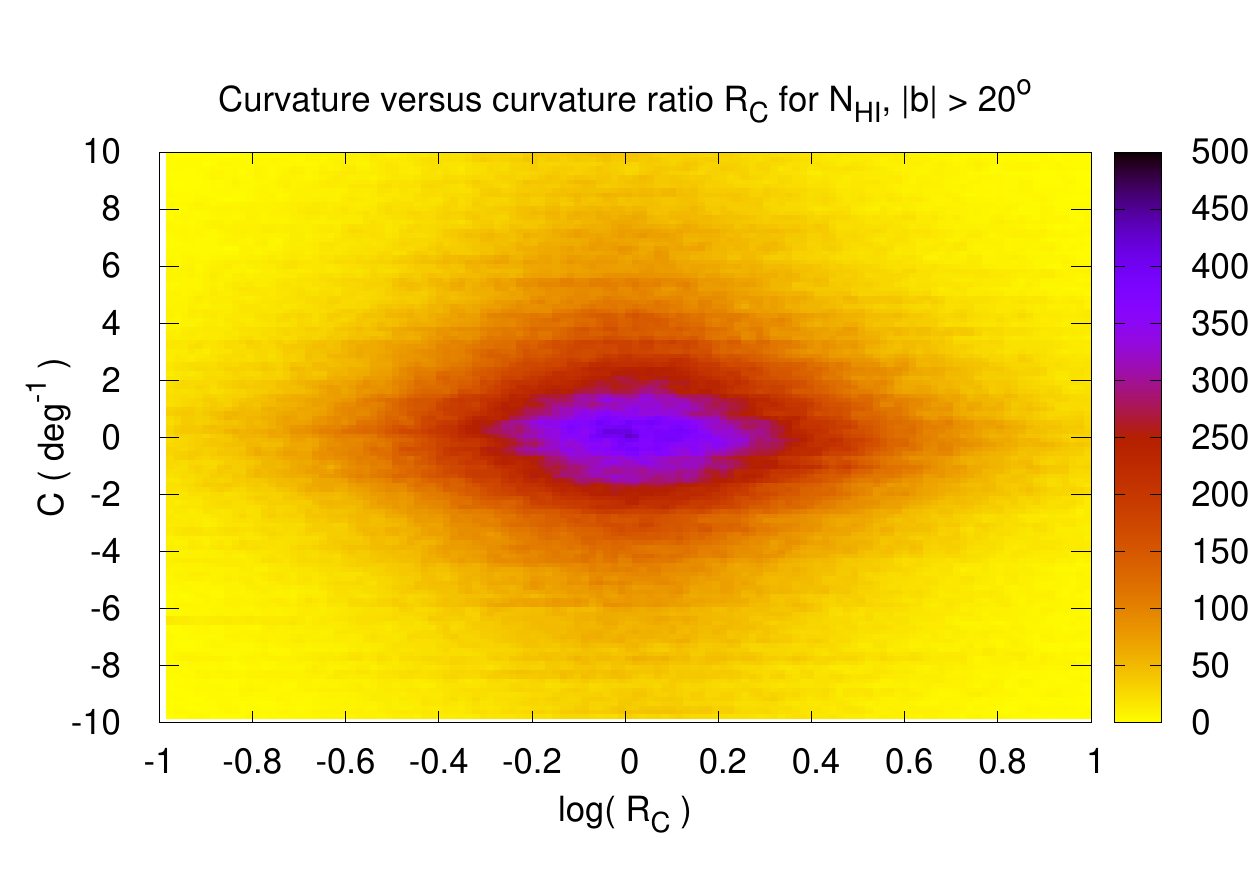}
   \includegraphics[width=7cm]{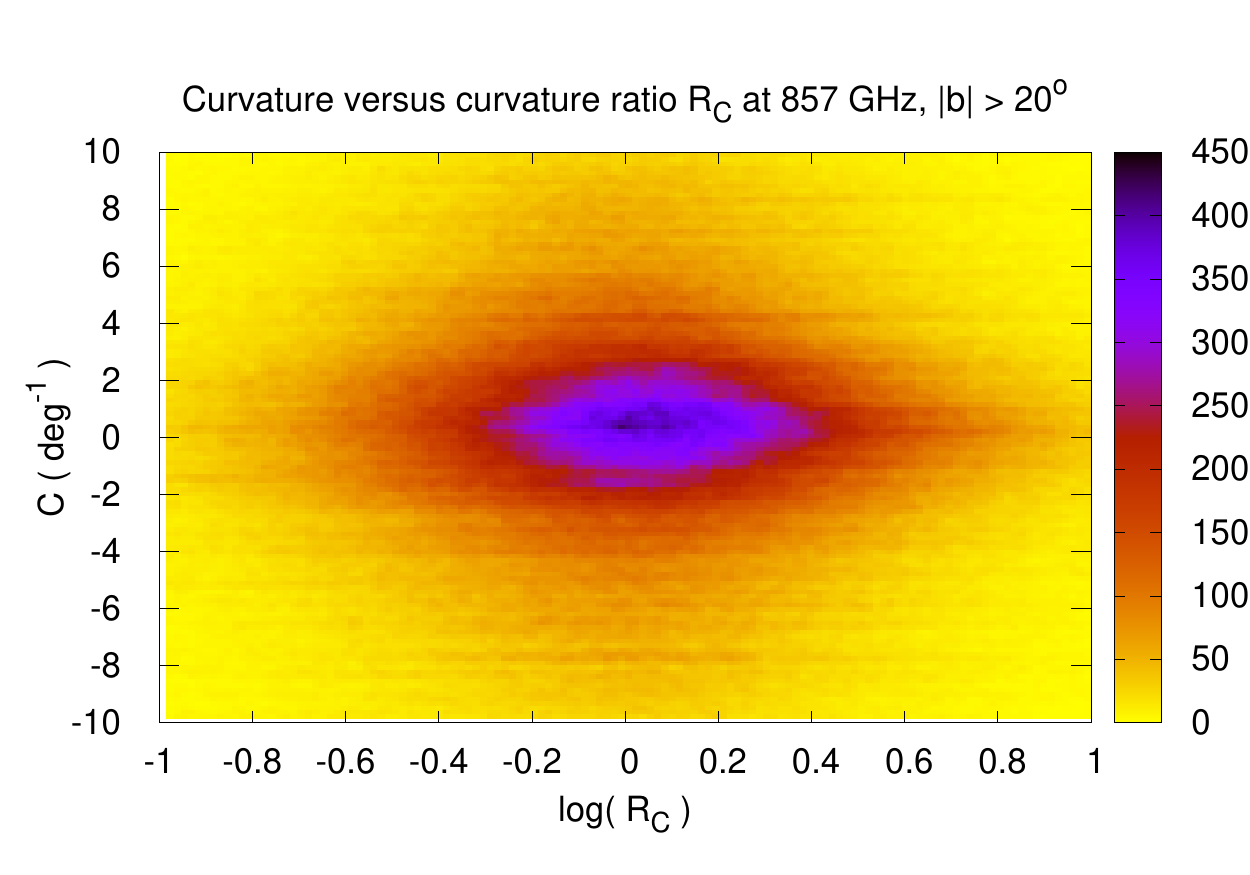}
   \includegraphics[width=7cm]{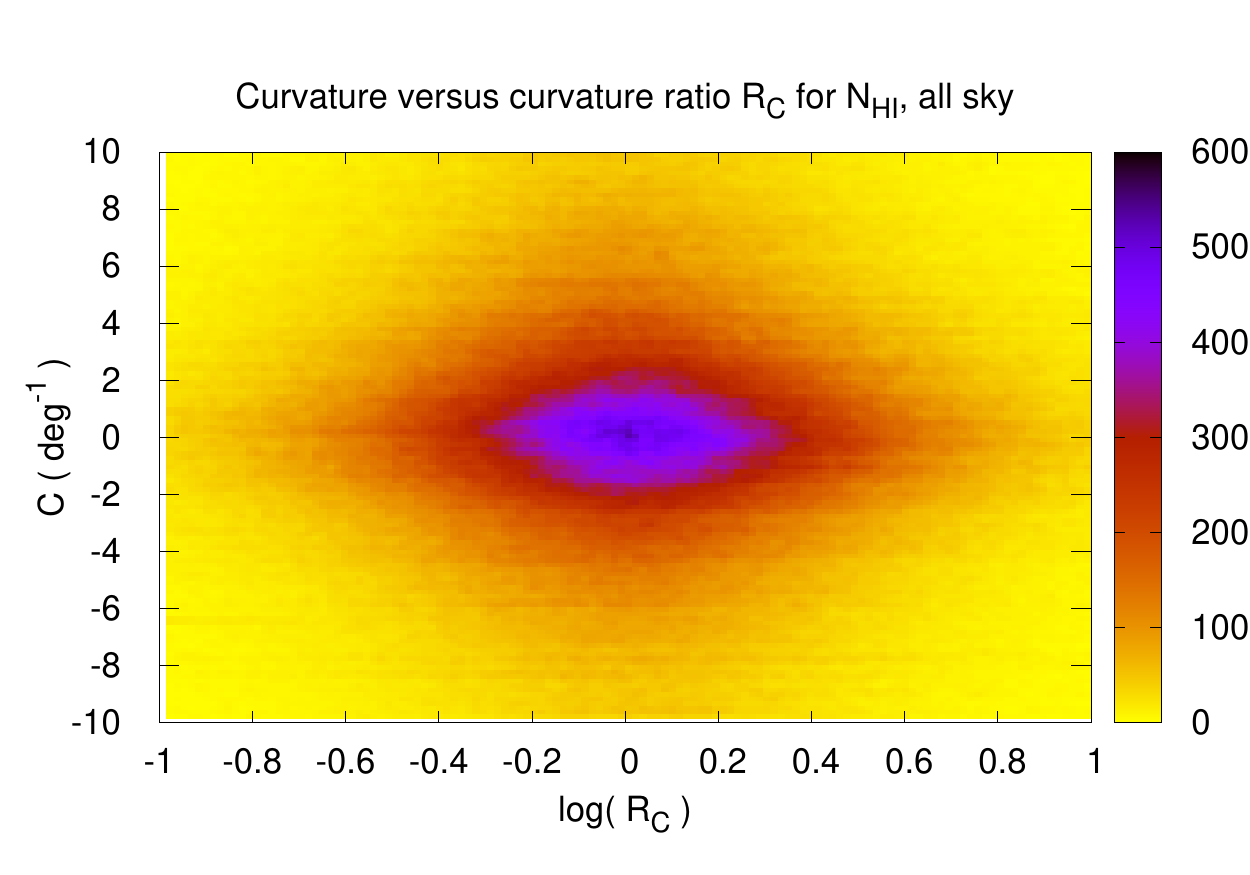}
   \includegraphics[width=7cm]{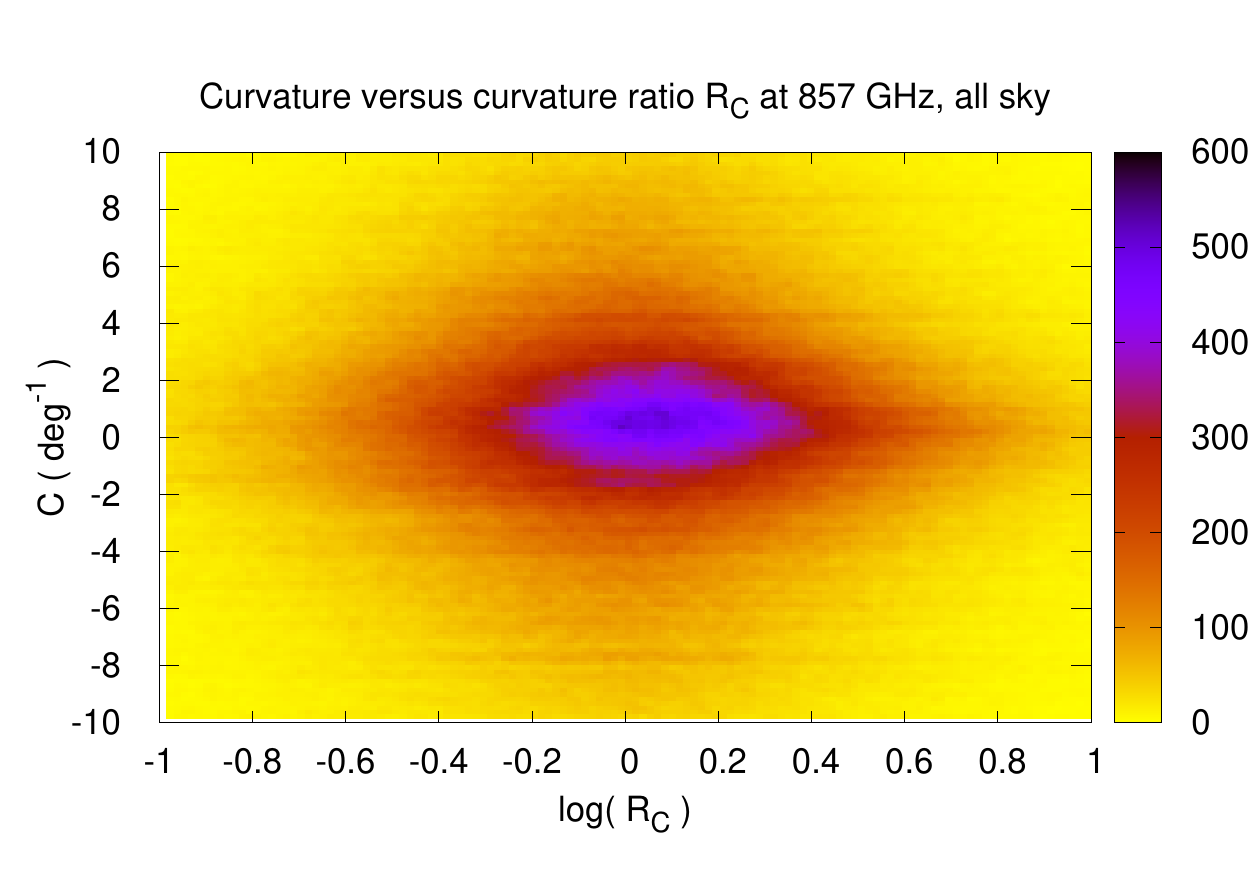}
    \caption{Two-dimensional histograms of filament curvatures $C$ in central parts
      of the filaments versus changes in curvature radius, characterized
      by the ratio $R_C$ for curvatures at the closest positions off the
      filaments. To the left we display ratios $R_C$ derived from
      \hi\ column densities, and to the right $R_C$ ratios are given for FIR
      intensities at 857 GHz. The top panels display distributions
      restricted to high Galactic latitudes, and the bottom panels are from
      all sky data. }
   \label{Fig_Curve_NH}
\end{figure*}

We extend the calculation of curvatures to all individual \hi\ channels
and consider at each filament position also the velocities
$v_{\mathrm{fil}} \pm \Delta v_{\mathrm{LSR}}$ for $|\Delta
v_{\mathrm{LSR}}| < 25$ \kms. To avoid any noise biases we use identical
criteria as in Sect. \ref{v_fil} for the significance of filamentary
structures. We characterize systematical changes in curvature between
envelope and center of the flux tube by the ratio $R_C(\Delta
v_{\mathrm{LSR}}) = C(v_{\mathrm{fil}} + \Delta v_{\mathrm{LSR}}) /
C(v_{\mathrm{fil}}) $. Figure \ref{Fig_Curve_depth_vel} displays two-dimensional
histograms of curvatures $C$ and curvature ratios $R_C$ at velocity
offsets of $\Delta v_{\mathrm{LSR}} = 1$, 4, and 16 \kms. We observe the
general trend that for increasing velocity offsets $\Delta
v_{\mathrm{LSR}}$ the curvature ratios $R_C$ increase. In other words,
the filament curvatures increase systematically when considering outer
filament layers that are characterized by increasing velocity offsets
$|\Delta v_{\mathrm{LSR}}|$. Considering the prognosis that the magnetic
field strength and curvature are for a small-scale dynamo
anticorrelated, $|B| \propto C^{-1/2}$ (\citet{Schekochihin2004},
\citet{Schekochihin2004}, and \citet{St-Onge2018}), we conclude that the
magnetic field of the low curvature segments in the centers of the flux
tubes must be enhanced relative to the more curled environments. These
central parts are most significant (see Sect. \ref{Curve_strength}). The
eigenvalues $\lambda_-$ increase systematically with increasing velocity
offsets and at $|\Delta v_{\mathrm{LSR}}| \sim 4$ \kms\ the number of
significant filamentary structures that can be analyzed drop by about
50\%. At $|\Delta v_{\mathrm{LSR}}| \ga 16$ \kms\ only a few
significant filamentary \hi\ structures remain.

\subsection{Structures offset in position}
\label{Envelope_pos}

Changes in filamentary structures for narrow \hi\ channels can also be
observed at positions offset from the central parts of the
filaments. However, the disadvantage of such an analysis is that we can
no longer track curvatures along the bones of the filaments. Curvatures
in the previous sections were determined along the filament ridges but
for offset positions this orientation is lost. We can only determine
average local curvatures that can be related to the curvature at the
closest filament position with distance $D \sim
\sqrt{((l-l_f)\cos(b))^2+(b-b_f)^2}$. It also can happen that the most
significant structure with the lowest $\lambda_-$ value at the position
$l,b$ is found at a velocity that deviates from $v_{\mathrm{fil}}$ at
the nearest position $l_f,b_f$. Accordingly we need to change definition
for the curvature ratios to $R_C(D) = C(l,b,v_{\mathrm{LSR}}) /
C(l_f,b_f,v_{\mathrm{fil}}) $. In Fig. \ref{Fig_Curve_depth_pos} we
display these ratios for \hi\ layers in the velocity ranges
$|v_{\mathrm{LSR}}-v_{\mathrm{fil}}| \le 1$, 4, and 16 \kms. $R_C(D)$
can in most cases only be traced to $D \la 1\degr$. Searching for
related filamentary structures up to distances $D \sim 10\degr$ needed
lots of CPU time but remained without success.

The plots in Fig. \ref{Fig_Curve_depth_pos} are not directly comparable
to those in Fig.  \ref{Fig_Curve_depth_vel} but they tell the
same. Curvatures for structures offset in velocity-position space from
the central parts of flux tubes increase significantly in comparison to
curvatures along central bones of the filaments. Increased curvatures
imply increased curling of magnetic field lines with decreasing field
strengths according to predictions for the small-scale dynamo. Such
structures have lower \hi\ intensities and the $\lambda_-$ values are
less significant.

\subsection{Structures in \hi\ column densities and FIR}
\label{Envelope_NH}

We repeated the analysis from the previous subsection but substituted
the individual channel maps with a single \hi\ column density map for $
|v_{\mathrm{LSR}}| < 50 $. The result is shown in
Fig. \ref{Fig_Curve_NH} on the left. In this case we find no evidence
for systematic changes of the curvature distributions with significant
preferences in curvature radius as observed in
Figs. \ref{Fig_Curve_depth_vel} and \ref{Fig_Curve_depth_pos}.
Averaging across numerous filaments smears out any information on the
curvature of the ISM. Structures along the line of sight but outside the
filaments are weaker, accordingly tangling causes de-correlation when
integrating in velocity. The same effect occurs if we
consider the distribution of FIR intensities. Figure \ref{Fig_Curve_NH}
shows that the curvature distributions in FIR (right) replicate the
distributions for $N_\mathrm{HI}$ on the left hand side.

As discussed in Sect. \ref {Vel_disp}, a coherent velocity field along
the filaments is only detectable if we recognize that the FIR filaments
are associated with \hi\ structures on small scales. We demonstrated in
Sect. \ref{Coherence} that these features are cold. Accordingly they
need to be analyzed with narrow channel spacings. Integrating in
velocity smears out structures caused by the CNM.  The differences
between Figs. \ref{Fig_Curve_depth_vel} and \ref{Fig_Curve_depth_pos} on
one side and Fig. \ref{Fig_Curve_NH} on the other side demonstrate
impressively observer's obligation to obey the sampling theorem
(\citet{Blackman1958} and \citet{Shannon1975}). Otherwise sensitive
information about cold turbulent structures on small scales, imprinted
on the velocity field, gets lost. This information is available in full resolution three-dimensional
spectral \hi\ data with sufficient sensitivity but not in two-dimensional {\it
  Planck} FIR observations or \hi\ data that are missing the third
dimension. The necessity to consider the three-dimensional distribution in the ISM was
also demonstrated recently by \citet{Pelgrims2021} when studying the
line of sight frequency decorrelation of polarized dust emission along
the line of sight. 

\section{Summary and discussion}
\label{Summary}

Filamentary \hi\ structures are known to be well aligned with the
plane-of-sky magnetic field orientation as measured with optical
starlight polarization and polarized thermal dust emission
\citep[e.g.,][]{Clark2018}. This tight correlation exists for narrow
velocity intervals only, and the orientation of these features is found
to be coherent in velocity space. \citet{Clark2018} and
\citet{Clark2019b} propose using this \hi\ coherence to define for such
structures an \hi\ polarization along the line of sight. Analogous to
optical polarization that can be produced by transition through aligned
structures in refracting media, this \hi\ polarization is indicative of
coherence and alignment in \hi\ filaments. It was shown by
\citet{Clark2019b} that \hi\ Stokes parameter maps are comparable to the
{\it Planck} 353 GHz Q and U maps of polarized dust emission. At a
resolution of 80\arcmin,\ many of the large-scale features are
reproduced in \hi.

Here we study the FIR and \hi\ coherence in more detail with a better
spatial resolution. We consider the more sensitive {\it Planck} 857 GHz
FIR data smoothed to 18\arcmin\ resolution. For the \hi\ we use
unsmoothed HI4PI data at resolutions of 10\farcm8 for EBHIS and
14\farcm5 for the GASS survey. Using the Hessian operator, we extract
filaments simultaneously for both data sets and determine alignment
angles along the filaments.  We consider several measures for the
alignment of the filamentary structures, and these indicate in all cases
that the alignment on small scales improves significantly with respect
to previous determinations on scales of around one degree. For \hi\ in
narrow velocity intervals of 1 \kms, we obtain an excellent alignment
between FIR and \hi\ filaments, indicating a close coherence between FIR
and CNM. These filamentary structures are also coherent in velocity
space and, in general, are cold. Following \citet{Clark2018} and
\citet{Clark2019b}, we determine in Sect. \ref{HI_pol} the
\hi\ polarization along the line of sight and find for filamentary
structures an average \hi\ polarization fraction of 30\% with peak
values around 80\%. This is a factor of two to three more than reported
in previous determinations. Since we did not apply any smoothing to the
\hi\ data, we probe, at an assumed distance of 100 pc
\citep[e.g.,][]{Sfeir1999}, regions of about 0.3 pc. 

We probed the velocity structure of the \hi\ counterparts for the FIR
filaments and discovered a well-defined coherent velocity field. Filaments
are local phenomena with center velocities around 0 \kms. The FWHM
width of this distribution is 16.6 \kms. Along the filaments the typical
velocity dispersion is $\mathcal{V} = 5.5 $ \kms. Most of the
\hi\ filaments are associated with the CNM. Toward the structures
investigated here we also observe Doppler
temperatures that are lower by up to a factor of six for the accompanied LNM and WNM gas, implying a CNM in cold
cores of a multiphase medium in phase transition. Low Doppler
temperatures imply that FIR and \hi\ coherence is best defined in narrow
velocity intervals and probably also in small and dense regions in the
centers of the filaments. The spatial distribution of the velocity field
in comparison to velocity dispersions $\mathcal{V}$ and harmonic mean
Doppler temperatures shows common structures for most of the prominent
filaments, indicating that velocity dispersions and Doppler temperatures
are correlated along these filaments.

The polarization angle dispersion $\mathcal{S}$ introduced in
\citet{Planck2015} measures fluctuations of the polarization angle along
FIR filaments. The total angular scale
probed by us is four times the 18\arcmin\ FIR resolution, hence five to six
times the \hi\ beam, or about 2 pc at an assumed distance of 100 pc. At
such a scale we find, for a large fraction of the positions, significant
fluctuations in the position angles relative to the center
position. Contrary to previous investigations by \citet{Clark2019b}, we
find no indications for a negative correlation between $\mathcal{S}$ and
the polarization fraction $p_\mathrm{HI}$. We confirm, however, the
inverse correlation between $\mathcal{S}$ and $p_\mathrm{353}$ at 353
GHz from Stokes parameters, derived previously mostly on angular scales
between 80\arcmin\ and 160\arcmin\ (\citealt{Planck2015},
\citealt{Planck2016}, \citealt{Clark2018}, \citealt{Clark2019b}, and
\citealt{Planck2020a}).

The high spatial resolution of our analysis allows us to parameterize
the bending of filamentary structures in terms of a curvature
distribution $P(C)$. We find a distribution that is well approximated by
curvatures, as predicted in the framework of a small-scale turbulent
dynamo by \citet{Schekochihin2002}. From our data we derive an excellent
agreement with the predicted curvature distribution
(\citealt{Schekochihin2002} and \citealt{Schekochihin2004}). This agreement
is, however, limited to curvatures $|C| \ga 1\ \mathrm{deg}^{-1}$,
corresponding to tangling radii of $R \la 1\degr $. While the
small-scale turbulent dynamo acts below this scale, the turbulent energy
on the largest scales must be fed by external sources, most probably by
supernovae. We confirm the prediction that for a small-scale dynamo
field strengths and curvatures are anticorrelated
(\citealt{Schekochihin2002} and \citealt{Schekochihin2004}). The most
prominent parts of the filaments have the lowest curvatures, and,
accordingly, these structures should have the strongest fields.
The curling in envelopes around the filaments increases significantly,
implying a decay of the field outside the central parts of the flux
tubes. Furthermore, we find that curvature is in general correlated with
filament coherence along the line of sight, characterized by the
\hi\ polarization fraction.

Our results on coherent FIR and \hi\ structures are for $|C| \ga
1\ \mathrm{deg}^{-1}$ incompatible with simulations of anisotropic MHD
turbulence in several of the fields selected by \citet{Planck2015b}. Throughout
those simulations there is a large-scale anisotropic component of the
magnetic field, as well as a turbulent component linked to the velocity
perturbations imposed on converging flows. From our results we conclude
that the spatial coherence of magnetized FIR and \hi\ filaments is shaped by
magnetic fields that were amplified by a small-scale turbulent
dynamo. Turbulence emerging from small scales can change the spectral
index of the turbulent flow at those velocities, which are characteristic
for FIR and \hi\ coherence.  In the framework of a small-scale turbulent
dynamo, it is also easy to support, on a Galactic scale, phase transitions
in a magnetically dominated state, as considered recently by
\citet{Falle2020}.

One of the main conclusions by \citet{Schekochihin2004} is that the
fully developed, forced, isotropic MHD turbulence is the saturated state
of the small-scale dynamo. In this case the kinetic energy spectrum is
dominated by the outer scale and has a steeper-than-Kolmogorov scaling
in the inertial range, while the magnetic energy is dominated by small
scales, at which it substantially exceeds the kinetic energy. The
strongest part of the local Galactic \hi\ emission extends over large
fractions of the sky at velocities close to zero. Assuming that this
emission is representative of a fully developed MHD turbulence driven
by a large-scale dynamo, we expect steep power spectra for this part of
the multiphase \hi. Focusing our attention on the coldest part of the
CNM, with volume filling factors at a 5\% level
\citep[e.g.,][]{Murray2020}, we find coherence between \hi\ structures
and the distribution of magnetic field lines. We find evidence that this
range is dominated by the magnetic energy and the small-scale
dynamo. Cold neutral medium  power spectra are expected to be shallow in comparison to
the large-scale, dominating part of the warmer \hi\ distribution with
excess power from structures on small scales.

In addition to filamentary structures, we find local condensations --
numerous blobs -- located along the filaments. These are cold coherent
FIR and \hi\ structures, point-like at 18\arcmin\ resolution, but they share
the properties of the filaments. This additional cold small-scale
emission is consistent with the excess power observed in power spectra
of the CNM at the highest spatial frequencies
\citep{Kalberla2019}. Since filaments are in general cold and restricted
locally to small scales, it is also plausible that the CNM power spectra
are shallower compared to multiphase power spectra since the power at
high spatial frequencies is increased by the small-scale dynamo. Cold
structures, driven by a small-scale dynamo, must evolve from small to
large scales. Turbulence in the local diffuse ISM has to be related to
phase transitions, but observational evidence currently appears to be
disregarded in many theoretical investigations \citep{Clark2019a}.

\begin{acknowledgements}
We acknowledge the referee for a careful reading and critical comments
that helped to improve the quality of the manuscript.  HI4PI is based on
observations with the 100-m telescope of the MPIfR (Max-Planck- Institut
für Radioastronomie) at Effelsberg and the Parkes Radio Telescope, which
is part of the Australia Telescope and is funded by the Commonwealth of
Australia for operation as a National Facility managed by CSIRO. This
research has made use of NASA's Astrophysics Data System.  Some of the
results in this paper have been derived using the HEALPix package.
   \end{acknowledgements}

\begin{appendix}  
  
  \section{The velocity decomposition algorithm} 
  \label{VDA}

In the main body of this publication we conclude that local \hi\ and
FIR filaments identified by the Hessian analysis are coherent structures
with well-defined radial velocities.  The interpretation of filamentary
structures as fibers in the sense of coherent density structures dated
back to \citet{Clark2014} and \citet{Clark2015} but was challenged by
\citet{Lazarian2018}. In the framework of velocity channel analysis
(VCA, \citet{Lazarian2000}) \hi\ structures seen
in individual channel maps are interpreted as velocity caustics,
intensity structures created by turbulent velocity fluctuations. The VCA
paradigm is that velocity crowding along the line of sight causes the
observed intensity enhancements that erroneously are interpreted as
density structures.  Thus caustics can mimic real physical
entities. \citet{Clark2019a} opposed and demonstrated that the
\hi\ filaments under discussion have enhanced FIR emission in comparison
to the \hi\ column densities, implying that this emission must originate
from a colder, denser phase of the ISM than the surrounding material.
The interpretation in terms of velocity caustics was reinforced by
\citet{Yuen2019} but \citet{Kalberla2020a} found it necessary to point
out that the small-scale \hi\ and FIR structures under debate are really
cold, caused by phase transitions rather than by velocity caustics.

After submission of this article  \citet{Yuen2021} derived the velocity
decomposition algorithm (VDA) with detailed recipes that intend 
to separate velocity and density contributions from observed channel
maps. This decomposition is closely related to VCA and based on the
fundamental VCA postulate \citep[][Sect. 6.3.1]{Lazarian2000} that
turbulent density and velocity fields are statistically uncorrelated for
the case of MHD turbulence. \citet{Yuen2021} use MHD simulations to
support the interpretation that intensity fluctuations are caused
predominantly by velocity caustics.  The VDA paper extends the VCA
paradigms and provides according to the authors the further theoretical,
numerical, and observational foundations for the theory describing the
statistics of the velocity caustics.

To decide whether dusty \hi\ filaments are density structures or rather
caustics we apply in the following VDA to the HI4PI data that have been
considered above. We repeat our analysis from Sect. \ref{Observations},
replacing the observed brightness temperature distribution with VDA
derived velocity and density fields.

\subsection{VDA definition}
\label{VDA_def}

Here we give only a very brief introduction to VDA, for detailed
explanations we refer to \citet{Yuen2021}. The basic postulate is that
in case of MHD turbulence the density and the velocity fluctuations are
statistically uncorrelated. Using VDA notations this is formulated as
$\langle p_v p_d \rangle = 0 $, for an ensemble average indicated by
$\langle ... \rangle$. The observed channel map $p$ (usually termed
brightness temperature $T_{\mathrm{B}}$) is decomposed in its velocity
contribution $p_v$ and density part $p_d$ according to
\begin{equation}
\begin{aligned}
  p_v &= p - \left( \langle pI\rangle-\langle p\rangle\langle I \rangle\right)\frac{I-\langle I\rangle}{\sigma_I^2}\\
  p_d &= p-p_v\\
  &=\left( \langle pI\rangle-\langle p\rangle\langle I \rangle\right)\frac{I-\langle I\rangle}{\sigma_I^2}
\end{aligned}
\label{eq:pvd}
\end{equation}
with $I = \int p(v) dv$ the total \hi\ intensity (or column density)
along the line of sight and $\sigma_I^2 = \langle (I - \langle I
\rangle)^2 \rangle$. The necessary condition is that the velocity width
$\Delta v$ of the PPV channel map is small in comparison to the
effective velocity width of the observed \hi\ gas.  As in
Sect. \ref{v_fil} we use $\Delta v = 1$ \kms. Typical cold
\hi\ structures have a velocity width of 3 \kms.

It is important to realize that VDA velocity and density fields
represent according to Eq. (\ref{eq:pvd}) velocity and density
fluctuations (positive or negative deviations from the mean) while the
observed \hi\ channel maps are density-weighted emission profiles,
differences between both approaches are explained in Fig. 2 of
\citet{Yuen2021}.

VDA velocity fluctuations are noticeable only in narrow velocity
intervals but vanish if velocity slices get thicker, $p_v = 0 $ for
$\Delta v\rightarrow \infty$. Per definition VDA density fluctuations
scale as $p_d\propto I$, in particular for low sonic Mach numbers $M_s
\ll 1$ \citep[][Sect. 3]{Yuen2021}. VDA densities, despite the naming as
densities, are not to be confused with volume densities; this can be
seen from Eq. (\ref{eq:pvd}); $p_v$, and $p_d$ have the same units as
$p$, usually termed brightness temperatures $T_{\mathrm{B}}$ in
K. Volume densities demand a definition of the volume that is occupied
by the \hi\ (or modeling an object considered as a cloud), such a
definition is not part of the VDA concept.

\subsection{Normalized covariance coefficients }
\label{NCC}

The correlation between velocity and density field can according to
\citet{Yuen2021} be best verified by using the normalized covariance
coefficient
\begin{equation}
  NCC(A,B) = \frac{\langle (A-\langle A\rangle)(B-\langle B\rangle)\rangle}{\sigma_A\sigma_B}
  \label{eq:NCC}
\end{equation}
to characterize correlations between two 2D maps $A$ and $B$; here we
use the same notations as \citet{Yuen2021}. This measure, also known as
Pearson product-moment correlation coefficient, is scale-invariant and
results in $NCC(A,B)\in[-1,1]$. The case $NCC(A,B) = 0 $ implies that
the two maps are statistically uncorrelated. A perfect correlation
requires that $|NCC(A,B)| = \pm 1$, the sign reflects the slope of the
linear regression that can be fitted in this case.

\begin{figure}[ht] 
   \centering
   \includegraphics[width=9cm]{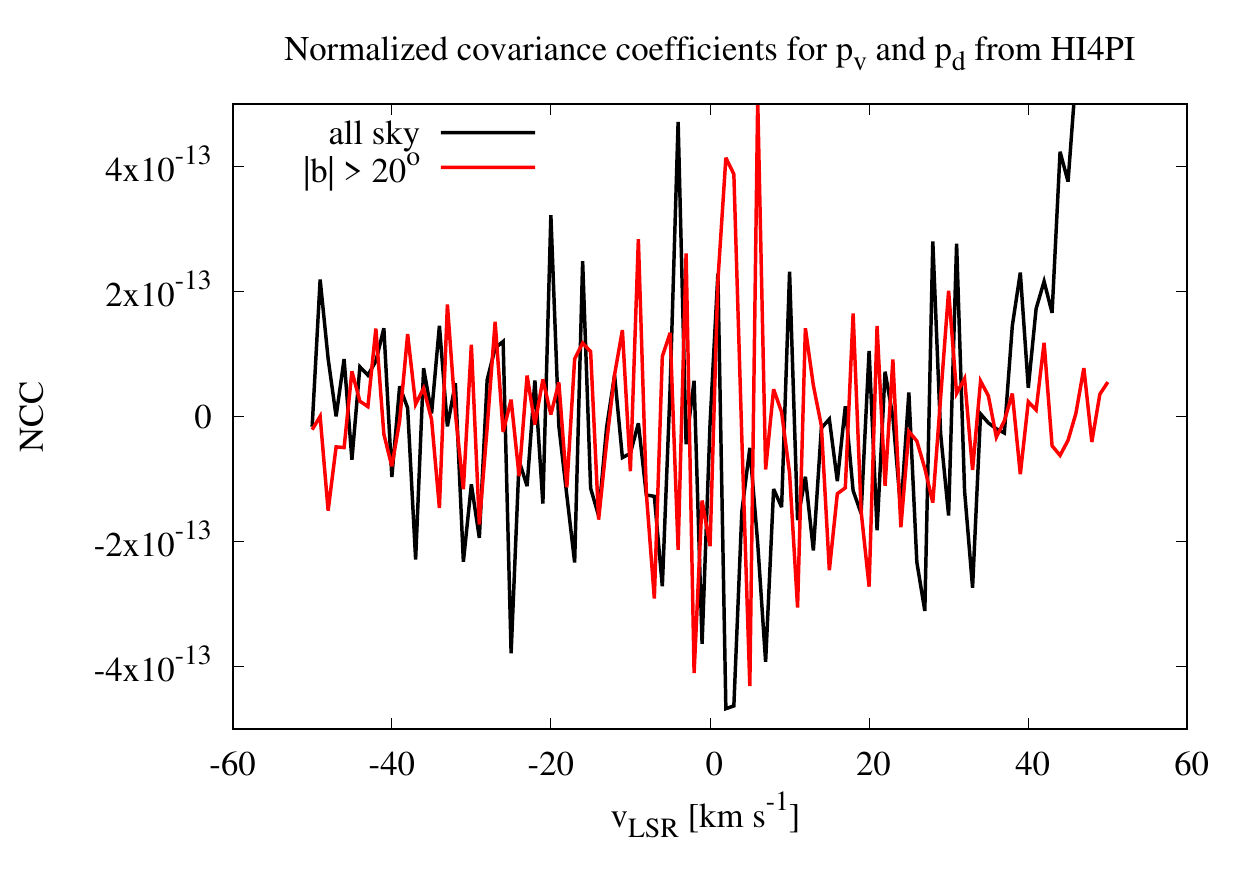}
   \caption{Normalized covariance coefficients derived from HI4PI data
     after VDA decomposition. }
   \label{Fig_NCC}
\end{figure}

Using Eq. (\ref{eq:pvd}) we decompose for each channel the observed
brightness temperature distribution $T_{\mathrm{B}}$ in VDA components. In the
following we use the notation $p = p_v + p_d$ that was proposed by
\citet{Yuen2021}. We verify that derived VDA velocity and density
components are statistically uncorrelated, $NCC(p_v,p_d) = 0$ see
Fig. \ref{Fig_NCC}. 

\begin{figure*}[thp] 
   \centering
   \includegraphics[width=9cm]{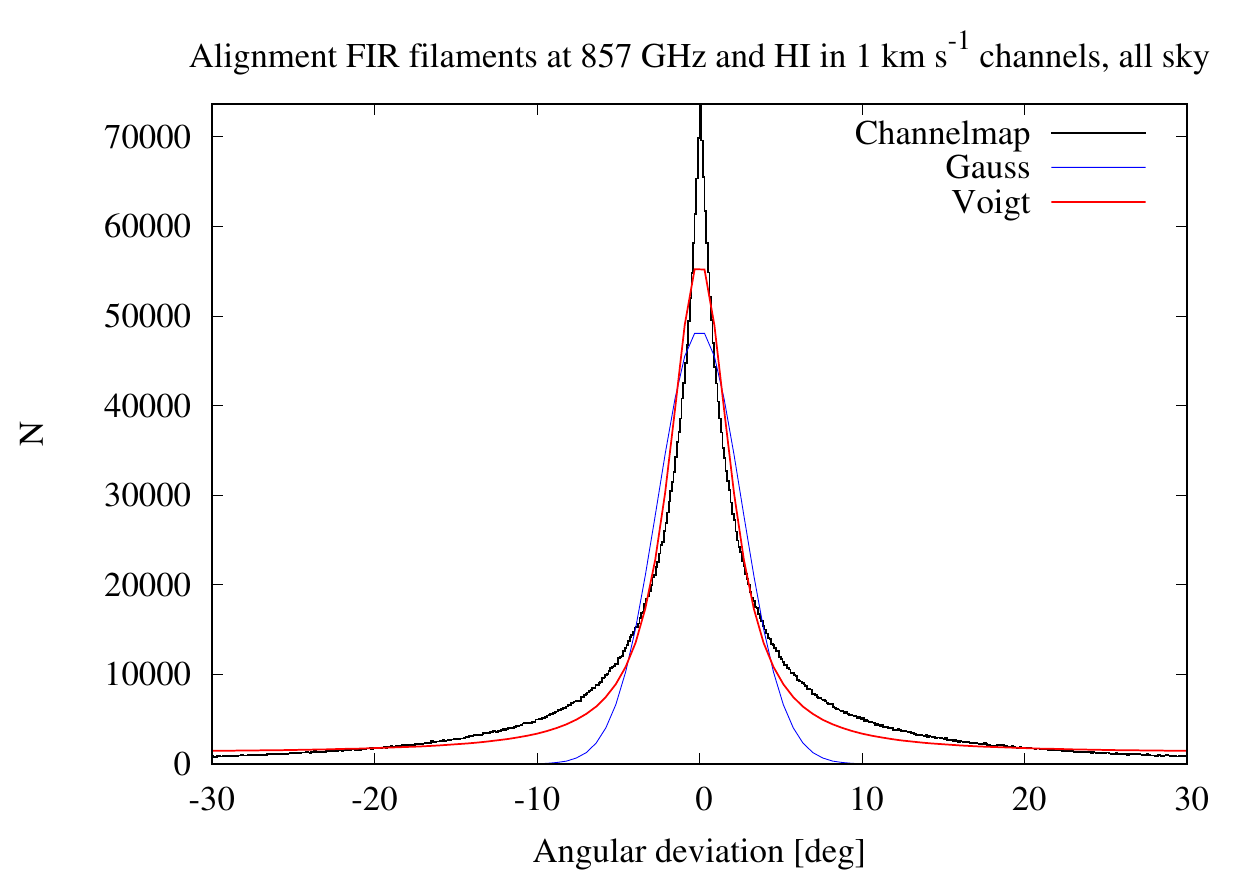}
   \includegraphics[width=9cm]{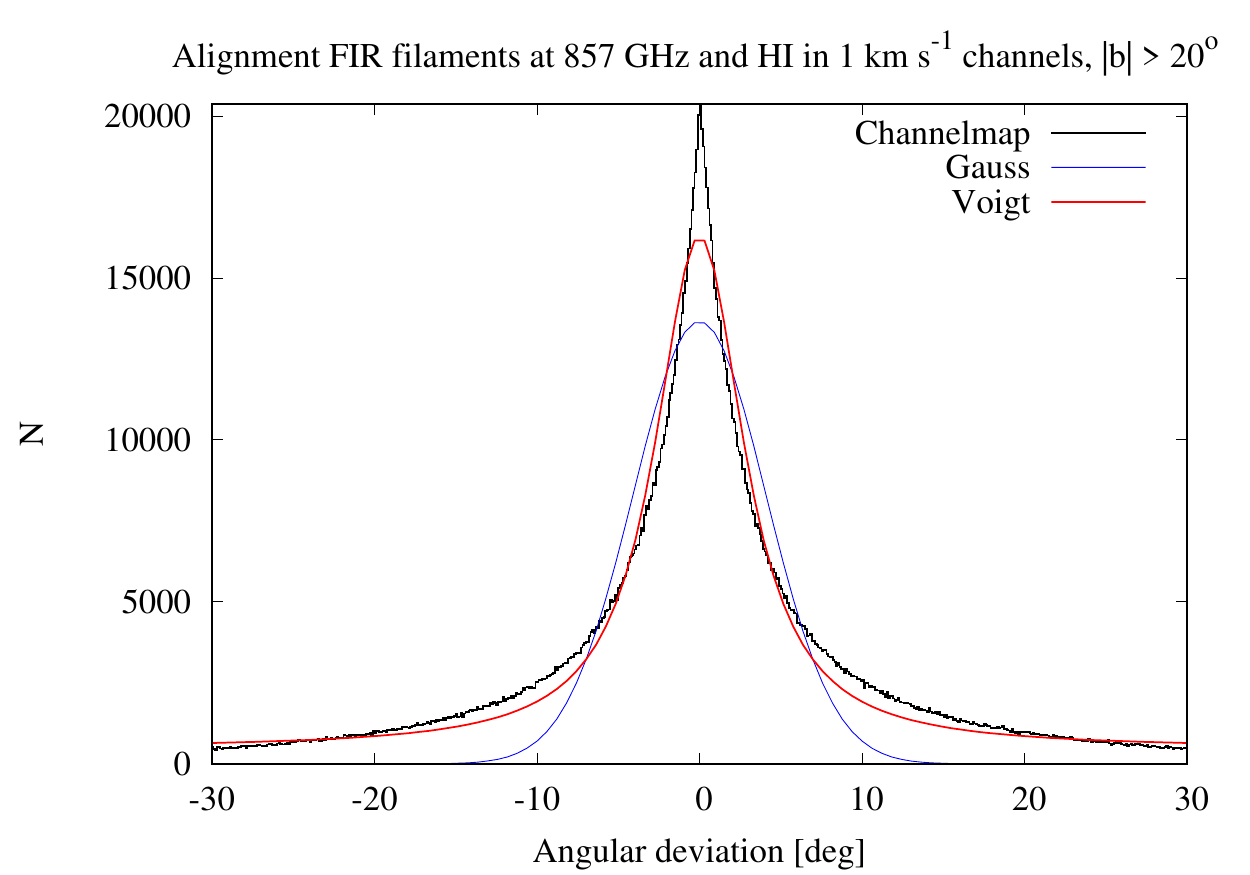}
   \includegraphics[width=9cm]{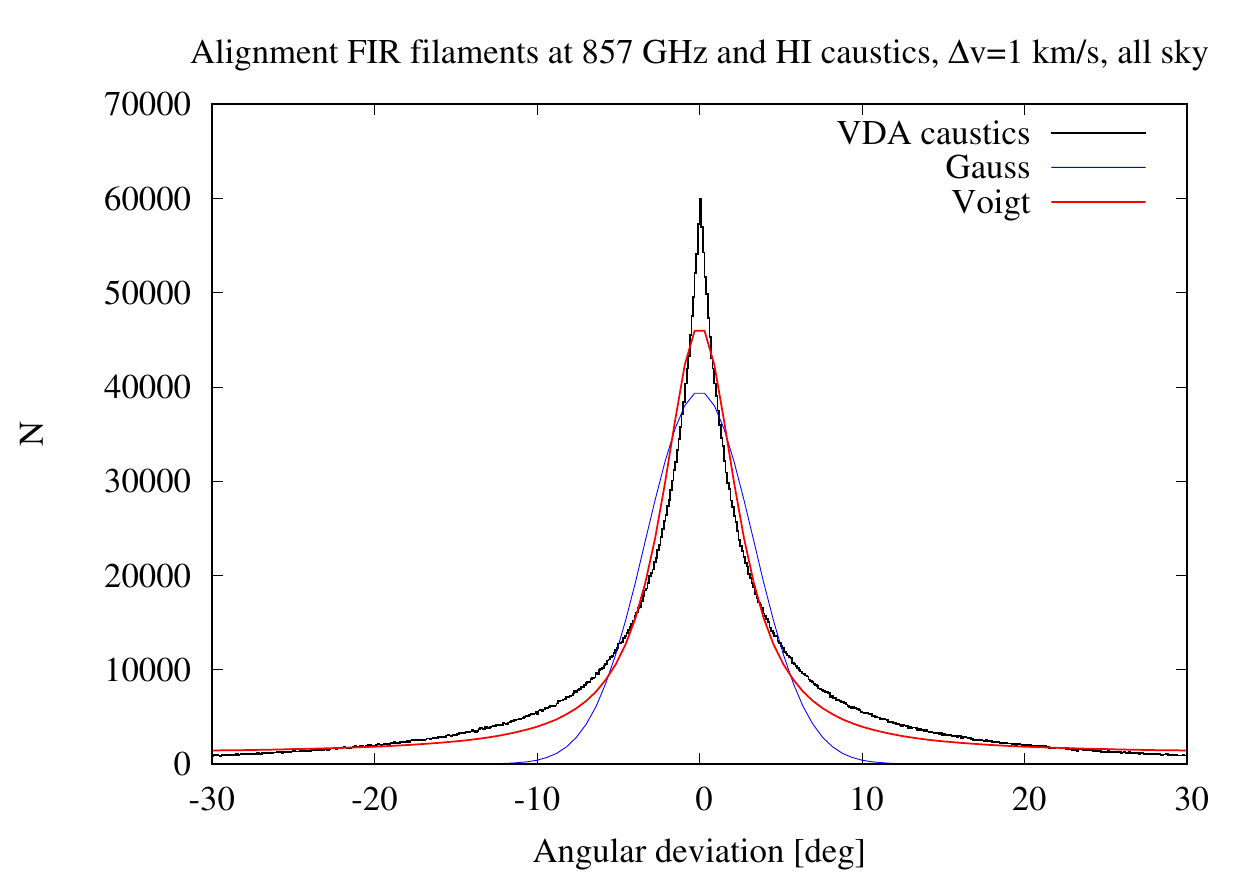}
   \includegraphics[width=9cm]{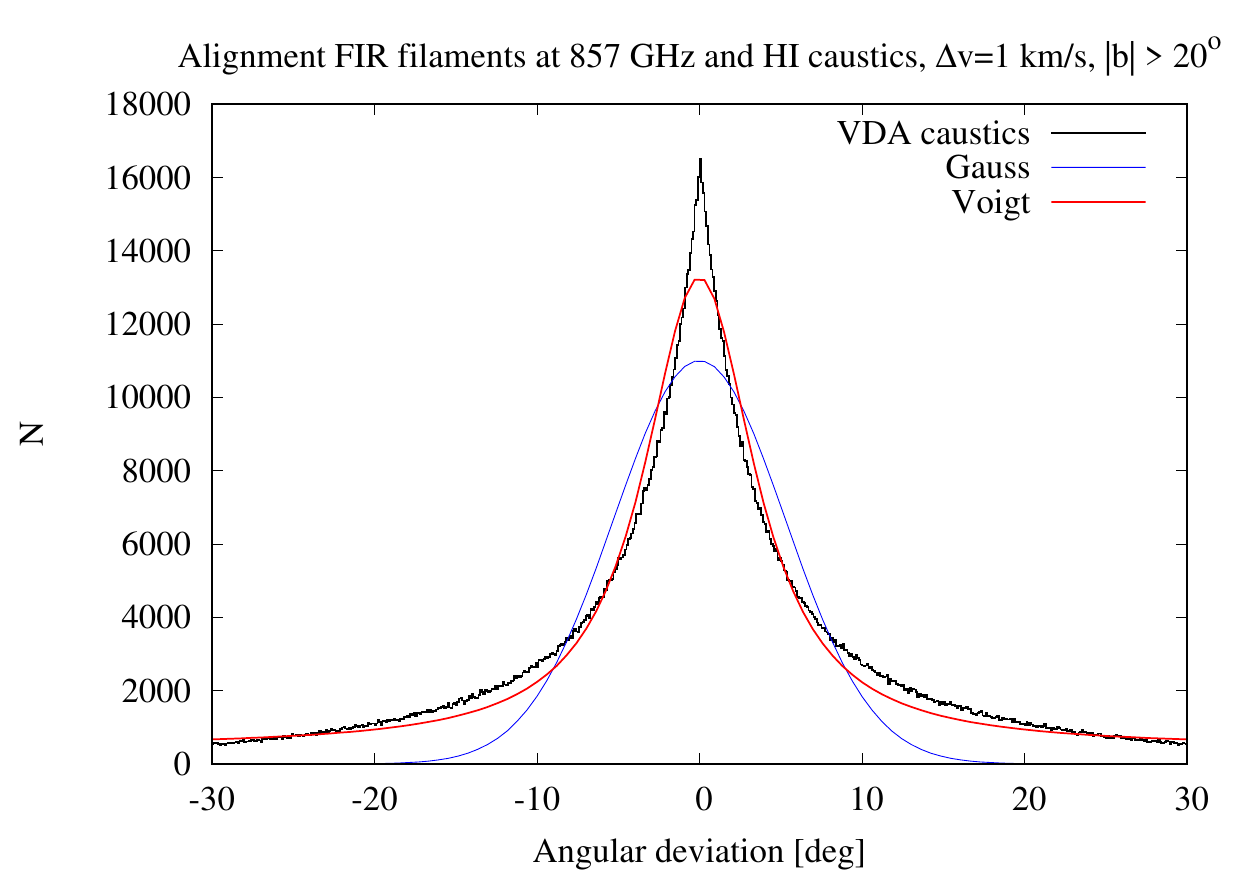}
   \includegraphics[width=9cm]{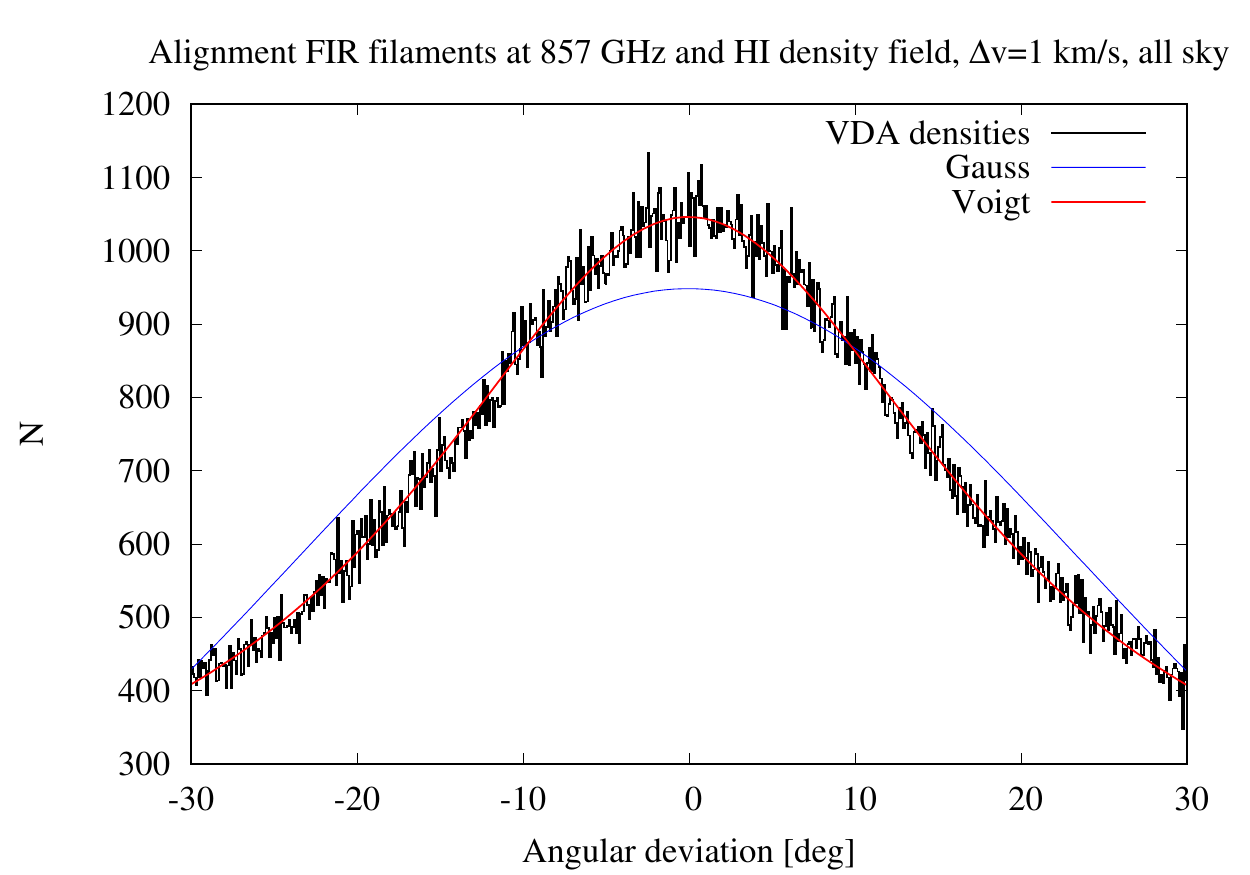}
   \includegraphics[width=9cm]{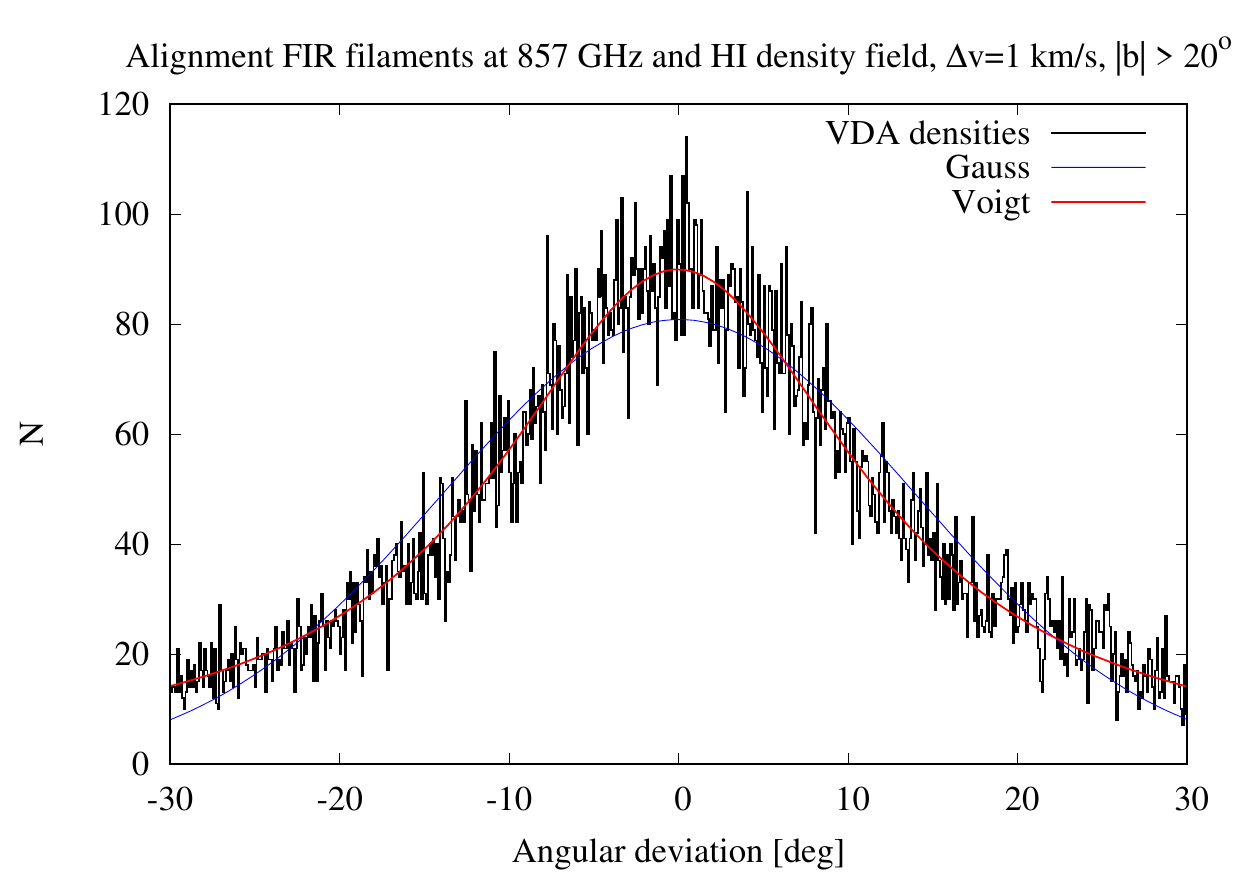}
   \caption{Histograms of angular alignment deviations according to
     Eq. (\ref{eq:angdif}) for filamentary structures. Top left: {\it
       Planck} 857 GHz compared with best fit single channel
     \hi\ filaments, all sky. Top right: {\it Planck} 857 GHz compared
     with best fit single channel \hi\ filaments, $|b| > 20
     \degr$. These two plots have been replicated from
     Fig. \ref{Fig_Aligne_857}. 
     Center left: {\it Planck} 857 GHz compared with best fit
     filaments from caustics $p_v$, all sky. Center right: {\it Planck} 857 GHz
     compared with best fit filaments from caustics $p_v$, $|b| > 20
     \degr$. Bottom left: {\it Planck} 857 GHz compared with best fit
     filaments from the VDA density distribution $p_d$, all sky. Bottom right:
     {\it Planck} 857 GHz compared with best fit filaments from the VDA
     density distribution $p_d$, $|b| > 20 \degr$. }
   \label{Fig_Aligne_VDA}
\end{figure*}

\begin{table*}
\caption{Filamentary alignment measures, for comparison with Table\ref{table:1}.}             
\label{table:A}      
\centering          
\begin{tabular}{c c c c c c c c c c }     
\hline\hline       
Data 1 & Data 2 & latitude & $f$ & \multicolumn{2}{c}{$\delta \theta$}  & $\xi$ & PRS & $\sigma_{\mathrm{PRS}}$ & Fig. \\ 
{\it Planck} 857 GHz & HI4PI & range & & $\sigma_{\mathrm{Gauss}}$ & $\sigma_{\mathrm{Voigt}}$ & & & \\
\hline                    
18\arcmin\ FWHM & \hi\ in 1 \kms\ channels &  all & .37 & 2\fdg6 & 1\fdg9
& 0.94 & 2873.67 & .17 & \ref{Fig_Aligne_VDA} top left\\  
18\arcmin\ FWHM &  VDA velocity field $p_v$ &  all & .37 & 3\fdg3 &
2\fdg5 & 0.94 & 2856.50 & .18 &  \ref{Fig_Aligne_VDA} middle left \\  
18\arcmin\ FWHM &  VDA density field $p_d$ &  all & .04 & 23\fdg8 & 20\fdg2
& 0.76 & 786.92 & .38 &  \ref{Fig_Aligne_VDA} bottom left \\
\hline
\hline
18\arcmin\ FWHM & \hi\ in 1 \kms\ channels &  $|b| > 20 \degr$ & .24 &
4\fdg1 & 3\fdg1 & 0.92 & 1817.64 & .23 & \ref{Fig_Aligne_VDA} top right \\
18\arcmin\ FWHM & VDA velocity field $p_v$ &  $|b| > 20 \degr$ & .23 &
5\fdg3 & 3\fdg9 & 0.91 & 1789.17 & .24 & \ref{Fig_Aligne_VDA} middle right \\
18\arcmin\ FWHM &  VDA density field $p_d$ &  $|b| > 20 \degr$ & .04 & 14\fdg0 &
13\fdg1 & 0.84 & 207.28 & .31 & \ref{Fig_Aligne_VDA} bottom right\\
\hline   
\end{tabular}
\end{table*}
%

\subsection{Alignment between FIR at 857 GHz and VDA derived structures}
\label{VDA_aligne}

We repeat the complete analysis described in Sect. \ref{Observations}
independently for VDA derived caustics $p_v$ and density structures
$p_d$ without modifying any of the program parameters. The basic
findings are summarized in Fig. \ref{Fig_Aligne_VDA} and Table
\ref{table:A}. To allow an easy comparison we replicate the previously
derived best fit entries from Fig. \ref{Fig_Aligne_857} and Table
\ref{table:1}.

The results are easy to summarize. The distributions of angular
alignment deviations according to Eq. (\ref{eq:angdif}) between
orientation angles of FIR filaments and \hi\ structures are very similar
if we use VDA caustics $p_v$ in place of \hi\ brightness
temperatures. The best fit result, presented in
Sect. \ref{Observations}, remains valid and is obtained by using
\hi\ channel maps with a velocity width of $\Delta v = 1$ \kms.  VDA
caustics $p_v$ have only a slightly ($\sim 30$\%) broader $\delta
\theta$ distribution compared to the best fit distribution obtained from
\hi\ channel maps, see the widths of the distributions in
Fig. \ref{Fig_Aligne_VDA} and the dispersions $\sigma_{\mathrm{Gauss}}$
and $\sigma_{\mathrm{Voigt}}$ in Table \ref{table:A}.  The angular
alignment deviations for the VDA density distribution $p_d$
(Fig. \ref{Fig_Aligne_VDA} bottom) are unacceptable large and the
filling factors $f$ in Table \ref{table:A} are very low. The VDA
densities $p_d$ are according to Eq. \ref{eq:pvd} derived from scaled
\hi\ column densities and we found already in Sect. \ref{Observations}
that integrating the \hi\ distribution does not lead to an improvement
in the angular alignment measures.

\begin{figure}[thp] 
   \centering
   \includegraphics[width=8.3cm]{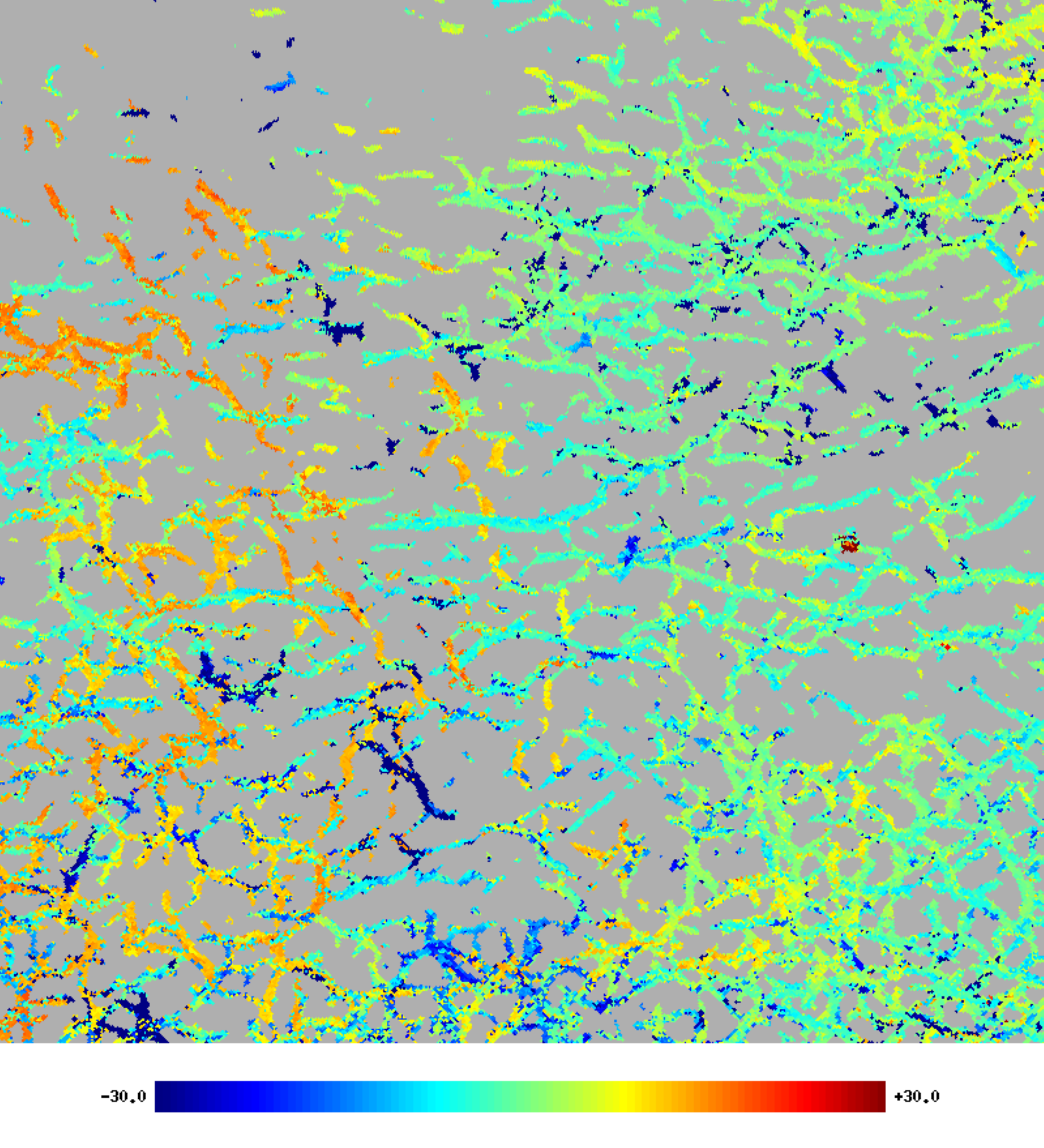}
   \includegraphics[width=8.3cm]{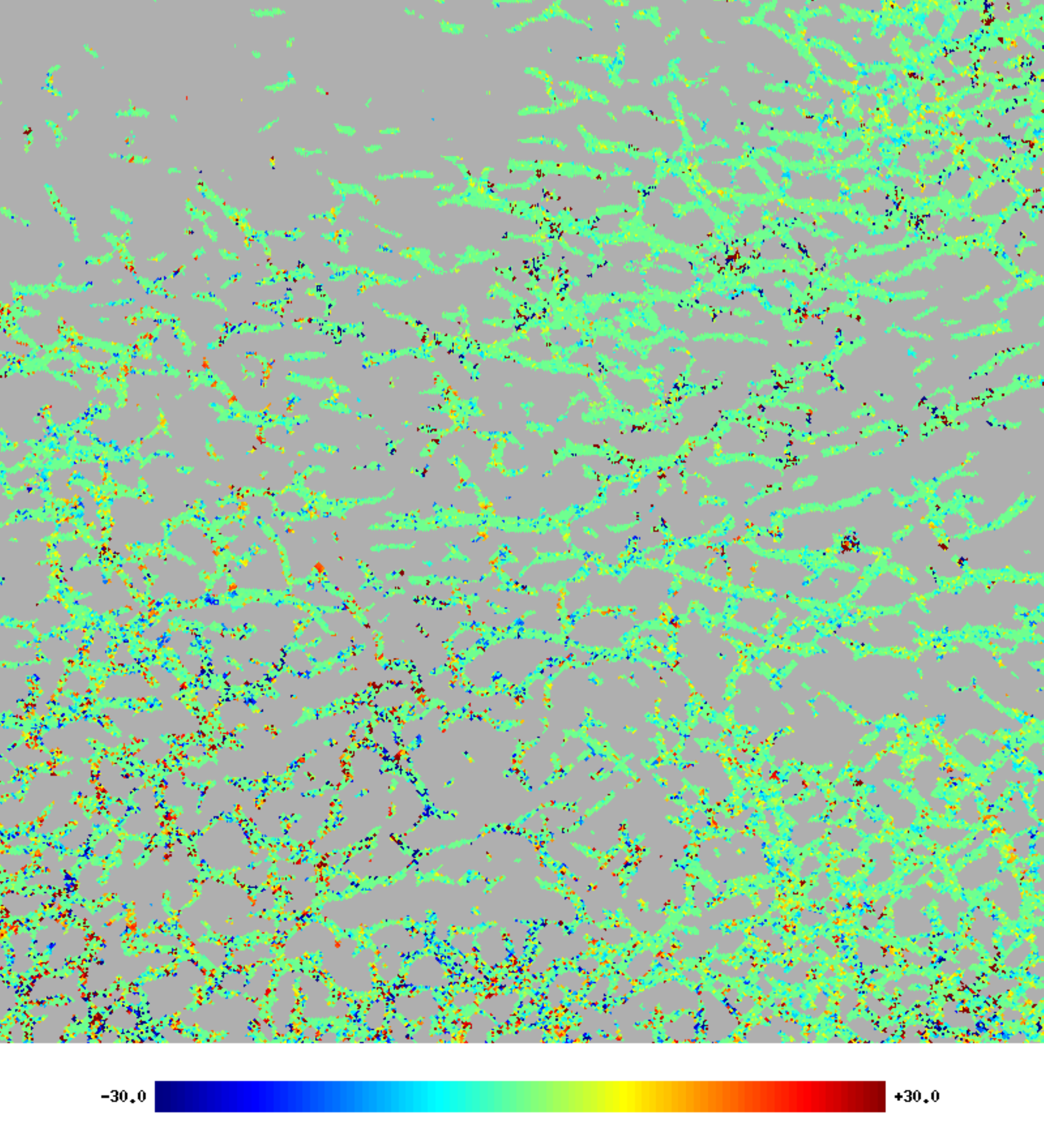}
   \caption{Top: Velocity field in the case of \hi\ filaments derived
     from velocity caustics $p_v$, to be compared with
     Fig. \ref{Fig_NoiseMaps}. Bottom: Velocity field of deviations
     between the solution described in Sect. \ref{v_fil} and the VDA $p_v$
     analysis in Sect. \ref{VDA_def}. }
   \label{Fig_NoiseMaps_VDA}
\end{figure}

The VDA based Hessian analysis from velocity caustics $p_v$ recovers in 97\%
of all cases identical positions for the derived filamentary
structures. For 43\% of the positions the filament velocities are
identical and for 53\% the differences are below 1 km/s. Allowing
uncertainties of $\mathcal{V} \la 6 $ \kms\ for the filament velocities
(see Sect. \ref{Vel_disp}) we get for 68\% of all filament positions
compatible velocities. To demonstrate VDA based velocity structures and
velocity uncertainties we display in Fig. \ref{Fig_NoiseMaps_VDA} on top
the velocity field for VDA caustics $p_v$ and below the deviations between the
previously derived velocity field shown in Fig. \ref{Fig_NoiseMaps} and
the one derived from caustics $p_v$. A close inspection of the lower plot
reveals that positions with significant differences in the derived
velocities are located predominantly on the filament outskirts. In
comparison to the filament centers these positions have lower S/N. We
notice also an increase of the velocity uncertainties toward the
Galactic plane, explainable with increasing confusion due to the
increasing complexity of the \hi\ emission.

\subsection{Implications}
\label{VDA_Implications}

Relating our results from the previous subsection to the main body of
this paper, we conclude that a Hessian analysis is able to detect
coherent \hi\ filaments in position and velocity that are associated
with FIR emission.  Our analysis leads to compatible results, regardless
whether \hi\ PPV channel maps $p$ with a channel width of 1 \kms\ are
used or VDA velocity caustics $p_v$ with a similar velocity
resolution. We infer from the fact that two different approaches lead
essentially to the same results that both methods necessarily must
describe the same objects, the \hi\ counterparts of FIR filaments.

Comparing interpretation and underlying assumptions of the different
methods we consider first the analysis of narrow \hi\ PPV channel
maps. Such data, all sky available from HI4PI, have been decomposed into
Gaussian components. The filaments are found to be associated with CNM,
implying that dusty filaments are cold. In pressure equilibrium CNM is
usually considered as a high volume density phase in equilibrium with a
surrounding low density WNM. The straight forward conclusion in the
framework of current equilibrium models \citep[e.g.,][]{Wolfire2003} is
that dusty filaments are local volume density structures, hence real
physical entities.  Such CNM structures can be considered as cold
filaments, embedded in a warmer medium \citep{Zucker2021}, objects with a
comparable depth to their width on the sky.  We refer in particular to
the interactive figures highlighting these
results\footnote[4]{\url{https://
 faun.rc.fas.harvard.edu/czucker/Paper_Figures/3D_Cloud_Topologies/gallery.html}}.
Distance determination, as emphasized by these authors, may be the key
to understand density structures.

A VDA decomposition intends to distinguish between velocity and density
structures in the turbulent \hi\ distribution. Velocity caustics $p_v$
are assumed to originate from velocity fluctuations along the line of
sight, causing a velocity crowding effect \citep{Lazarian2018}. Such
fluctuations can mimic structures that can be misinterpreted as actual
real physical objects in density. Filamentary structures may accordingly
not be interpreted as actual real physical objects. It is expected that
``structures in the channel maps are coming from contributions from
clouds in different physical locations along the line of sight''. A
corollary of this assertion is that caustics are not expected to be
associated with a particular \hi\ phase, temperature effects are not
expected. VDA density fields $p_d$ are theoretically expected to be
badly representative for filamentary structures \citep{Lazarian2018}.

\begin{figure}[thp] 
   \centering
   \includegraphics[width=8.3cm]{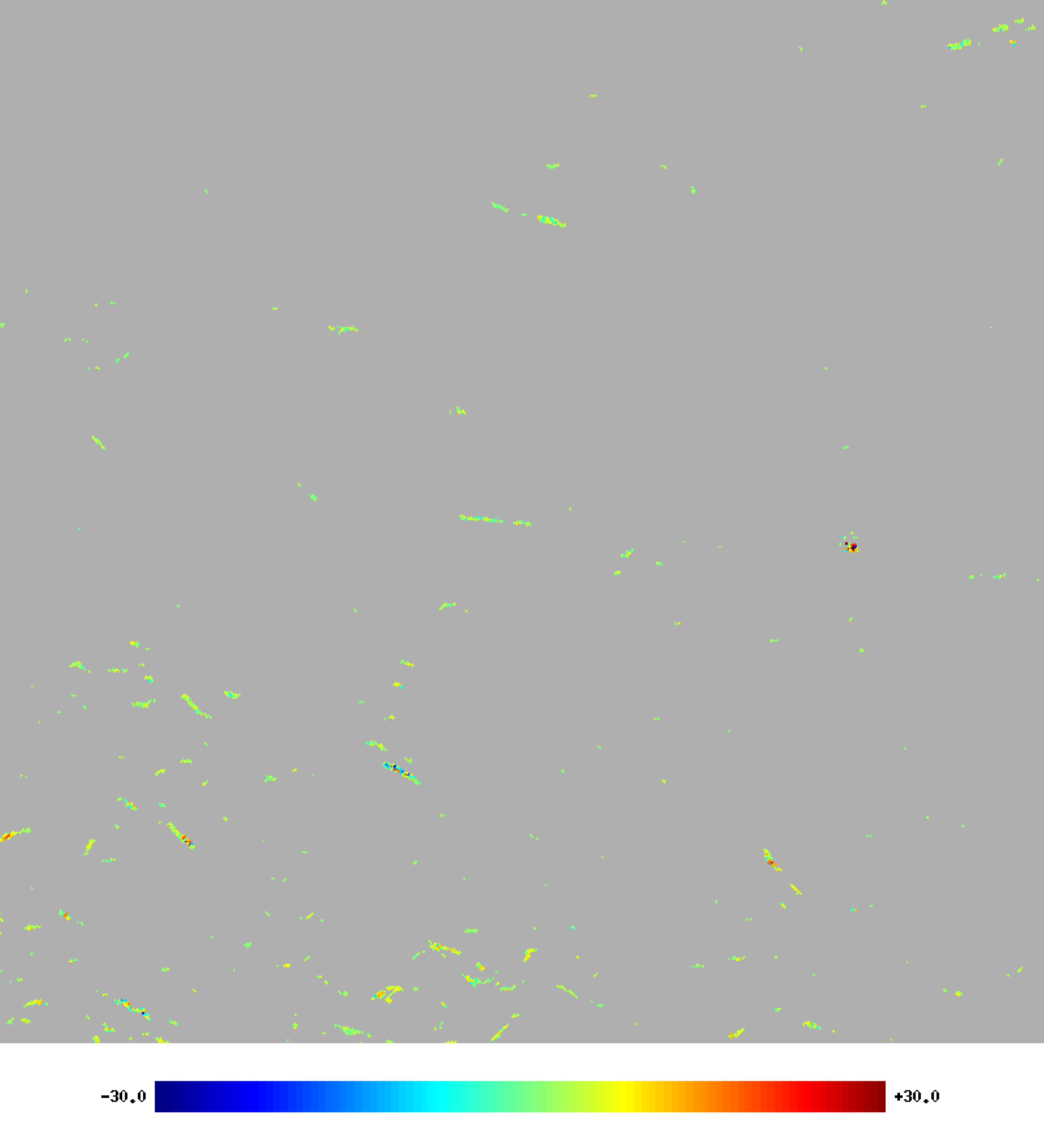}
   \caption{Filamentary structures and velocity field derived from VDA
     densities. To be compared with VDA caustics shown in
     Fig. \ref{Fig_NoiseMaps_VDA} top. }
   \label{Fig_NoiseMaps_VDA_dens}
\end{figure}

The VDA density field is per definition velocity independent. These
structures are according to Eq. \ref{eq:pvd} derived from scaled
\hi\ column densities and it is known that \hi\ column density
structures can reproduce the observed FIR emission distribution only
very crudely \citep{Planck2014}. We found already in
Sect. \ref{Observations} that integrating the \hi\ distribution leads to
a degradation of the angular alignment measures. Figure
\ref{Fig_NoiseMaps_VDA_dens} shows consistently that only 2.5\% of all
filament positions are recovered from the $p_d$ distribution.

Both remaining approaches - FIR filaments associated with \hi\ filaments
or alternatively caustics - are incompatible and should lead to
distinguishable results. The filament positions are however identical on
a 97\% level. It was emphasized by \citet{Yuen2021} in their Fig. 2 that
velocities of VDA caustics must differ from velocities of the
\hi\ emission enhancements, as well as from velocities of associated VDA
density structures. Contrary to this expectation we derive a consistent
and coherent velocity field along the filaments with insignificant
velocity deviations between both approaches (Fig. \ref{Fig_NoiseMaps_VDA}).
  
The theory in \citet{Lazarian2000} suggests that most of the structures
visible in ``thin channel maps'' (at low velocity width) should be due
to velocity caustics. Alternatively ``thick'' maps are assumed to be
dominated by density variations. In this situation it is necessary to
study what is happening with the filaments as the channel map thickness
increases. We considered this case in Sect. \ref{Observations}.
Correlations between FIR emission and \hi\ brightness temperatures are
clearly confined to narrow velocity intervals. The analytical study in
\citet{Lazarian2000,Lazarian2004} revealed also that ``channel maps are
sensitive only to velocity fluctuations if the corresponding turbulent
density fluctuations are dominated by large-scale contributions'' (for
detailed discussions see \citet{Lazarian2018}). Below we consider this
condition in some detail.

The obvious discrepancies between the interpretation of caustics
(velocity crowding) and real physical objects (local volume densities)
may be related to assumptions concerning the linear scales of the
fluctuations (sizes of eddies) under consideration. As discussed in
Sect. \ref{Curvature_dist}, our current assumption is that HI4PI
observations can resolve structures down to a lower linear scale of
approximately 1 pc, corresponding to the observed typical filament width
observable with a resolutions of 10\farcm8 for EBHIS and 14\farcm5 for
the GASS survey (see also \citet{Clark2014} and
\citet{Kalberla2016}). The curvature distribution of a small-scale
dynamo from Eq. (\ref{eq:kin}) can be fitted up to scales of
approximately 2\fdg5 (see Fig. \ref{Fig_C2}) and filamentary structures
are found to decouple in position on scales of $\ga 1\degr$\ 
(Sect. \ref{Envelope_pos}). A pc scale belongs therefore clearly to the regime
that is dominated by the fluctuation dynamo.

The estimate of scales dominated by velocity caustics was derived by
\citet[][Sect. 11.1]{Lazarian2018} from large-scale turbulence,
cascading down from an injection scale of 100 pc. The underlying
assumption is that turbulence is long wave dominated with a steep
velocity spectrum and a spectral index $\gamma \la -3 $. A short wave
dominated or shallow velocity spectrum with $\gamma \ga -3 $ is
considered not to be physically motivated \citep{Chepurnov2009}. Based
on that \citet{Lazarian2018} conclude that on scales larger than 3 pc
the intensity of fluctuations in the channel maps is produced to a
significant degree by velocity caustics rather than real physical
entities, i.e., filaments. A direct application of this model assumption
leads to a statistical distribution of \hi\ emitters along the line of
sight on scales that are large in comparison to the observed narrow
diameters of the filamentary structures (hence ``different physical
locations along the line of sight''). Such a model assumption
invalidates a description of filaments with an average radial volume
density profile around 3D spines as recently advocated by
\citep{Zucker2021}.

\begin{figure}[thp] 
   \centering
   \includegraphics[width=9cm]{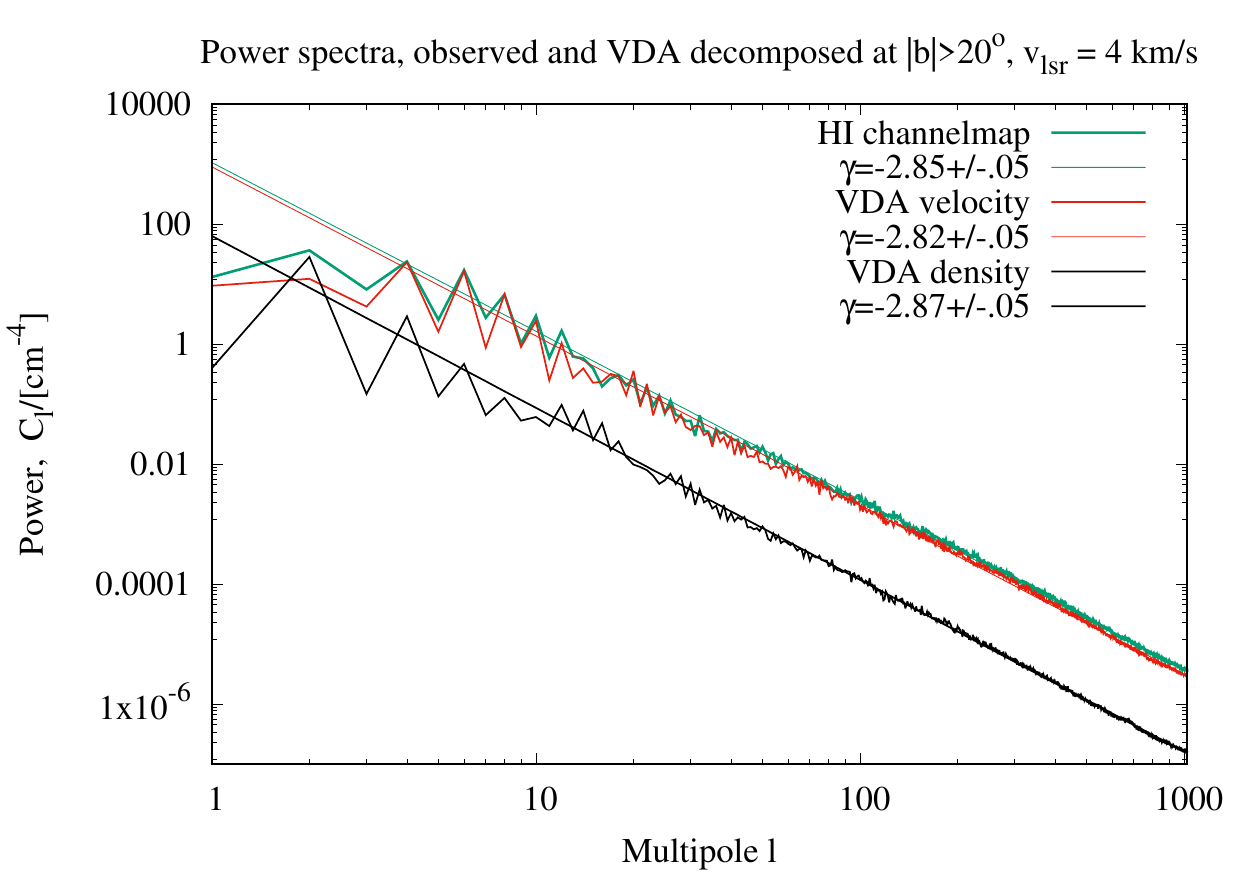}
   \caption{Turbulent power spectra for the observed \hi\ emission at
     high Galactic latitudes at a velocity of $ v_{\mathrm{LSR}} = 4 $
     \kms, compared with power spectra derived from VDA velocity and
     density contributions.}
   \label{Fig_power}
\end{figure}

To test whether observed power spectra are short or long wave dominated
we calculate turbulence power spectra and search for the velocity with
steepest velocity power spectrum derived from HI4PI data (see
\citet{Kalberla2019}, Fig. 9). For the VDA velocity power spectrum we
obtain $\gamma_{\mathrm{v}} = -2.82 \pm 0.05 $ at a velocity of $
v_{\mathrm{LSR}} = 4 $ \kms, almost identical with
$\gamma_{\mathrm{chan}} = -2.85 \pm 0.05 $ from the observed
\hi\ channel map at the same velocity, see Fig. \ref{Fig_power}. The
spectral index $\gamma_{\mathrm{d}} = -2.87 \pm 0.05 $ for the VDA
density is per definition velocity independent. We conclude that HI4PI
and VDA power spectra of the multiphase \hi\ medium are all shallow or
short wave dominated, $\gamma \ga -3 $. Filamentary structures are
dominated by the CNM and the spectral indices in this regime are even
shallower, $\gamma_{\mathrm{CNM}} \ga -2.2 $ \citep{Kalberla2019}. Dust
models for the Galactic pole regions result in spectral indices
$\gamma_{\mathrm{dust}} \sim -2.4 $ \citep{Ghosh2017,Adak2020}. Shallow
indices with $\gamma \ga -3 $ appear to be typical for the ISM at high
Galactic latitudes.

The shallow CNM spectral index implies that there is more turbulent
energy associated with filaments on small scales, in clear contradiction
of the long wave dominance assumed by \citet{Chepurnov2009} and
\citet{Lazarian2018}.

In the main body of this paper we found that FIR emission is dominated
by the CNM. This CNM is expected to be coupled to ions and the magnetic
field on scales larger than 0.04 pc for slow modes or 0.005 pc in case
of Alfv\'en modes \citep{Xu2017}. This is far below the HI4PI resolution
limit. Our current results support model assumptions with filamentary
\hi\ structures shaped by dynamo-generated small-scale fields. This
differs from large-scale dynamo theories that are usually handled in the
mean-field framework \citep[e.g.][]{Parker1979}. Mean-field theories
assume that small-scale magnetic fluctuations result from the shredding
of the mean field by the turbulence. The Kolmogorov type cascade assumed
by \citet{Lazarian2018} is not relevant for regions dominated by a small
scale dynamo \citep[][Sect. 1.2]{Schekochihin2004}. Strong fluctuations
within the CNM, causing caustics, are rather expected on small scales.

In case of the small-scale dynamo the magnetic field is amplified on
scales below the viscous scale (where the turbulent kinetic energy is
dissipated by viscous forces, we refer to Fig. 1 of
\citet{Schekochihin2004}). This assertion is confirmed by
\citet{Kriel2022}; the peak magnetic field scale depends linearly on the
resistive magnetic field scale. In consequence the magnetic field shapes
the CNM even on scales far below the HI4PI angular resolution limit. A
velocity crowding on such small scales would not be in conflict with
HI4PI observations. Emitters at different physical positions along the
line of sight but with the same radial velocities would then be located
within the filaments. The turbulent motions are in this case confined to
the dense CNM. Turbulent motions affect the observed velocity
distribution, a process that is usually described as turbulent line
broadening (see Eq. (4) but also Sect. 6 of \citep{Yuen2021}).  A close
agreement of the angular alignment between FIR and both, \hi\ emission
and VDA caustics, would be a natural consequence.

In this interpretation VDA caustics are related to real \hi\ structures
and cause just a modification of the observed density weighted emission
from the filaments. Random motions cause some scatter of the VDA derived
orientation angles $\Theta$, in consequence the angular alignment
between FIR and \hi\ filaments is degraded. Setting the focus on VDA
caustics does not change the results significantly. The limited 30\%
broadening of the angular alignment distribution from VDA velocity
fields in comparison to \hi\ channelmap data (Fig. \ref{Fig_Aligne_VDA})
implies that the VDA derived velocity field is spatially well correlated
with the filaments in \hi\ and FIR but does not dominate the observed
\hi\ structures. VDA caustics share positions and velocities of FIR
filaments almost as precise as \hi\ filaments do. 

Filamentary structures, shaped by a small-scale dynamo, may be
responsible for the so far unexplained tiny-scale atomic structures
(TSAS). These structures appear to be overdense and overpressured
\citep{Stanimirovic2018}, with densities and pressures that are orders
of magnitude higher than the theoretical values based on the heating and
cooling balance. \citet{Deshpande2000} tried to explain TSAS by an
extension of the power law spectrum known on large scales. The shallow
power law index of $\gamma_{\mathrm{TSAS}} = -2.75 $, determined over
scales between 0.02 to 4 pc by \citet{Deshpande2000} fits well to the
steepest average index of $\gamma \ga -2.87 $ from HI4PI.

\subsection{Summary}
\label{VDA_Summary}

Our analysis, using \hi\ channel maps and alternatively VDA velocity
fields, leads to comparable results. Observed intensity structures,
common in FIR and \hi, can be interpreted either as real physical
entities (filaments) or as spurious structures, caused by velocity
crowding effect. The first interpretation is based on model assumptions
concerning structure formation in a multiphase medium, the second one on
turbulence models. Using \hi\ channel maps or VDA derived velocity
components in our analysis leads however to essentially identical
results.  Both approaches describe the same objects, \hi\ counterparts
of FIR filaments. Thus caustics are linked to real physical entities
with properties described in the main body of this paper. 

We explain the apparent discrepancies, discussed by \citet{Clark2019a}
and \citet{Lazarian2018}, with a predominance of small scale
fluctuations caused by a fluctuation dynamo. Restricting turbulent
fluctuations to small scales within the filaments confines velocity
crowding effects to CNM structures. Caustics represent in this case
predominantly fluctuations within a denser \hi\ medium. Velocity
crowding thus causes a modulation of observed column densities but does
not generate significant spurious filamentary structures.  Caustics from
distinct different physical locations along the line of sight, advocated
by \citet{Lazarian2018}, are inconsistent with our results.  We question
the interpretation that caustics are caused by fluctuations on large
scales, associated with steep spectral indices. Velocity crowding from
large scale caustics cannot be made responsible for an enhanced FIR
emission in \hi\ filaments with well defined coherent velocity
structures. VDA applications, discussed in this Appendix, do not alter
any of the results and conclusions from the main body of this paper.

\end{appendix}

\end{document}